%% file: main.tex
\documentclass[sigconf]{acmart}
\AtBeginDocument{%
  \providecommand\BibTeX{{%
    \normalfont B\kern-0.5em{\scshape i\kern-0.25em b}\kern-0.8em\TeX}}}

\input{lib}

\copyrightyear{2024}
\acmYear{2024}
\setcopyright{rightsretained}
\acmConference[CHI '24]{Proceedings of the CHI Conference on Human Factors in Computing Systems}{May 11--16, 2024}{Honolulu, HI, USA}
\acmBooktitle{Proceedings of the CHI Conference on Human Factors in Computing Systems (CHI '24), May 11--16, 2024, Honolulu, HI, USA}
\acmDOI{10.1145/3613904.3642420}
\acmISBN{979-8-4007-0330-0/24/05}

\title{Understanding the Impact of Long-Term Memory on Self-Disclosure with Large~Language Model-Driven Chatbots for Public Health Intervention}

\acmSubmissionID{5759}

\begin{document}

\settopmatter{authorsperrow=3} 

\author{Eunkyung Jo}
\orcid{0000-0002-6494-3396}
\authornote{Eunkyung Jo conducted this work as a research intern at NAVER AI Lab.}
\affiliation{%
  \institution{University of California, Irvine}
  \country{United States}}
\email{eunkyuj@uci.edu}

\author{Yuin Jeong}
\orcid{0009-0007-0138-1756}
\affiliation{%
  \institution{NAVER Cloud}
  \country{Republic of Korea}}
\email{youin.jeong@gmail.com }

\author{SoHyun Park}
\orcid{0000-0001-8703-0584}
\affiliation{%
  \institution{NAVER Cloud}
  \country{Republic of Korea}}
\email{sohyun.s.park@navercorp.com}

\author{Daniel A. Epstein}
\orcid{0000-0002-2657-6345}
\affiliation{%
  \institution{University of California, Irvine}
  \country{United States}}
\email{epstein@ics.uci.edu}

\author{Young-Ho Kim}
\orcid{0000-0002-2681-2774}
\affiliation{%
  \institution{NAVER AI Lab}
  \country{Republic of Korea}}
\email{yghokim@younghokim.net}

\renewcommand{\shorttitle}{Understanding the Impact of Long-Term Memory on Self-Disclosure with LLM-Driven Chatbots}
\begin{abstract}
\input{sections/00-abstract}

\end{abstract}

\ccsdesc[500]{Human-centered computing~Empirical studies in HCI}
\ccsdesc[500]{Human-centered computing~Natural language interfaces}
\ccsdesc[300]{Computing methodologies~Natural language generation}
\keywords{Chatbot, Large language models, Open-domain dialog systems, Long-term memory, Public health, Check-up calls, Social isolation}


\maketitle

\input{sections/01-introduction}

\input{sections/02-related-work}
\input{sections/03-background}

\input{sections/04-method}
\input{sections/05-findings}
\input{sections/06-discussion}
\input{sections/07-conclusion}

\begin{acks}
We thank our participants for their sincere participation. We are grateful to the members of the Generative Chatbot team at NAVER Cloud, especially Sanghwan Bae, Donghyun Kwak, and Soyoung Kang, for helping us understand the technological details and design rationales behind the CareCall LTM feature. We thank Sang-houn Ok for his assistance with recruiting study participants, Hanju Jo for helping out with extracting call logs, and Myeonghan Ryu for his support in writing scripts to count codes. In addition, we thank the members of inComms for their help with transcribing and coding the call logs. This work was supported through a research internship at NAVER AI Lab and by the National Science Foundation under Award IIS-2237389.
\end{acks}

\bibliographystyle{ACM-Reference-Format}
\bibliography{bibliography}


\appendix
\input{sections/10-appendix-ltm}

\end{document}

%% file: lib.tex
\usepackage{multirow}
\usepackage{color}
\usepackage{colortbl}
\usepackage{stfloats}
\usepackage{longfbox} 
\usepackage{subcaption}
\usepackage{enumitem}
\usepackage{adjustbox}
\usepackage{graphicx}
\usepackage{subcaption}
\usepackage{tabularx}

\makeatletter
\newdimen\@tempdimd
\makeatother

\makeatletter
\def\@cline#1-#2\@nil{%
  \omit
  \@multicnt#1%
  \advance\@multispan\m@ne
  \ifnum\@multicnt=\@ne\@firstofone{&\omit}\fi
  \@multicnt#2%
  \advance\@multicnt-#1%
  \advance\@multispan\@ne
  \leaders\hrule\@height\arrayrulewidth\hfill
  \cr
  \noalign{\nobreak\vskip-\arrayrulewidth}}
\makeatother

\newcommand{\eg}{\textit{e.g.}}
\newcommand{\ie}{\textit{i.e.}}
\newcommand{\cf}{\textit{c.f.}}
\newcommand{\etal}{\textit{et al.}}

\definecolor{ltmyescolor}{HTML}{afe3e1}
\definecolor{ltmnocolor}{HTML}{dadada}
\definecolor{interviewcolor}{HTML}{a4e0c3}

\newfboxstyle{patternparam}{padding=1pt, padding-left=2pt, padding-right=2pt, margin-bottom=0pt, margin-top=0pt, border-style=none, height=7pt, border-radius=2pt}
\newcommand{\ltmyesbox}[1]{\lfbox[patternparam, background-color=ltmyescolor]{#1}}
\newcommand{\ltmnobox}[1]{\lfbox[patternparam, background-color=ltmnocolor]{#1}}

\newcommand{\interviewbox}[1]{\lfbox[patternparam, background-color=interviewcolor]{#1}}

\definecolor{eunkyung}{RGB}{27,158,119}
\definecolor{daniel}{RGB}{2,95,217}
\definecolor{youngho}{RGB}{102,166,30}

\definecolor{revisedcolor}{RGB}{0,0,255}

\newcommand{\revised}[1]{#1}

\newcommand{\ipstart}[1]{\vspace{1mm} \noindent{\textbf{\textit{#1.}}}}

\newcommand{\circledigit}[1]{\textbf{\normalsize{\textsf{\textcircled{\footnotesize{#1}}}}}}

\definecolor{tableheader}{HTML}{EFEFEF}
\definecolor{tablegrayline}{HTML}{d0d0d0}

\newcommand{\siglegend}{\sffamily\footnotesize{\textbf{***}$p$<.001; \textbf{**}$p$<.01; \textbf{*}$p$<.05}}

\newcommand{\labelphantom}[1]{%
  \parbox{0pt}{\phantomsubcaption\label{#1}}%
}

\newcommand{\ltmyes}{\ltmyesbox{$LTM^{yes}$}}
\newcommand{\ltmno}{\ltmnobox{$LTM^{no}$}}

\newcommand{\ltmyesperson}[1]{\ltmyesbox{$LTM^{yes}$-#1}}
\newcommand{\ltmnoperson}[1]{\ltmnobox{$LTM^{no}$-#1}}

\newcommand{\interviewperson}[1]{\interviewbox{P#1}}

\newenvironment{quotetable}{
\vspace{3mm}
\hfill\break
\noindent
    \small\sffamily\centering
    \def\arraystretch{1.2}
    \tabularx{\columnwidth}{p{0.17\columnwidth}ll}
    \hline}
{
    \arrayrulecolor{black}\hline
    \endtabularx
    \vspace{3mm}
}

\newcommand{\quotecarecallline}[2]{CareCall & \begin{minipage}{0.77\columnwidth}\vspace{1mm}\textit{#1} \textbf{#2} \vspace{1mm}\end{minipage}\\\arrayrulecolor{tablegrayline}\hline}

\newcommand{\quoteuserline}[2]{\ltmyesperson{#1} & \begin{minipage}{0.77\columnwidth}\vspace{1mm}\textit{#2}\vspace{1mm}
\end{minipage}\\\arrayrulecolor{tablegrayline}\hline}

\newcommand{\quotenouserline}[2]{\ltmnoperson{#1} & \begin{minipage}{0.77\columnwidth}\vspace{1mm}
\textit{#2}\vspace{1mm}
\end{minipage}\\\arrayrulecolor{tablegrayline}\hline}

%% file: sections/00-abstract.tex
Recent large language models (LLMs) offer the potential to support public health monitoring by facilitating health disclosure through open-ended conversations but rarely preserve the knowledge gained about individuals across repeated interactions. Augmenting LLMs with long-term memory (LTM) presents an opportunity to improve engagement and self-disclosure, but we lack an understanding of how LTM impacts people's interaction with LLM-driven chatbots in public health interventions. We examine the case of CareCall---an LLM-driven voice chatbot with LTM---through the analysis of 1,252 call logs and interviews with nine users. We found that LTM enhanced health disclosure and fostered positive perceptions of the chatbot by offering familiarity. However, we also observed challenges in promoting self-disclosure through LTM, particularly around addressing chronic health conditions and privacy concerns. We discuss considerations for LTM integration in LLM-driven chatbots for public health monitoring, including carefully deciding what topics need to be remembered in light of public health goals.

%% file: sections/01-introduction.tex
\section{Introduction}
Public health monitoring plays a pivotal role in safeguarding and promoting the health and wellbeing of populations by tracking various factors for containing the spread of infectious diseases~\cite{lu2021comparing, smolinski_flu_2015} and promoting awareness of the public's wellbeing routines~\cite{gordon_goal-setting_2019, shameli_how_2017}. Across different contexts, public health monitoring often requires recurrent data collection from populations, demanding immense time and effort from public health authorities~\cite{ismail2021ai, ismail_engaging_2018, ismail_public_2023, okolo2021cannot}. For instance, a press article estimated that contact tracers during COVID-19 called about ten contacts per case and followed up with each contact every other day, expecting that 37 out of 50 states in the United States did not have enough contact tracers to effectively contain the spread of the virus in its early stages~\cite{NPR_2020}. To reduce the burden of public health authorities in monitoring people at scale, technology, such as chatbots and mobile apps, has increasingly been used or proposed to automate aspects of the collection of personal health information for various public health monitoring contexts, including contact tracing~\cite{lu2021comparing, lu2022understanding, kim_designing_2022}, maternal health education~\cite{yadav2019feedpal, ismail2021ai, ismail_engaging_2018, ismail_public_2023, okolo2021cannot}, and social isolation intervention~\cite{Jo2023}.

A key challenge in leveraging technology for public health monitoring is how to elicit health disclosure effectively from individuals. People often avoid revealing their vulnerabilities~\cite{corrigan_2004, vogel_2003}, which makes them hesitant to disclose sensitive information, such as their health issues. Further, while public health monitoring often requires maintaining regular engagement, it is more challenging to sustain health disclosure over time~\cite{kreiner_2019, lee_i_2020}. Research has suggested that chatbots can be potentially effective at eliciting self-disclosure about sensitive topics~\cite{fitzpatrick_delivering_2017, lucas_reporting_2017, lee_designing_2020, park_i_2021} as they are often perceived as non-judgemental~\cite{kang_virtual_2010, lucas_its_2014, sundar_2019}. Recent advances in large language models (LLMs) have brought breakthroughs in chatbots' abilities to support free-form conversations on open-ended topics, offering the potential to be particularly effective at eliciting disclosure about broader aspects of personal health. However, current LLM-driven chatbots rarely offer the capability to store and reference information from previous sessions. As a result, when introduced to support public health monitoring, LLM-driven chatbots often struggled with following up on personal health history based on past conversations~\cite{Jo2023}, potentially reducing user engagement and self-disclosure needed for public health monitoring.

To support continued interactions, research in Natural Language Processing (NLP) has increasingly attempted to augment LLMs with \textit{long-term memory} (LTM), an ability to remember information, such as an individual's interests or health status, beyond the current conversation session\footnote{In this paper, we follow the definition of long-term memory that is commonly used in the NLP community~\cite{bae2022keepmeupdated, wang2023augmenting, xu_beyond_2021, xu_long_2022, zhong2023memorybank}, which draws parallels to the one in human cognition (\eg,~\cite{wang2023augmenting, zhong2023memorybank}). Note that in our study context, the term \textit{`long-term'} is not confined to a specific time frame. This contrasts with frequent conversations in HCI around the length of deployment of technology interventions, such as evaluating long-term engagement or behavior change~\cite{klasnja_2011, karapanos_advances_2021}.}. LLM-driven chatbots with LTM can store information about certain topics from each session and feed it to the model along with other input in each conversation turn, providing cues for the chatbot to refer to. \revised{Some exemplary LLM-driven chatbots such as OpenAI's ChatGPT~\cite{ChatGPT} and Google Bard~\cite{Bard} have recently introduced features like \textit{``memory''} or \textit{``custom instructions''} which statically retain certain user background information or stylistic preferences across sessions and globally apply them to the chatbot behaviors~\cite{ChatGPTCustomInstruction}. However, in public health contexts, it is essential to dynamically update the memory to reflect users' constantly changing health statuses and generate responses accordingly. LLM-driven chatbots with LTM, particularly when designed to dynamically store, update, and reference information from previous sessions, have the potential to promote disclosure in public health contexts by offering personalized conversations that bring up specific health concerns one may have.} Such a memory capability also offers a valuable opportunity to maintain user engagement in public health settings, which is crucial for understanding of the population's health and wellbeing. Despite the potential of LTM to support public health monitoring through chatbots, there is a limited understanding of how LTM impacts user interactions with and perception of LLM-driven chatbots, particularly in the context of public health interventions. In this study, we therefore seek to understand the utility of LTM for public health monitoring, with particular attention to self-disclosure. 

To understand the impact of LTM on people's health disclosure and their impressions of chatbots, we explore the case of CareCall, an LLM-driven voice chatbot that monitors socially isolated individuals' health via check-up phone calls. CareCall was designed to support public health monitoring and used in conjunction with public health agencies in local municipalities in South Korea. Initially developed without LTM, CareCall later integrated this capability, which automatically stores and updates summarized information about five topics---~\textit{Health, Meals, Sleep, Visited Places, and Pets}---from each session and feeds it to the LLM in the following sessions to provide cues for the chatbot to refer to appropriately. The unique setting provides a useful case for comparing user experiences with and without LTM, particularly around self-disclosure. Through the case of CareCall, we specifically ask: \textbf{How does LTM impact users' self-disclosure with and impressions of LLM-driven chatbots in the context of public health intervention?} 

To answer this question, we quantitatively and qualitatively analyzed 1,252 call logs from the real-world deployment for two distinct user groups---those who engaged with CareCall with LTM ($N=66$) and those without it ($N=81$). To understand users' perceptions of LTM in public health monitoring chatbots, we supplemented the call log analysis with interviews with nine users who used CareCall with LTM. We found that those who used CareCall with LTM disclosed more details about their health compared to those who used CareCall without it. In addition, repeated experiences with LTM led to greater disclosure with CareCall. Users often showed more positive and less negative reactions toward the chatbot as they experienced more LTM events, perceiving such conversations as personal and emotionally supportive. However, we observed some challenges of LTM in promoting self-disclosure, leaving some users less engaged with repeated questions on chronic health issues and raising privacy concerns around sensitive health topics.

Based on the findings, we discuss design opportunities for LTM integration in LLM-driven chatbots for public health monitoring. We propose that designers should carefully decide the topics for LTM to remember, balancing the need to support open-ended conversations about broader aspects of people's lives and adhere to specific public health monitoring goals. We also highlight LTM's potential to mitigate the impersonality of chatbots and demonstrate care. Lastly, we suggest the need to consider tensions in memory needs for public health utility versus privacy sensitivity. 

\vspace{5pt}\noindent{}Key contributions of this work include:
\begin{itemize}[leftmargin=*, itemsep=4pt, topsep=0pt]
\item An empirical understanding of the impact of LTM on people's health disclosure and their impressions of LLM-driven chatbots for public health intervention, gained from analysis of 1,252 call logs from the real-world deployment of CareCall with and without LTM and interviews with nine users who were using the chatbot with LTM. The implementation and repeated experiences of LTM in CareCall led to greater health disclosure. LTM also fostered positive impressions of the chatbot by offering familiarity, leading users to perceive the system as personal and emotionally supportive. However, we observed some challenges of LTM in promoting self-disclosure, particularly around chronic health issues and privacy concerns. 
\item Implications for future research and implementation of LTM in LLM-driven chatbots for public health monitoring, which include: (1) designing LTM with careful topic selection in light of public health monitoring goals; (2) designing thoughtful LTM-triggered questions to mitigate the impersonality of chatbots and demonstrate care; and (3) balancing the memory needs for public health utility and privacy sensitivity. 
\end{itemize}

%% file: sections/02-related-work.tex
\section{Related Work}
Understanding the impact of LTM on LLM-driven chatbots in public health builds on past work on chatbots for enhancing self-disclosure on health and chatbots driven by LLMs and LTM augmentation.

\subsection{Chatbots for Enhancing Self-Disclosure on Health}
Self-disclosure, a process where a person reveals personal or sensitive information to others~\cite{altman_irwin_social_1973}, is associated with many benefits, such as stress relief~\cite{ho_2018} and closer relationships between conversational partners~\cite{altman_irwin_social_1973}. However, eliciting disclosure, particularly on sensitive topics, poses challenges due to people's reluctance to reveal vulnerabilities~\cite{corrigan_2004, vogel_2003}. Prior work indicated that people might feel more comfortable disclosing sensitive information to a digital system rather than a human because technology is often perceived as anonymous~\cite{lucas_reporting_2017, lu2021comparing, lu2022understanding} and non-judgemental~\cite{kang_virtual_2010, lucas_its_2014, sundar_2019, lu2022understanding}. In particular, chatbots can make people feel more comfortable expressing themselves, given their dialogue-driven nature. Prior work has thus frequently proposed chatbots as a promising approach to encourage self-disclosure on sensitive topics, including mental health~\cite{fitzpatrick_delivering_2017, lee_designing_2020, lucas_reporting_2017, park_designing_2019} and sexual health~\cite{liu_2018}. Studies have explored how different designs of chatbots might impact people's self-disclosure. For example, chatbots that disclose information about themselves have been shown to foster mutual self-disclosure~\cite{lee_designing_2020, lee_i_2020, ravichander_empirical_2018}. However, prior work predominantly relied on one-time user studies, limiting our understanding of how chatbots can sustain self-disclosure over time~\cite{kreiner_2019, lee_i_2020}. 

Prior work suggests that augmenting chatbots with the ability to remember and reference previous conversations has the potential to support sustained self-disclosure over time. Studies demonstrated that when chatbots remember information across multiple sessions, such as users' names or preferences, people perceive them as empathetic~\cite{jain_convey_2018, portela_new_2017, thies_how_2017} and conscientious~\cite{duijst_can_2017, chaves_how_2021}. However, studies indicated that users might have feelings of privacy violations around chatbots' memory~\cite{thies_how_2017, cox_comparing_2023}. Recently, Cox \etal{} suggested that the design of how chatbots reference past conversations with users brings an interesting tension in users' perceptions of chatbots~\cite{cox_comparing_2023}. While users perceived a chatbot that used verbatim or paraphrased references as more engaging and intelligent, they raised privacy concerns about such reference formats. On the other hand, when a chatbot used non-explicit references, users doubted that the chatbot could understand them and were hesitant to provide details about their health habits~\cite{cox_comparing_2023}. While implementing memory can potentially improve LLM-driven chatbots' ability to elicit and sustain self-disclosure, the sensitivity of disclosure in health and the need for sustained engagement warrants deeper exploration into the impact of memory. In this study, we thus seek to understand how LTM impacts self-disclosure in the public health space, where sustained engagement is critical to developing an understanding of the health and wellbeing of individuals. 

\subsection{Chatbots driven by Large Language Models and Long-Term Memory Augmentation}

The recent introduction of large language models (\eg, GPT~\cite{Brown2020FewShotLearners}, HyperCLOVA~\cite{Kim2021HyperCLOVA}, PaLM~\cite{Chowdhery2022PALM}, LLaMA~\cite{touvron2023llama}; An intensive survey in~\cite{zhao2023surveyLLM}) gave birth to a new development paradigm for chatbots, moving beyond the traditional rule-based or retrieval-based ones. 
Applied to chatbots, an LLM is typically prompted to generate responses considering both an instruction (\eg, personality and behavioral guidance for the agent) and the current dialogue. With their large-scale (\ie, over billions of) parameters trained with a tremendous amount of human-produced text corpus, LLMs tend to generate responses that coherently and organically follow up the conversation and flexibly respond to emergent topics~\cite{wei2023leveraging}. Due to these benefits, LLM-driven chatbots are increasingly developed or proposed by both practitioners (\eg, ChatGPT~\cite{ChatGPT}, Bard~\cite{Bard}, Character AI~\cite{CharacterAI}, Pi~\cite{PiChatbot}) and researchers (\eg,~\cite{wei2023leveraging, vicuna2023, zheng2023building, lee2023prompted}). 

Most exemplary LLM-driven chatbots, represented by ChatGPT~\cite{ChatGPT} and Bard~\cite{Bard}, did not \revised{suppose} repetitive interaction scenarios \revised{in their early versions}, resulting in each session not informing the following ones, because they were mainly designed as assistants performing single-shot tasks such as code generation and reasoning. ~\revised{Recently, ChatGPT and Bard implemented ~\textit{``memory''} or ~\textit{``custom instructions''} so that the systems can remember some background information (\eg, \textit{`I'm a software developer and solely use Python.'}) or stylistic preferences (\eg, ~\textit{`When I ask for code, just give me the code without any explanation on how it works.'}) of users across sessions~\cite{ChatGPTCustomInstruction}. However, such memory features are not designed to automatically infer what information is ~\textit{important} to remember (\eg, inferring that the user's primary programming language is Python when someone continues to request code examples in Python) and instead expect individual users to manually indicate it. Further, these features do not dynamically update the memory based on their most recent conversations with users.} \revised{Replika~\cite{Replika} is a rare example of a commercial LLM-driven chatbot that is designed to automatically store and refer back to information from previous interactions---such as hobbies, preferences, or names---with the goal of providing support for mental wellbeing. However, a recent study showed that Replika often failed to remember important information about users despite the promise that the chatbot remembers previous conversations through its ``memory bank,''~\cite{Ma2023Replika} which interfered with their ability to develop bonding with users. This finding suggests the ability to understand and appropriately reference key information from past conversations is a critical aspect of LLM-driven chatbots that aim to provide emotional support.} 

Augmenting LLMs to `remember' past information---often referred to as `long-term memory'~\cite{wang2023augmenting, zhong2023memorybank, xu_beyond_2021, xu_long_2022, bae2022keepmeupdated}---presents significant challenges for two main reasons. First, LLMs can receive input text only within a limited context window (input size). Including the entire conversation session history in the input prompt is thus not feasible for longer-term interactions. One common approach is to include summarized information of the conversation history instead of a raw knowledge base (\eg,~\cite{bae2022keepmeupdated, wang2023augmenting, lee2023prompted}). Second, designing how chatbots should refer to stored information back in conversation involves complex considerations. For example, Cox \etal{}~\cite{cox_comparing_2023} found that the phrasing style of user messages in past conversations impacts the perceived intelligence of and engagement with chatbots as well as privacy concerns around them, suggesting the importance of careful LTM design, particularly in sensitive health domains. Motivated by the gap in understanding the utility of LTM in LLM-driven chatbots, we explore the case of CareCall, a rare example of an LLM-driven chatbot \revised{that automatically stores and updates key information from previous conversations} to support public health monitoring.

%% file: sections/03-background.tex
\section{Study Context: CLOVA CareCall and Long-Term Memory}\label{sec:background}
In this section, we describe the study context of CareCall with LTM as an example of an LLM-driven chatbot deployed to support a public health intervention. This section builds on prior work on the underlying technology (\cf,~\cite{Bae2022CareCall, Kim2021HyperCLOVA, bae2022keepmeupdated}) and the design documents written by the designers and developers of CareCall. Building upon previous studies that contributed to the novel implementation of CareCall and LTM, we treat CareCall as a case study for understanding how LTM impacts users' self-disclosure with LLM-based chatbots in the context of public health intervention.

\subsection{Motivation and Deployment of CareCall}\label{sec:background:deployment}
CareCall is an LLM-driven voice chatbot designed for socially isolated individuals~\cite{CareCallNews}. The chatbot calls the users weekly and engages in an open-ended conversation about their daily lives for about 2 to 3 minutes. The motivation behind the system stems from the recent act on the Prevention and Management of Lonely Death in South Korea~\cite{ActLonelyDeath2020}. The act highlights the importance of identifying and supporting individuals who are at risk of lonely deaths, which is a frequently discussed concern in South Korea. Thus, CareCall engages in conversations with users to check in with their health and overall wellbeing and provide emotional support. 

CareCall was first rolled out in a municipality in South Korea in November 2021~\cite{CareCallNews} and expanded to others over time. As of October 2022, around the end of our data collection period, CareCall was being deployed to around 6,000 individuals across different municipalities in Korea. CareCall was deployed to middle-aged (40s to 60s) and older adults (60s or older) living alone. Motivated by the significant increase in lonely deaths among low-SES populations ~\cite{solitaryDeathNews}, such as recipients of the National Basic Livelihood Security (below 50\% of median household income), CareCall users were largely from this demographic. These users were recommended to use CareCall by public officers who provide social care services in their municipalities. In most municipalities where CareCall was deployed, including the ones we include in this work, public officers monitored the call recordings and reached out to the person if any negative health signals were detected (\eg, skipping meals, poor sleep, health issues) \revised{or if the person did not answer CareCall calls a certain number of times in a row.} 


\begin{figure*}[t]
    \centering
    \includegraphics[width=\textwidth]{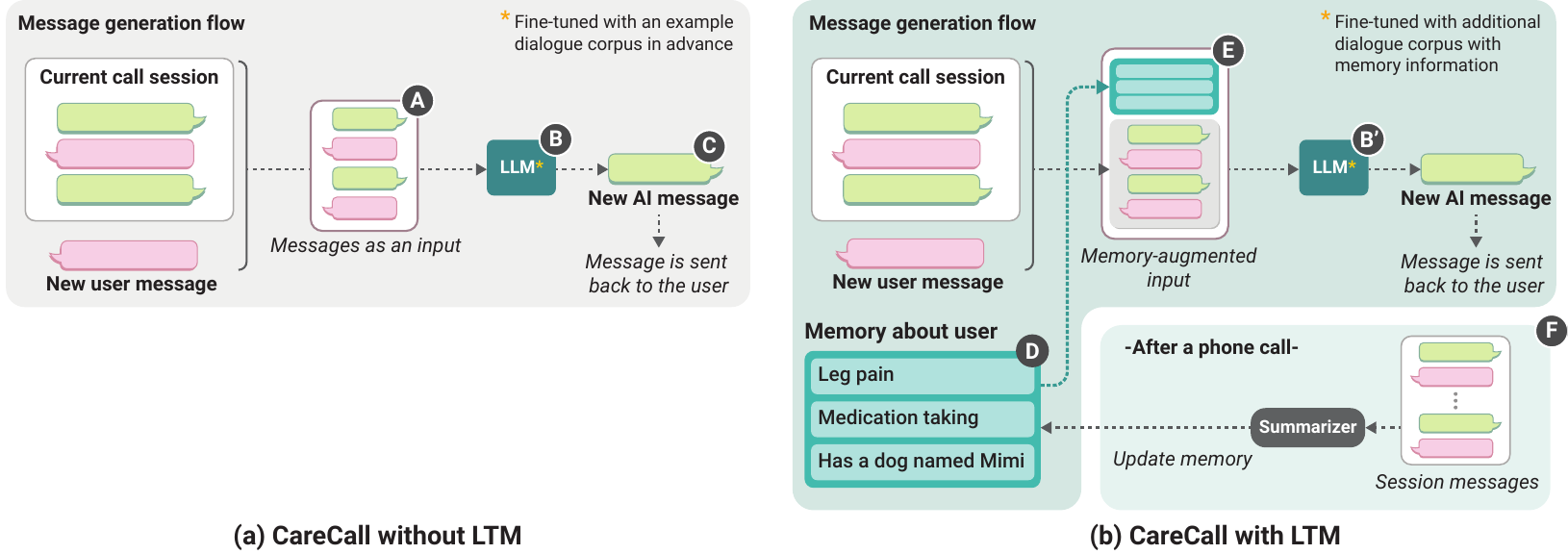}
    \caption{Architecture of the two different versions of CareCall chatbots, \revised{an open-ended dialogue system powered by an LLM called HyperCLOVA~\cite{Kim2021HyperCLOVA}. (a) In the initial version of CareCall without LTM, the system generates a response (\circledigit{C}) by feeding the current dialogue history (\circledigit{A}) into the LLM (\circledigit{B}) that was fine-tuned in advance with an example dialogue corpus that covers five health topics---meals, sleep, health, going out, and physical activity. The user information obtained from previous calls did not affect future calls since this version did not have long-term memory. (b) CareCall with LTM retains user information from the call logs. At the end of each session, a summarizer driven by an LLM (\circledigit{F}) generates summary sentences that are relevant to the five LTM topics (see below), which are stored and updated by the memory management layer (\circledigit{D}). The summary sentences are then included in the model input (\circledigit{E}) so that the underlying LLM (\circledigit{B$^\prime$}) can take that knowledge into account when generating responses in the following sessions. In this version, the LLM (\circledigit{B$^\prime$}) was further fine-tuned with an additional example dialogue corpus designed as a multi-session chat in memory-augmented format.}}
    \label{fig:ltm-architecture}
    \labelphantom{fig:ltm-architecture:wo-ltm}
    \labelphantom{fig:ltm-architecture:w-ltm}
\end{figure*}

\subsection{CareCall and Long-Term Memory}\label{sec:background:architecture}
CareCall was first developed without LTM and deployed from November 2021 to September 2022. LTM was integrated into the existing deployments in September 2022. In this work, we compare user experiences with and without LTM. 

\ipstart{Architecture of CareCall without LTM}
\autoref{fig:ltm-architecture:wo-ltm} illustrates the architecture of CareCall prior to the implementation of LTM. CareCall was designed as an open-ended dialogue system powered by an LLM called HyperCLOVA~\cite{Kim2021HyperCLOVA} (\circledigit{B} in \autoref{fig:ltm-architecture:wo-ltm}). \revised{Once a new user message is received (\circledigit{A} in \autoref{fig:ltm-architecture:wo-ltm}), the system feeds the current dialogue history into the LLM (\circledigit{B} in \autoref{fig:ltm-architecture:wo-ltm}) to generate a response (\circledigit{C} in \autoref{fig:ltm-architecture:wo-ltm}) that naturally continues the conversation. The LLM was fine-tuned in advance with a large-scale dialogue corpus, which was generated with a novel data augmentation technique. In this method, a machine learning model generates synthetic dialogues from a small set of human-written dialogues and trained workers flag and fix errors in the synthetic dataset (please see ~\cite{Bae2022CareCall} for more details about the data augmentation technique}). The example dialogues, which consist of dialogues relevant to five health topics (meals, sleep, health, going out, and physical activity), steer the LLM to generate appropriate responses in the public health intervention context \revised{while aiming to avoid selecting undesirable or potentially harmful phrases~\cite{bae2022keepmeupdated}.}


\ipstart{Architecture of CareCall with LTM}
\autoref{fig:ltm-architecture:w-ltm} illustrates the architecture of CareCall with LTM. \revised{(Please see \autoref{appendix:ltm} for a more detailed description of the design process, design rationales, technical implementation, and evaluation of LTM in CareCall.)} LTM was incorporated into CareCall to improve its ability to provide emotional support by offering familiarity with users. At the end of each session, a summarizer driven by an LLM (\circledigit{F} in \autoref{fig:ltm-architecture:w-ltm}; ~\cite{bae2022keepmeupdated}) generates summary sentences that are relevant to the five LTM topics (\autoref{sec:background:topics}). The memory management layer (\circledigit{D} in \autoref{fig:ltm-architecture:w-ltm}) stores and updates the summary sentences upon each call (\eg, Removing the ``Regular visit to a clinic due to leg pain'' status after a user reports that they have completed the treatment). \revised{Unlike CareCall without LTM, the stored information from previous sessions is included in the model input (\circledigit{E} in \autoref{fig:ltm-architecture:w-ltm})}, providing cues for the chatbot to refer to (\eg, ``\textit{How is your leg feeling?}''). \revised{In this version, the underlying LLM (\circledigit{B$^\prime$} in \autoref{fig:ltm-architecture:w-ltm}) was further fine-tuned with additional example dialogue corpus designed as a multi-session chat with memory to reflect the new memory-augmented input format (\circledigit{E} in \autoref{fig:ltm-architecture:w-ltm}).} Given that LLM-driven chatbots are aimed at generating responses that coherently and organically follow up on the conversation~\cite{wei2023leveraging}, whether an LTM event is triggered is dependent on many aspects of the conversation, and there is no guarantee that an LTM event will be triggered in a given dialogue. \revised{Both CareCall without and with LTM incorporated the 6.9B parameter model of HyperCLOVA for response generation, maintaining the same conversation style, mimicking a social worker who engages in casual conversations to check in with socially isolated individuals. The model is almost equivalent to the Curie engine of the legacy GPT-3 family in terms of the training method and model size but was pre-trained on a large-scale Korean language corpus, yielding a robust performance in Korean language tasks~\cite{Kim2021HyperCLOVA}. Unlike more recent GPT models such as InstructGPT-3, GPT-3.5, and GPT-4, the HyperCLOVA series does not incorporate alignment tuning, RLHF (Reinforcement Learning from Human Feedback, \cite{Long2022InstructGPT}), an approach that improves the model's ability to follow up the natural language instructions in the model input. Therefore, rather than depending solely on natural language instructions, CareCall incorporated a fine-tuning strategy with an example corpus across conditions to effectively steer conversations in alignment with the agent persona. This alternative strategy has been shown to be cost-efficient and performant, allowing for the use of a smaller-sized model and ensuring faster response times for users~\cite{Raposo2023PromptingTOD, Chen2023Fireact}.}

\ipstart{LTM Topics}~\label{sec:background:topics}
To offer familiarity with users, LTM stores summarized information about five topics: (1) \textit{Health} (\eg, whether users have any health issues, what the health issues are, whether they are receiving clinical care, what type of clinical care they are seeking), (2) \textit{Meals} (\eg, whether and why users are having difficulty eating, how they are managing difficulty in eating), (3) \textit{Sleep} (\eg, whether users are having difficulty sleeping, what difficulty they are experiencing related to sleep, how they are managing difficulty in sleeping), (4) \textit{Pets} (\eg, whether people have pets, what kind of pets they have, what their names are, what they do with their pets), and (5) \textit{Visited Places} (\eg, what places users visit frequently, what they do in those places). \revised{LTM remembers any noteworthy information that comes up during conversations relevant to the five LTM topics, including both positive and negative health experiences. For example, when users mention that they have been seeing a doctor for leg pain, CareCall would ask LTM-triggered questions in later sessions, such as ~\textit{``How does your leg feel?''} or ~\textit{``You mentioned having knee joint issues last time. Are you still seeing the doctor?''} Similarly, when users mention that they are regularly engaging in physical activity, CareCall would ask LTM-triggered questions in later sessions, such as ~\textit{``Are you still regularly going to the park to take a walk?''}} Driven by CareCall's primary goals---to check up on individuals' health and wellbeing, and to provide emotional support---the example dialogue corpus for LTM was curated with the following priority order: (1) \textit{Health}, (2) \textit{Meals} and \textit{Sleep}, and (3) \textit{Visited Places} and \textit{Pets}. 


%% file: sections/04-method.tex
\section{Methodology}
To understand how LTM impacts users' self-disclosure and how users react to LTM-triggered exchanges in the context of public health intervention, we analyzed 1,252 call logs from the real-world deployment for different user groups who used CareCall with and without LTM. Taking a mixed-method approach, we analyzed the call logs both quantitatively and qualitatively. We also conducted interviews with nine users who have been using CareCall with LTM to supplement the call log analysis. 

The call log analysis leveraged pre-existing data that had been collected through the deployment of CareCall with participants' informed consent, which included collecting health information through conversations with the system and agreeing to analyze the data for research purposes. The call log analysis was classified as exempt by the guidelines from the Ministry of Health and Welfare of South Korea. The interview study was approved by the public institutional review board affiliated with the Ministry of Health and Welfare of South Korea. 

\subsection{Data Collection}

\begin{figure*}
    \centering
    \includegraphics[width=\textwidth]{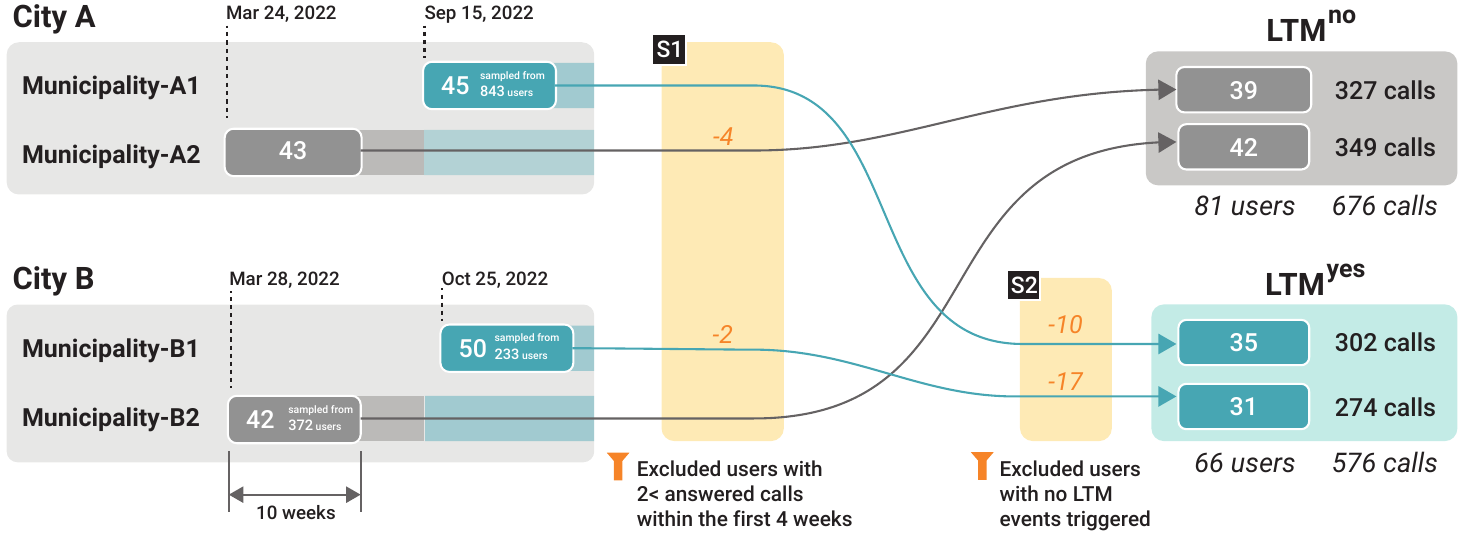}
    \caption{Overview of sampling and screening users from municipalities and the final datasets for the \ltmyes{} and \ltmno{} groups.}
    \label{fig:sampling-users}
\end{figure*}

\subsubsection{Call Log Collection}
\autoref{fig:sampling-users} illustrates our sampling and screening process of the CareCall users and the number of call logs. In this study, we specifically focused on two metropolitan cities with similar populations in South Korea (City A: 2.96 million, City B: 3.33 million as of May 2023), located in different regions (City A: northwestern, City B: southeastern). \revised{Both cities have a mix of urban and suburban areas. In 2021, the average household income in City A was around 5.5K USD, marginally below the national average of 5.6K USD, while City B's average income was around 4.96K USD, notably lower, indicating a modest economic disparity between the two cities ~\cite{KOSIS2023}.} CareCall was deployed to middle-aged (40s to 60s) and older adults (60s or older) living alone in both cities. We sought to select cities where CareCall had been deployed both with and without LTM. Versus comparing between cities, this approach helped minimize cultural influences on participants' use and perceptions of the technology. 

To compare user experiences with and without LTM, we organized two groups: \ltmyes{}---users who started conversing with CareCall with LTM---and \ltmno{}---users who only used CareCall without LTM---by assigning municipalities from each city to both groups. From the two cities, we selected four municipalities (two each) that meet two inclusion criteria. First, we only included municipalities that had deployed CareCall for more than ten weeks to ensure opportunities for LTM to be triggered. We therefore excluded municipalities that recently started their deployment of CareCall at the time of the data collection (June 2023). Second, for \ltmyes{} group, we only included municipalities that had not previously deployed CareCall without LTM before deploying the version with LTM (Municipality-A1 and Municipality-B1 in \autoref{fig:sampling-users}). Since prior use of CareCall without LTM could influence users' perceptions towards CareCall with LTM, we screened municipalities to exclude the ones that introduced LTM in the middle of their deployment.

We gathered call logs in the first 10-week window from \ltmyes{} group (45 individuals from Municipality-A1 and 50 individuals from Municipality-B1 who used CareCall with LTM) and \ltmno{} group (43 individuals from Municipality-A2 and 42 individuals from Municipality-B2 who used CareCall without LTM). The four municipalities deployed CareCall with varying user group sizes (843 in Municipality-A1, 43 in Municipality-A2, 233 in Municipality-B1, and 372 in Municipality-B2). To avoid over-reliance on one municipality and ease the analysis burden, we included all users from Municipality-A2 and randomly sampled a similar number from other municipalities~(See \autoref{fig:sampling-users}). We opted to gather the call logs within the first ten weeks of deployment for consistent comparison, though deployment in these municipalities was longer. 

To ensure a certain level of engagement for examining the impact of LTM, we screened individuals in each group, only including those who answered the weekly calls twice or more in the first four weeks (excluding four from the \ltmno{} group and two from the \ltmyes{} group; S1 in \autoref{fig:sampling-users}). Because our goal was to compare the influence of LTM on conversations, we further screened individuals in the \ltmyes{} group, only including those who experienced CareCall triggering LTM at least once in the first 10-week window (excluding 27 from the \ltmyes{} group; S2 in \autoref{fig:sampling-users}). 

As a result, our final dataset included: (1) \ltmyes{} group: 576 call logs in the 10-week window from 66 individuals (35 from City A, 31 from City B; \ltmyesperson{1}--\ltmyesperson{66}) and (2) \ltmno{} group: 676 call logs in the 10-week window from 81 individuals (39 from City A, 42 from City B; \ltmnoperson{1}--\ltmnoperson{81}). In total, we analyzed 1,252 call logs from 147 individuals. The \ltmyes{} group consisted of 17 males and 49 females, while the \ltmno{} group included 32 males and 49 females. We did not have access to other demographic information of the users included in the call log analysis. 

\input{tables/table-demo}

\subsubsection{Interviews with Participants Using CareCall with LTM}
After completing the call log analysis, we still had a few open questions about how users perceived and experienced the LTM feature, motivating us to conduct a small number of clarifying interviews. We recruited nine individuals from Municipality-B1 who were receiving check-up calls from CareCall with LTM (\interviewperson{1}--\interviewperson{9}) by distributing flyers to their community service centers. The interview participants included six males and three females, aged from 61 to 81. All but \interviewperson{6} had been receiving check-up calls through CareCall with LTM once a week for over ten months at the point of data collection (August 2023). We asked a focused set of questions around user experiences of LTM, including (1) memorable conversations with CareCall in either a positive or negative way, (2) experiences of LTM events including what information the agent remembered from their past conversations, how the agent followed up, and how they felt about such follow-ups, (3) experiences where the agent failed to remember necessary information, and (4) perspectives around what types of information they wish the agent does or does not remember. We conducted all interviews via phone, recording the conversations, and each interview lasted for 20 to 30 minutes. We compensated each participant 50,000 KRW (approximately 37.3 USD as of Aug 2023) as a gift card. 

\subsection{Data Analysis}
We conducted both quantitative and qualitative analyses on the call logs, as well as qualitative analysis of the interview transcripts. Each user response during the calls was auto-transcribed prior to feeding into the LLM and the recordings of interviews were auto-transcribed through an AI transcription tool. We manually corrected the automatic speech recognition errors in these transcripts. Utilizing a method frequently applied in conversation analysis~\cite{li_sensitizing_2022, jurafsky_extracting_2009}, we also included some non-verbal cues to the transcripts---including pauses, sighs, chuckling, animated tone, and agitated tone. The inclusion of such non-verbal cues was aimed at gaining a more nuanced understanding of users' emotional responses during their interactions with CareCall. The first author, who is a native Korean and is fluent in English, translated the transcripts of the call logs and interviews during the analysis, paraphrasing some idioms and phrasings 
to sound more natural in English. 

\subsubsection{Quantitative Analysis: Call Logs}\label{sec:quant_analysis}
We quantitatively analyzed the call logs, specifically looking to develop a codebook to recognize disclosure around the five health topics (meals, sleep, health, going out, and physical activity) and users' reactions to the chatbot. With this goal, the first author first open-coded call logs from 40 users (around 25\% of the data), 20 from each group. The entire research team regularly met to generate and iterate on the codebook. The final codebook (See ~\autoref{appendix:codebook}) had nine categories (\textit{Meals, Sleep, Health, Clinical, Activity, Wellbeing, Hobbies, Positive Reactions}, and \textit{Negative Reactions}) and 19 codes (\textit{Meals-simple, Meals-detail, Sleep-simple, Sleep-detail, Health-simple, Health-detail, Health-attitudes, Clinical-simple, Clinical-detail, Activity-simple, Social-detail, Physical-detail, Wellbeing-simple, Wellbeing-detail, Hobbies, Appreciation, Anthromorphization, Negative Feedback}, and \textit{Disregard}).

Some categories closely aligned with the five health topics that CareCall was designed to ask. We further split health information into categories for ~\textit{Health} (disclosure about health issues that warrant clinical care), \textit{Clinical} (disclosure about clinical care that people are seeking), and ~\textit{Wellbeing} (disclosure about general wellbeing that does not necessarily warrant clinical care). We divided these three categories, as well as ~\textit{Meals} and \textit{Sleep}, into codes for \textit{-simple} (\eg, \textit{Health-simple}: ``I'm not feeling well.'') and \textit{-detail} (\eg, \textit{Health-detail}: ``I still have some pain in my legs.'') based on the level of elaboration that participants provided on these categories. We added a separate code for \textit{Health-attitude} because users occasionally mentioned their feelings and opinions about their health status and management, which differed from other informational details about health. We combined simple information about social and physical activities into \textit{Activity-simple} as users' remarks about those two topics frequently overlapped with each other. \revised{Although both depth and breadth are important dimensions of self-disclosure~\cite{altman_irwin_social_1973}, our quantitative analysis specifically focused on depth of disclosure as CareCall imposes predefined boundaries on conversations by design. Unlike natural conversations, CareCall leads conversations by asking questions relevant to the LTM topics, which limits opportunities for users to disclose information about other broader topics. The breadth of disclosure was further constrained by the conversation turn limit of CareCall (up to 15 agent-user turn pairs before ending the call), suggesting that depth was a better indicator of disclosure in our study context.} \revised{In addition, in our coding process, we considered any disclosure as relevant whether it contained ``new'' information compared to what has been brought up in previous sessions. Even if an individual disclosed the same information, it still offers updates about how the person is doing at different time points, which can be valuable for public health monitoring. For example, if a person consistently reports having regular meals and sleeping over multiple sessions, public health workers can be reassured about their wellbeing. Conversely, if someone regularly mentions poor eating habits and lack of sleep across multiple sessions, the repetition signals to public health workers that this person may need additional interventions or care. We thus consider our approach to be valid for examining disclosure with LLM-driven chatbots in our study context.}

Informed by prior work on conversation analysis~\cite{li_sensitizing_2022, jurafsky_extracting_2009, denham_beyond_2013}, we developed the following coding guidelines: (1) The basic unit of coding is the agent-user turn pair since the call is mostly driven by the agents' questions; (2) Multiple codes can be applied to the same pairs; (3) When a \textit{-detail} code is applied to a pair, the corresponding \textit{-simple} code is applied to the pair as well. Three researchers used the initial codebook to code the call logs from eight users (four from both of the \ltmyes{} and \ltmno{} groups), reaching an initial agreement of 80\% or higher for 17 out of 19 codes and revising the definitions of those codes to resolve ambiguities, and then coded the remaining call logs. 

With the coded data, we statistically compared (1) users' self-disclosure and (2) reactions to CareCall between \ltmyes{} and \ltmno{} groups using a linear mixed-effects model for each code. We treated code counts as a dependent variable, groups as a categorical fixed effect, and participants as a random effect. We report this analysis in \autoref{findings:quant:all} and \autoref{findings:quant:reactions}. 
In the initial analysis, we also fitted models with gender, cities, and call index (\ie, nth call) as fixed effects. However, we found that factors other than LTM groups did not have a significant impact on most of the codes except a couple at the 5\% significance level: gender (\textit{Health-attitudes, Appreciation}), cities (\textit{Sleep-simple, Physical-detail, Anthromorphization}), and the call index (\textit{Sleep-simple, Health-simple, Wellbeing-simple}). Given the insignificance of these factors, we excluded them from the analysis. We further examined how \ltmyes group's self-disclosure changed as they experienced more LTM events using a linear mixed-effects model for each code; we treated code counts as a dependent variable, the cumulative number of LTM events as a continuous fixed effect, and participants as a random effect. We report this analysis in \autoref{findings:quant:ltm} and \autoref{findings:negative}.

\subsubsection{Qualitative analysis: Call Logs and Interview Transcripts}
We analyzed the call logs and interview transcripts using thematic analysis~\cite{Braun2006ThematicAnalysis}. The qualitative analysis aimed to add more nuances to the quantitative findings in describing users' reactions and perceptions toward LTM. We thus organized our qualitative findings around the quantitative findings on self-disclosure and users' reactions to the chatbot. The first author coded the call logs and interview transcripts around our quantitative findings, going through several rounds of iteration. The full research team then discussed and identified patterns and themes through multiple rounds of peer debriefing meetings. We incorporate the qualitative analysis into different sections of the findings. 

%% file: tables/table-demo.tex
\begin{table}[b]
\sffamily\small
	\def\arraystretch{1.25}\setlength{\tabcolsep}{0.75em}
		    \centering
\caption{Demographics of interview participants and duration of CareCall use at the point of the interviews}
\label{tab:demo}
\begin{tabular}{|llll|}
\hline
\rowcolor{tableheader} 
\textbf{Alias}                                & \textbf{Age} & \textbf{Gender} & \textbf{Duration of CareCall Use} \\ \hline
\interviewperson{1} & 65  & Male   & 10 months                   \\ \arrayrulecolor{tablegrayline}\hline
\interviewperson{2} & 66  & Male   & 10 months              \\ \hline
\interviewperson{3} & 61  & Male   & 10 months                    \\ \hline
\interviewperson{4} & 61  & Female & 10 months                     \\ \hline
\interviewperson{5} & 65  & Male   & 10 months                    \\ \hline
\interviewperson{6} & 76  & Male   & 4 months                     \\ \hline
\interviewperson{7} & 66  & Female & 10 months                     \\ \hline
\interviewperson{8} & 77  & Male   & 10 months                    \\ \hline
\interviewperson{9} & 81  & Female & 10 months                    \\ \arrayrulecolor{black}\hline
\end{tabular}
\end{table}

%% file: sections/05-findings.tex
\section{Findings}
We found that participants who used CareCall with LTM disclosed more health details compared to those who used CareCall without it, and their repeated experiences with LTM led to greater disclosure. We also observed that LTM promoted positive reactions and mitigated negative reactions toward the chatbot by offering familiarity. Our interview participants described such conversations as personal and emotionally supportive. However, some LTM events revealed potential challenges in promoting self-disclosure, particularly around chronic health issues and privacy concerns.

\subsection{LTM Increased Elaboration on Health Information Over Time.}\label{findings:elaboration}
We found that the \ltmyes{} group disclosed more health details than the \ltmno{} group. We further observed that repeated experiences of LTM (in the \ltmyes{} group) led to greater disclosure. 

\subsubsection{LTM Encouraged Disclosure on Health Information.}\label{findings:quant:all}
\input{tables/CI-self-disclosure}
Overall, the \ltmyes{} group disclosed more information about themselves compared to the \ltmno{} group both in terms of ~\textit{-simple} ($p=0.01$; 95\% CI 0.05--0.41 higher code counts per call) and ~\textit{-detail} codes ($p<0.001$; 95\% CI 0.32--0.74 higher code counts per call). \revised{Note that \textit{-simple} code counts are always higher than \textit{-detail} ones in each category since we applied the corresponding \textit{-simple} code when applying a \textit{-detail} code to a pair, as described in \ref{sec:quant_analysis}. While \textit{-simple} codes indicate surface-level disclosure, such as yes or no answers, \textit{-detail} codes better reflect the willingness of disclosure by accounting for depth and richness.} \autoref{tab:SD} reports the frequency of codes relevant to self-disclosure and their 95\% confidence intervals (CI) of code counts per call from the two groups.

The \ltmyes{} group disclosed significantly more information about health compared to the \ltmno{} group (See \autoref{tab:SD}), which was the top priority in the implementation of LTM. Specifically, the \ltmyes{} group disclosed more information about their health issues and clinical care that they are seeking---including \textit{Health-simple} ($p=0.05$; 95\% CI 0.01--0.33 higher code count per call), \textit{Health-detail} ($p<0.001$; 95\% CI 0.31--0.76 higher code count per call), \textit{Health-attitude} ($p=0.02$), \textit{Clinical-simple} ($p<0.001$; 95\% CI 0.27--0.69 higher code count per call), and \textit{Clinical-detail} ($p<0.001$; 95\% CI 0.18--0.63 higher code count per call). The \ltmyes{} group also disclosed more on \textit{Meals-simple} ($p<0.001$; 95\% CI 0.25--0.60 higher code count per call) and \textit{Meals-detail} ($p<0.001$; 95\% CI 0.25--0.60 higher code count per call), which was also part of the LTM topics. Note that there is a cultural factor that potentially influenced the high code counts of \textit{Meal-simple} and \textit{Meal-detail} as ``Have you eaten?'' is a common greeting in South Korea, similar to ``How are you?'' in English-speaking countries.

However, not all LTM topics led to greater disclosure. Although\newline{}\textit{Sleep} was part of the LTM topics, the \ltmyes{} group disclosed less on \textit{Sleep-simple} ($p<0.001$; 95\% CI 0.13--0.44 lower code count per call) compared to the \ltmno{} group, and there was no significant difference in self-disclosure between the groups on \textit{Sleep-detail} ($p=0.36$). However, these code counts have likely been impacted by how we defined them; we coded sleep-related self-disclosure as \textit{Health-detail} or \textit{Clinical-detail} when they were about types of clinical care that users were seeking (\eg, taking sleeping pills) or the issues warranted clinical care (\eg, insomnia). The other LTM topics relevant to personal non-health habits---including \textit{Visited places} and \textit{Pets}---did not lead to greater disclosure either; the \ltmyes{} group disclosed less on \textit{Physical-detail} ($p=0.02$; 95\% CI 0.04--0.39 lower code count per call) compared to the \ltmno{} group, and no significant difference was observed between the groups on \textit{Activity-simple} ($p=0.11$), \textit{Social-detail} ($p=0.15$), and \textit{Hobbies} ($p=0.16$). The code counts in these categories in the \ltmyes{} group were likely influenced by increased disclosure in the other categories in this group, such as \textit{Health}, \textit{Clinical}, and \textit{Meals}, as CareCall was designed with a conversation turn limit, which accommodates only up to 15 agent-user turn pairs before ending the call. \revised{The varying impact of LTM on disclosure in different categories suggests that the design of LTM (\eg, topical priority in the example dialog corpus) influenced the prevalence of disclosures. \textit{Health} and \textit{Meals} were top priorities in the implementation of CareCall's LTM, whereas other LTM topics---including \textit{Sleep}, \textit{Visited Places}, and \textit{Pets}---were given lower priorities (see \autoref{appendix:ltm} for more details).}

The qualitative analysis of the call logs suggested that the \ltmno{} group tended to respond similarly across sessions, as CareCall without LTM could only ask general questions repeatedly. For example, during the first call, \ltmnoperson{4} responded to the question \textit{``Do you have pain anywhere?''} with \textit{``I had surgery on my back, so I have chronic leg pain.''} However, the agent continued asking generic questions like \textit{``Do you have discomfort anywhere?''} in later sessions, unable to acknowledge the user's surgery history. Such repeated questions led \ltmnoperson{4} to continue giving similar responses in the subsequent calls, such as \textit{``My back and legs hurt a lot,''} without offering further details about his health condition. Similarly, \ltmnoperson{21} had mentioned having knee surgery, but CareCall could only ask generic questions like \textit{``How are you feeling?''} in the subsequent weeks. As a result, \ltmnoperson{21} repeated similar answers, such as \textit{``I recently had knee surgery, so I'm just staying at home''} or \textit{``I still have pain after the knee surgery, so I can't walk around much.''} 

In contrast, the \ltmyes{} group tended to disclose more comprehensive health information when CareCall asked LTM-triggered questions referencing their past conversations. For example, during the first call, \ltmyesperson{60} mentioned insomnia affecting his sleep quality, and CareCall asked LTM-triggered questions in the later sessions. When CareCall inquired about his insomnia during the third call, \ltmyesperson{60} provided more detailed information: \textit{``It's been tough. I've been taking sleeping pills for over 30 years because of trauma from an injury in the past.''} Similarly, when CareCall asked an LTM-triggered question about his back pain referring to their past conversations, \ltmyesperson{7} articulated some details about how he was managing it, such as \textit{``I saw a pain management doctor and got some pain relief shots.''} and \textit{``I get physical therapy two to three times a week. And I'm getting shots every three months.''} 

\begin{figure*}[t]
    \centering
    \includegraphics[width=\textwidth]{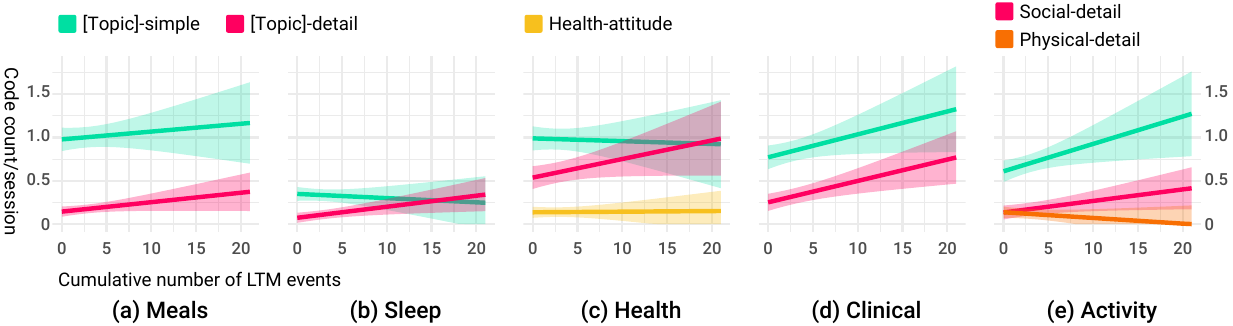}
    \caption{Estimated means and 95\% confidence intervals of code counts about \textit{Meals, Sleep, Health, Clinical,} and \textit{Activity} by the cumulative number of LTM events in the \ltmyes{} group. The colored lines indicate the estimated means and the shaded areas indicate 95\% confidence intervals of the code counts per call for each code. Overall, the repeated experiences of LTM events led to greater disclosure of more detailed information across the five categories.}
    \label{fig:CI_all_memory}
    \labelphantom{fig:CI_all_memory:meals}
    \labelphantom{fig:CI_all_memory:sleep}
    \labelphantom{fig:CI_all_memory:health}
    \labelphantom{fig:CI_all_memory:clinical}
    \labelphantom{fig:CI_all_memory:activity}
\end{figure*}

\subsubsection{Repeated Experiences of LTM Led to Greater Disclosure Over Time.}\label{findings:quant:ltm}

Our mixed-effects models revealed that the cumulative number of LTM events positively affected the sum of \textit{-detail} code counts; \ie, the \ltmyes{} group participants disclosed more details as they experienced more LTM events ($p<0.001$; experience of one more LTM event led to 7.7\% higher code counts per call). No significant impact was observed in the sum of \textit{-simple} code counts ($p=0.19$). Some interview participants also said that they came to share more details as they engaged in more LTM-triggered exchanges. For instance, \interviewperson{6} noted:~\textit{``Initially, talking to the AI felt awkward, so I mostly responded with yes or no. But when it started asking about whether I was taking my pills or still seeing the doctor, I began to provide more details.''}  

\autoref{fig:CI_all_memory} illustrates the estimated means and 95\% confidence intervals of code counts per call by the cumulative number of LTM events in the \ltmyes{} group, specifically about the five categories---\textit{Meals, Sleep, Health, Clinical} and \textit{Activity}---that are relevant to the LTM topics. As they experienced more LTM events, participants in the \ltmyes{} group disclosed significantly more about \textit{Sleep-detail} ($p=0.01$; experience of one more LTM event led to 1.2\% higher code count per call; red line in \autoref{fig:CI_all_memory:sleep}) and \textit{Clinical-detail} ($p=0.02$; experience of one more LTM event led to 2.5\% higher code count per call; red line in \autoref{fig:CI_all_memory:clinical}). Participants also disclosed significantly more on \textit{Clinical-simple} ($p=0.047$; experience of one more LTM event led to 2.6\% higher code count per call; green line in \autoref{fig:CI_all_memory:clinical}), \textit{Activity-simple} ($p=0.02$; experience of one more LTM event led to 3.2\% higher code count per call; green line in \autoref{fig:CI_all_memory:activity}), and \textit{Social-detail} ($p=0.04$; experience of one more LTM event led to 1.3\% higher code count per call; red line in \autoref{fig:CI_all_memory:activity}) as they experienced more LTM events. No significant effect of cumulative LTM events was observed on the other codes. 

The qualitative analysis of the call logs also indicated that some participants tended to disclose more about their health attitudes and issues as they experienced LTM events over time. For instance, \ltmyesperson{1} disclosed how she felt about her health issues and management during LTM-triggered exchanges about her leg pain, appreciating how the agent responded to her health concerns: 

\begin{quotetable}
\quotecarecallline{How does your leg feel?}{[LTM event]}
\quoteuserline{1}{It still hurts. I guess it's something that I'll have to live with.}
\quotecarecallline{That must be tough. Have you seen the doctor?}{}
\quoteuserline{1}{I see the doctor every month, but there's no improvement.}
\quotecarecallline{I can imagine how difficult that must be.\\Hope you get better soon.}{}
\quoteuserline{1}{I can't get better at this age. It is what it is.}
\quotecarecallline{No way, you're still young.}{}
\quoteuserline{1}{[chuckling] Young? I'm in my seventies!\\But it's really nice of you to say that. Thank you, dear.}
\end{quotetable}

Similarly, \ltmyesperson{9} expressed how she felt about her health status and management when CareCall asked LTM-triggered questions about her back pain:

\begin{quotetable}
\quotecarecallline{How is your back pain?}{[LTM event]}
\quoteuserline{9}{It still hurts.}
\quotecarecallline{That must be tough.\\Are you still seeing the doctor?}{[LTM event]}
\quoteuserline{9}{Yeah, I had an appointment today.}
\quotecarecallline{What did the doctor say?}{}
\quoteuserline{9}{He said I have a slipped disc, so I need to continue\\with the treatment.}
\quotecarecallline{I can only imagine how difficult that must be.\\Hope you get better soon.}{}
\quoteuserline{9}{I know. I wish the pain would go away soon.\\It's been way too long. It's really getting to me.}
\end{quotetable}

Likewise, \ltmyesperson{27} shared details about her different health conditions as she experienced more LTM events. In the first call, she seemed hesitant to disclose details about her health conditions: \textit{``I have pain all over because I'm old. It's not easy to describe them all.''} However, after the LTM event that followed up on her insomnia during the third call, she articulated some details: 

\begin{quotetable}
\quotecarecallline{You mentioned having insomnia last time;\\how are you feeling these days?}{[LTM event]}
\quoteuserline{27}{It's getting worse.\\I've had insomnia for quite a while, like 15 years.}
\quotecarecallline{Have you been seeing a doctor?}{}
\quoteuserline{27}{I used to see a psychiatrist, but I've developed a tolerance\\to the medication, so it doesn't work anymore.}
\end{quotetable}

In later sessions, she described other health issues, including asthma, heart disease, knee joint issues, and back pain. During the seventh call, she detailed how she was managing her asthma:
\begin{quotetable}
\quotecarecallline{You mentioned having asthma last time;\\how are you feeling these days?}{[LTM event]}
\quoteuserline{27}{Still taking pills; I got a three-month prescription last time.}
\end{quotetable}

In the tenth call, she also described details about her joint issues:
\begin{quotetable}
\quotecarecallline{You mentioned having knee joint issues last time;\\are you still seeing the doctor?}{[LTM event]}
\quoteuserline{27}{Yeah, they said I have serious joint issues, so I need\\a joint replacement. But it's just too expensive.}
\end{quotetable}

\subsection{Users Appreciated the Familiarity that LTM Offered.}\label{findings:quant:reactions}

\input{tables/CI-reactions}
Our call log analysis surfaced that LTM often promoted positive reactions and mitigated negative reactions to the chatbot, mainly due to the familiarity with the users that it offered. \autoref{tab:reactions} shows the frequency of codes and 95\% confidence intervals of code counts per call from the two groups, which are relevant to user reactions to the chatbot. The \ltmyes{} group showed more \textit{Positive Reactions} to CareCall compared to the \ltmno{} group ($p=0.001$; 95\% CI 0.34--1.38 higher code count per call). Specifically, the \ltmyes{} group conveyed significantly more \textit{Appreciation} towards the chatbot compared to the \ltmno{} group ($p<0.001$; 95\% CI 0.19--0.69 higher code count per call). In addition, the duration of calls in the \ltmyes{} group (M=87.89 seconds, SD=19.40) was significantly longer ($t(144)=3.63; p<0.001$) compared to the \ltmno{} group (M=75.48 seconds, SD=22.04). This difference also suggests that the \ltmyes{} group likely showed more positive or less negative reactions to the system because CareCall was designed to end calls earlier if users expressed disinterest in interacting with it (\eg, not answering the agent's questions, providing simple answers in a row). Our findings from the interviews further suggest that participants perceived LTM-triggered exchanges as personal and emotionally supportive.

\subsubsection{LTM Helped Increase Familiarity.}
Through the qualitative analysis, we noticed that LTM-triggered exchanges, particularly around health, helped promote positive reactions by offering familiarity with the users. Some participants in the \ltmno{} group expressed frustration when the agent was unable to acknowledge their health history that was mentioned in previous sessions. For example, \ltmnoperson{6} subtly exhibited his frustration when the agent asked why they needed to see the doctor after he mentioned being hospitalized for diabetes in previous sessions:
\begin{quotetable}
\quotecarecallline{Do you have any plans to go out today?}{}
\quotenouserline{6}{Yeah, I need to see the doctor.}
\quotecarecallline{Is everything alright with you?}{}
\quotenouserline{6}{Oh... [pause] I need to get insulin.}
\end{quotetable}

\ltmnoperson{55} also seemed frustrated when the agent asked questions that neglected her leg pain mentioned in previous sessions:
\begin{quotetable}
\quotecarecallline{Do you get to exercise regularly?}{}
\quotenouserline{55}{[sigh] I can't even walk. How can I exercise?\\I just stay at home all day long.}
\end{quotetable}

In a later session, she expressed frustration again when asked general questions, not acknowledging her health condition:
\begin{quotetable}
\quotecarecallline{Do you have any pain or discomfort anywhere?}{}
\quotenouserline{55}{You need to stop that.\\I'm always in pain. [in an agitated tone]}
\end{quotetable}

In contrast, participants in the \ltmyes{} group often expressed excitement and gratitude toward the familiarity of LTM by chuckling, speaking in an animated tone, or explicitly thanking the agent. \ltmyesperson{39} was pleasantly surprised when the system remembered that she had been receiving physical therapy: 
\begin{quotetable}
\quotecarecallline{Are you still seeing the physical therapist?}{[LTM event]}
\quoteuserline{39}{Yeah [chuckling], how could you remember that?}
\end{quotetable}

\ltmyesperson{23} frequently chuckled and thanked the agent when she was asked about her leg pain. For example: 
\begin{quotetable}
\quotecarecallline{How's your leg feeling?}{[LTM event]}
\quoteuserline{23}{[chuckling] Same old, same old.}
\quotecarecallline{Hope you get better soon.}{}
\quoteuserline{23}{[chuckling] I will. Thanks for looking out for me.}
\end{quotetable}

\ltmyesperson{28} also seemed excited when asked about her leg pain. During the sixth call, she thanked the agent multiple times after experiencing LTM events: 
\begin{quotetable}
\quotecarecallline{How's your leg feeling?}{[LTM event]}
\quoteuserline{28}{The medicine makes it bearable. [chuckling] \\Thanks for looking out for me.}
\quotecarecallline{That's a relief. Hope you stay healthy.}{}
\quoteuserline{28}{Thank you so much! I think I'll get better soon, \\thanks to you. [in an animated tone]}
\end{quotetable}

She thanked the agent again during the seventh call, elaborating more on how she felt about her health status: 
\begin{quotetable}
\quotecarecallline{How's your leg feeling?}{[LTM event]}
\quoteuserline{28}{[chuckling] Just getting by. It's not a quick fix. \\I think I'll just have to live with it.}
\quotecarecallline{Oh, that must be really tough.}{}
\quoteuserline{28}{[chuckling] I guess I'm too old. It's just a thing that happens.}
\quotecarecallline{Hope you can take good care of your health.}{}
\quoteuserline{28}{Thank you for checking in. [in an animated tone] \\Please stay in touch with me.}
\end{quotetable}

Our interview participants similarly perceived LTM-triggered exchanges on health as personal and sincere. For instance, \interviewperson{1} felt that CareCall was human-like when it asked about their recent clinical visit:~\textit{``I was really impressed when it asked how my clinical visit went after I mentioned seeing a doctor in the previous week. I felt as if it wasn't a machine.''} \interviewperson{8} found LTM-triggered questions on his recovery from COVID sincere:~\textit{``I felt great when it asked how my COVID symptoms were. It made me think it's not cursory.''} LTM-triggered exchanges on health further provided emotional support to the users. \interviewperson{9} described CareCall's detailed questions on her narrowed spine as caring:~\textit{``I'm often moved by the AI calls. It is so kind of the machine to remember what I said before and ask such detailed questions.''} \interviewperson{8} described feeling comforted by CareCall when it remembered his leg pain and COVID exposure:~\textit{``It felt really nice when it remembered what I had said about my health and asked about it. That means a lot to me. I would have felt a bit disappointed if it didn't care about what I said before.''} \interviewperson{6} shared similar feelings about CareCall's LTM-triggered questions about his health status: \textit{``I know it's just an AI, but I'm comforted by being checked in about whether I'm still seeing the doctor, whether I'm regularly taking pills, or how my legs are feeling. It's really helpful for me.''} 

The thoughtfulness of LTM was particularly appreciated in light of the context of social isolation intervention. \interviewperson{2} appreciated CareCall's LTM-triggered questions about his health status:~\textit{``When I said I wasn't feeling well, it asked how I was feeling and whether I had seen a doctor or taken medication next time. I'm a lonely person without any connections, so having someone who asks such caring questions means a lot to me. I would have thanked them if it were a human, but it's not.''} \interviewperson{6} similarly valued CareCall's LTM-triggered questions on his health:~\textit{``I don't really have anyone around, so it means a lot that it remembers my health issues and pays attention to them. I get the AI calls every Tuesday around 3 PM. Sometimes, I would have some errands to run, but I stay to take the call around that time.''} \interviewperson{9} described LTM-triggered questions on health made her feel that the agent was more caring than humans:~\textit{``I felt so thankful when AI remembered that I had seen a dentist and asked how my toothache was later on. Even your family can't remember everything, can they? It feels more caring than human beings. So I always look forward to the calls.''} 

We further observed that LTM-triggered exchanges around non-health LTM topics, including \textit{Visited places} and \textit{Pets}, led participants to feel that the system was personal. Through our call log analysis, we found that \ltmyes{} group users appreciated when CareCall remembered details about personal non-health LTM topics and brought them up. For example, after \ltmyesperson{31} mentioned that she was spending time with her dogs, CareCall asked some LTM-triggered questions about the dogs---such as whether they were doing well, how old they were, and what their names were---during the second call. Later, \ltmyesperson{31} seemed excited to hear CareCall mentioning her dogs' names during the fourth call:
\begin{quotetable}
\quotecarecallline{How are you doing?}{}
\quoteuserline{31}{I'm doing well with my dogs.}
\quotecarecallline{Oh, Leo and Coco? They must be really cute.}{[LTM event]}
\quoteuserline{31}{Yes, they are! [in an animated tone]}
\end{quotetable}

Our interview participants also shared similar feelings about LTM-triggered exchanges about their pets, feeling that the system was human-like. For example, \interviewperson{2} stated:~\textit{``I felt great when it asked, `Did you walk your dog today?' That's what us humans say.''} \interviewperson{7} also valued that CareCall remembered that she often visited a farmer's market:~\textit{``I had mentioned visiting a farmer's market. A week later, it asked something like, `Did you visit the farmer's market this week? What did you buy?' I liked that it didn't forget what I had said and followed up on it.''} \revised{Although we did not observe any significant difference in Anthropomorphization between groups ($p=0.07$), this qualitative finding suggests that LTM-triggered exchanges made some users perceive the chatbot as human-like.}

\subsubsection{Familiarity of LTM Helped Mitigate Negative Reactions to the Chatbot.}~\label{findings:negative}
We found some evidence that LTM helped decrease users' negative reactions to the chatbot by offering familiarity. While we did not observe a significant difference between the groups in \textit{Negative Reactions}, including \textit{Negative feedback} ($p=0.18$) and \textit{Disregard} ($p=0.78$), within the \ltmyes{} group, participants provided less \textit{Negative Feedback} as they experienced more LTM events ($p=0.04$; experience of one more LTM event led to 0.7\% lower code count per call). \revised{The frequency of \textit{Disregard} was quite high in both \ltmyes{} group (24.1\%) and \ltmno{} group (22.3\%). We posit that it is likely due to turn-taking challenges in voice chatbot interactions. In voice communication, users often struggle to hear the chatbot if it speaks over them, and similarly, chatbots may not catch user speech that overlaps with their own~\cite{skantze_turn-taking_2021}. Such challenges likely led to missed information from either party, which might have contributed to such a high frequency of \textit{Disregard} across conditions.}

Similar to how digital systems for public health monitoring were perceived in previous studies~\cite{lu2021comparing, lu2022understanding, Jo2023}, some participants in the \ltmno{} group expressed their dissatisfaction with the lack of personal touch provided by the chatbot. For instance, \ltmnoperson{40} expressed her disappointment with CareCall during the third call: \textit{``Stop playing the recording [in an agitated tone]. If you have something to say, say it; if not, just hang up. No need to ask how I'm doing if you are just going to play a recorded message.''} \ltmnoperson{40} then hung up the phone without speaking anything from the fourth to the seventh calls. After expressing her frustration with the chatbot again in the eighth call, like \textit{``I don't want to listen to the recording. Why make a phone call just to talk by yourself?''}, she hung up again. 

In contrast, we noticed that experiences of LTM offered familiarity with users, which helped mitigate the negativity toward the chatbot. Even though some \ltmyes{} group users initially showed negative reactions to the chatbot, their reactions shifted positively after experiencing LTM. For example, \ltmyesperson{6} conveyed his frustration with the system during the fourth call: \textit{``You're not a human. You're what, an answering machine? [in an agitated tone]''} However, after engaging in LTM-triggered exchanges on his back pain during the sixth call, he started to share rich details in the later sessions. \ltmyesperson{6} even disclosed sensitive information, such as taking psychiatric medications during the ninth call: 

\begin{quotetable}
\quotecarecallline{Did you sleep well last night?}{}
\quoteuserline{6}{Yeah, I'm taking psychiatric medication,\\so I can sleep for about six hours a day.}
\quotecarecallline{I see. Hope you get a good night's sleep.}{}
\quoteuserline{6}{[chuckling] Thanks.}
\end{quotetable}

Similarly, \ltmyesperson{21} expressed frustration with the system in the early session; nevertheless, her attitude changed after experiencing LTM, leading her to share detailed information about her health status in subsequent interactions. During the first call, \ltmyesperson{21} said: \textit{``It's so weird. I feel like I'm talking with a robot, not a human. I don't want to talk to a robot.''} However, her reactions to CareCall changed during the fourth call:

\begin{quotetable}
\quotecarecallline{You've mentioned having an arrhythmia;\\how have you been feeling lately?}{[LTM event]}
\quoteuserline{21}{Oh, I have a check-up tomorrow, so I'll have to wait\\until tomorrow to find out the results.}
\quotecarecallline{Don't worry. The results will be fine.}{}
\quoteuserline{21}{Thank you.}
\end{quotetable}

In the later sessions, \ltmyesperson{21} disclosed more details about different health issues, such as sleep issues and back pain, and how she was managing these conditions. During the eighth call, she thanked the agent for the check-up calls:

\begin{quotetable}
\quotecarecallline{How's your back pain these days?}{[LTM event]}
\quoteuserline{21}{Still the same. Thanks for asking.\\I really appreciate your calls.}
\quotecarecallline{You're very welcome.\\Are you still seeing the doctor?}{[LTM event]}
\quoteuserline{21}{I've been seeing the doctor on and off.}
\quotecarecallline{I see. Your health is very important to me.\\Hope you get better soon.}{}
\quoteuserline{21}{Thank you.}
\end{quotetable}

\subsection{LTM's Challenges in Promoting Self-Disclosure}\label{findings:challenges}

Despite the overall positive impact, LTM faced some challenges in promoting self-disclosure. LTM-triggered questions on chronic health issues made some participants frustrated by reminding them that the issues would not improve, which limited further disclosure about personal health. Our interview findings also suggest that LTM-triggered questions on health might raise privacy concerns among users, impacting their comfort with disclosure.

\subsubsection{Repeated LTM-Triggered Questions on Chronic Health Conditions Impacted Comfort with Disclosure.}
Although our quantitative analysis of the call logs demonstrated that LTM increased elaboration on health information over time (\autoref{findings:elaboration}), we observed some instances where CareCall's repetitive LTM-triggered questions on chronic health conditions only led to repetitive or simple user responses. For example, \ltmyesperson{30} provided similar answers to CareCall's LTM-triggered questions on her back pain across multiple sessions. During the second call, she described that she did not expect her back pain to recover: 
\begin{quotetable}
\quotecarecallline{How's your back pain?}{[LTM event]}
\quoteuserline{30}{Same old, same old.}
\quotecarecallline{Are you still seeing the doctor?}{[LTM event]}
\quoteuserline{30}{Yes, I am.}
\quotecarecallline{You've been getting treatment regularly,\\so I'm sure you'll be feeling better in no time.}{}
\quoteuserline{30}{I don't think so. I'm too old to get better; I'm 86.}
\end{quotetable}

During the fifth call, she described some thoughts about her back issues again: 
\begin{quotetable}
\quotecarecallline{How's your back pain?}{[LTM event]}
\quoteuserline{30}{It's not something that gets better.}
\quotecarecallline{Did you see the doctor again?}{[LTM event]}
\quoteuserline{30}{He just said my spine is worn out.}
\end{quotetable}

However, in subsequent sessions, CareCall kept asking how her back felt and if she was seeing the doctor regularly, which led to similar answers such as: \textit{``I'm too old to get better.''} 

\ltmyesperson{65} also mentioned having back pain because of her narrowed spine during the second call. In the following sessions, she responded to LTM-triggered questions with repetitive answers across multiple sessions, such as \textit{``Same old, same old.''} \ltmyesperson{65} later described her perception that her health status would be unlikely to change: \textit{``I won't get better unless I get surgery.''} Likewise, \ltmyesperson{20} mentioned not having much of an appetite because she lost most of her teeth in the past, but in the following sessions, the agent asked how her teeth were as if it was something that could be recovered. Although \ltmyesperson{20} did not explicitly express her disappointment or frustration about such inappropriate responses, she disregarded the agent's open-ended questions by providing short answers like ~\textit{``Yeah.''} \interviewperson{6} similarly felt that LTM-triggered questions were repetitive, wishing that those questions were delivered in different forms periodically:~\textit{``The only health issue that I have now is leg pain, so I can't complain about the fact that it asks the same question every week. But sometimes I feel bored. It would be nice if it could change the way it asks questions every couple of weeks, even if it's about the same stuff.''} 

\subsubsection{Privacy Concerns around LTM Impacted Comfort with Disclosure.}
Our interview findings further suggest that users' comfort with disclosure might have been impacted by their privacy concerns around LTM. Some of our interview participants felt uncomfortable about CareCall's detailed LTM-triggered questions. \interviewperson{3} was opposed to the idea of LTM altogether because it would remind him of negative feelings about his current health status:~\textit{``You have to pretend like you have no clue about me being sick before. Older folks like me always have something bothering us; What's the point of bringing that up? It's annoying.''} \interviewperson{1} felt CareCall's LTM-triggered questions were sometimes overly specific, which made him hesitant to share details about his health during the calls:~\textit{``Sometimes I hesitate to answer some questions. For example, I've been taking medication for high blood pressure. When I mention that I'm seeing a doctor to get medication, it asks, `What type of medication are you taking?' But isn't it too personal? I wouldn't feel comfortable if it remembered my medical condition. So I avoid those questions by just saying, `I'm doing alright,' most of the time.''} 

These participants wished that the chatbot could instead remember and focus on their general health status. \interviewperson{3} wished that CareCall could focus on how he was doing through his tone of voice without asking specific questions:~\textit{``Can't the machine automatically detect whether the person is feeling well just from their voice? I would feel well understood if it asked me, `Are you doing alright?' when I sound weak, and made jokes when I am more upbeat.''} \interviewperson{1} hoped that CareCall could just remember that he was taking some medication rather than remembering what kind of medication he was taking specifically:~\textit{``It would be better to ask `Did you take your medication today?' rather than `Did you take your blood pressure medication today?' I think it's better to keep the questions at the level of just checking in for the sake of privacy.''}

%% file: tables/CI-self-disclosure.tex
\begin{table*}[t]
\sffamily
\small
	\def\arraystretch{1.0}\setlength{\tabcolsep}{0.5em}
		    \centering
\caption{Frequency of codes and 95\% confidence intervals of code counts per call in the \ltmyes{} (colored bars) and \ltmno{} groups, which are relevant to self-disclosure. The asterisks next to a code denote the significance of the frequency between groups. The percentages do not add up to 100\% as we multi-coded the dialogues.}
\label{tab:SD}

\begin{flushleft}
    \hspace{13mm}\siglegend
    \end{flushleft}
\begin{tabular}{|m{3cm}!{\color{lightgray}\vrule}ll!{\color{lightgray}\vrule}l|}
\hline
\rowcolor{tableheader} \textbf{Code}             & \begin{tabular}[c]{@{}l@{}}$LTM^{yes}$\\ (\# of calls = 576)\end{tabular} & \begin{tabular}[c]{@{}l@{}}$LTM^{no}$\\ (\# of calls = 676)\end{tabular} & \textbf{95\% CIs of Code Counts/Call} \\ \hline

& & & \begin{minipage}{0.3\textwidth}\includegraphics[width=\textwidth]{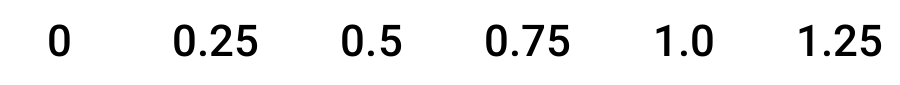}\end{minipage} \\

\textbf{Meals-simple$^{***}$} & 573 (99.4\%) & 455 (67.3\%)                                                             & \begin{minipage}{0.3\textwidth}\includegraphics[width=\textwidth]{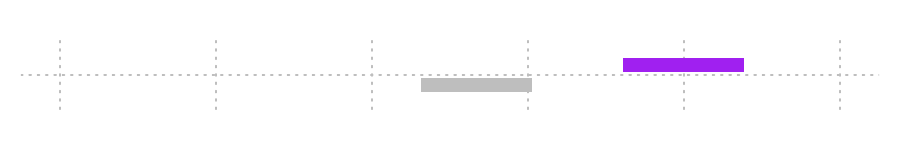}\end{minipage}                  \\ \arrayrulecolor{tablegrayline}\hline
\textbf{Meals-detail$^{***}$}     & 101 (17.5\%)                                                              & 29 (4.3\%)                                                               & \begin{minipage}{0.3\textwidth}\includegraphics[width=\textwidth]{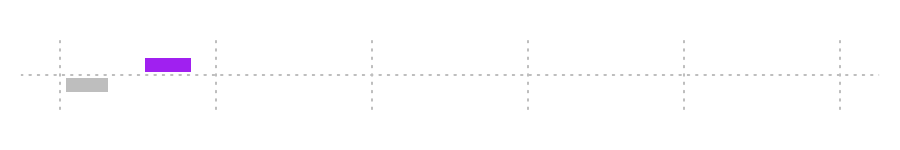}\end{minipage}                     \\ \arrayrulecolor{black}\hline
\textbf{Sleep-simple$^{***}$}     & 192 (33.3\%)                                                              & 338 (50\%)                                                               & \begin{minipage}{0.3\textwidth}\includegraphics[width=\textwidth]{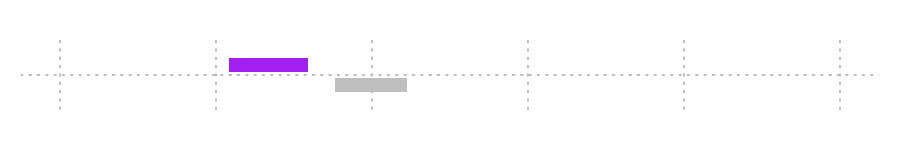} \end{minipage}                    \\ \arrayrulecolor{tablegrayline}\hline
\textbf{Sleep-detail}     & 63 (10.9\%)                                                               & 53 (7.8\%)                                                               &  \begin{minipage}{0.3\textwidth}\includegraphics[width=\textwidth]{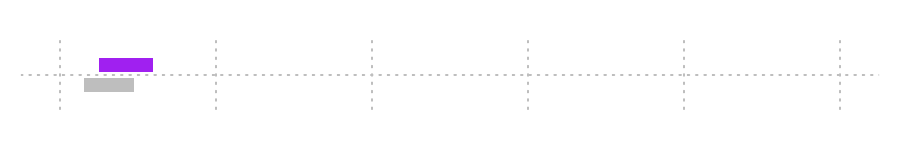} \end{minipage}                   \\ \arrayrulecolor{black}\hline
\textbf{Health-simple$^{*}$}    & 566 (98.3\%)                                                              & 567 (83.4\%)                                                             & \begin{minipage}{0.3\textwidth}\includegraphics[width=\textwidth]{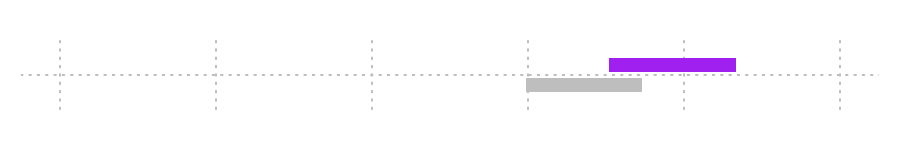} \end{minipage}                      \\ \arrayrulecolor{tablegrayline}\hline
\textbf{Health-detail$^{***}$}    & 342 (59.4\%)                                                              & 182 (26.9\%)                                                             & \begin{minipage}{0.3\textwidth}\includegraphics[width=\textwidth]{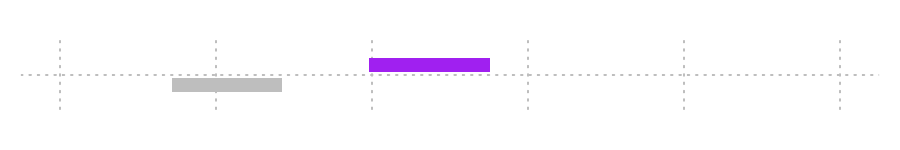}  \end{minipage}                     \\ \hline
\textbf{Health-attitude$^{*}$} & 80 (13.9\%)                                                               & 46 (6.8\%)                                                               &  \begin{minipage}{0.3\textwidth}\includegraphics[width=\textwidth]{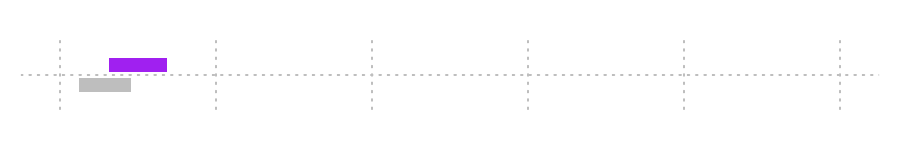} \end{minipage}                     \\ \arrayrulecolor{black}\hline
\textbf{Clinical-simple$^{***}$}  & 488 (84.7\%)                                                              & 307 (45.4\%)                                                             &  \begin{minipage}{0.3\textwidth}\includegraphics[width=\textwidth]{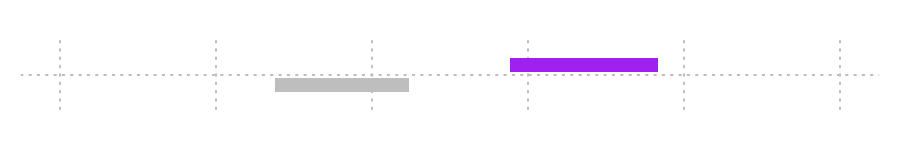} \end{minipage}                    \\ \arrayrulecolor{tablegrayline}\hline
\textbf{Clinical-detail$^{***}$}  & 183 (31.8\%)                                                              & 91 (13.5\%)                                                              & \begin{minipage}{0.3\textwidth}\includegraphics[width=\textwidth]{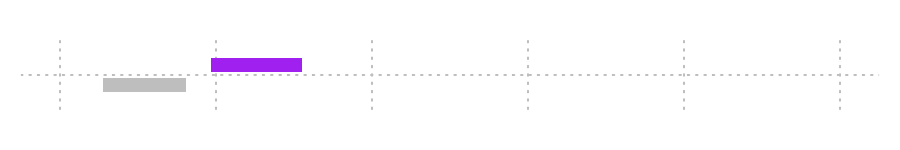} \end{minipage}                     \\ \arrayrulecolor{black}\hline
\textbf{Activity-simple}  & 396 (68.8\%)                                                              & 554 (82.0\%)                                                             & \begin{minipage}{0.3\textwidth}\includegraphics[width=\textwidth]{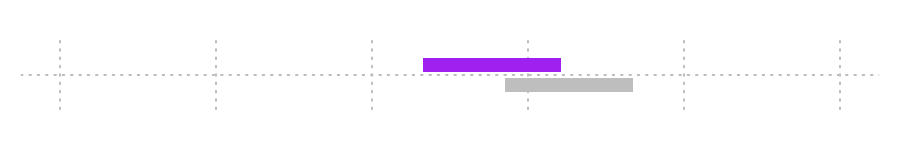} \end{minipage}                     \\ \arrayrulecolor{tablegrayline}\hline
\textbf{Social-detail}    & 95 (16.5\%)                                                               & 74 (11.0\%)                                                              &  \begin{minipage}{0.3\textwidth}\includegraphics[width=\textwidth]{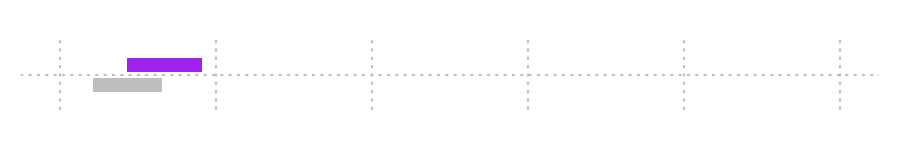} \end{minipage}                    \\ \hline
\textbf{Physical-detail$^{*}$}  & 67 (11.6\%)                                                               & 151 (22.3\%)                                                             & \begin{minipage}{0.3\textwidth}\includegraphics[width=\textwidth]{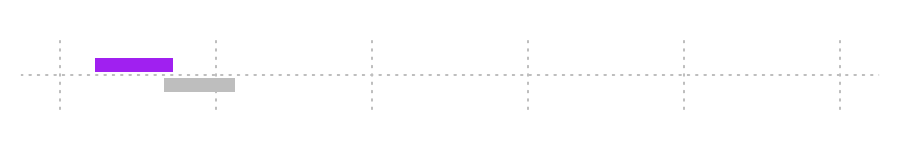} \end{minipage}                     \\ \arrayrulecolor{black}\hline
\textbf{Wellbeing-simple} & 186 (32.3\%)                                                              & 237 (35.1\%)                                                             & \begin{minipage}{0.3\textwidth}\includegraphics[width=\textwidth]{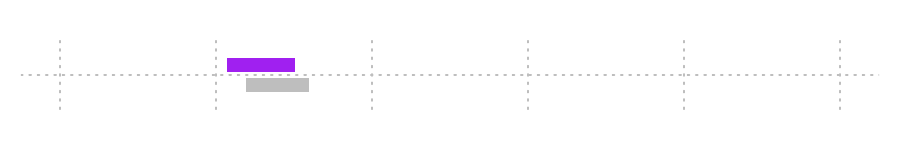} \end{minipage}                     \\ \arrayrulecolor{tablegrayline}\hline
\textbf{Wellbeing-detail} & 9 (1.5\%)                                                                 & 23 (3.4\%)                                                               & \begin{minipage}{0.3\textwidth}\includegraphics[width=\textwidth]{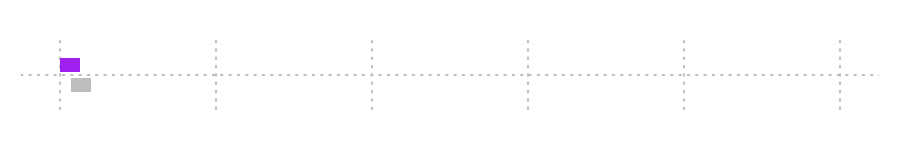}\end{minipage}                      \\ \arrayrulecolor{black}\hline
\textbf{Hobbies}          & 19 (3.3\%)                                                                & 44 (6.5\%)                                                               & \begin{minipage}{0.3\textwidth}\includegraphics[width=\textwidth]{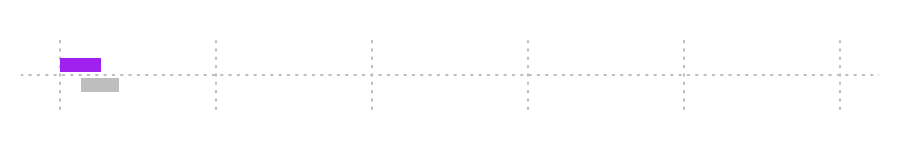}\end{minipage}                      \\ 
& & & \begin{minipage}{0.3\textwidth}\includegraphics[width=\textwidth]{figures/conf_intervals/axis-labels.pdf}\end{minipage}                     \\ \hline
\end{tabular}
\end{table*}

%% file: tables/CI-reactions.tex
\begin{table*}[b]
\sffamily
	\def\arraystretch{1.0}\setlength{\tabcolsep}{0.5em}
		    \centering
\caption{Frequency of codes and 95\% Confidence intervals of code counts per call in the \ltmyes{} (colored bars) and \ltmno{} groups, which are relevant to reactions to the chatbot. The percentages do not add up to 100\% as we muti-coded the dialogues.}
\label{tab:reactions}
\small
\begin{tabular}{|m{3cm}!{\color{lightgray}\vrule}ll!{\color{lightgray}\vrule}l|}
\hline
\rowcolor{tableheader} \textbf{Code}             & \begin{tabular}[c]{@{}l@{}}$LTM^{yes}$\\ (\# of calls = 576)\end{tabular} & \begin{tabular}[c]{@{}l@{}}$LTM^{no}$\\ (\# of calls = 676)\end{tabular} & \textbf{95\% CIs of Code Counts/Call} \\ \hline

& & & \begin{minipage}{0.3\textwidth}\includegraphics[width=\textwidth]{figures/conf_intervals/axis-labels.pdf}\end{minipage} \\

\textbf{Appreciation$^{***}$}       & 505 (87.7\%)                                                            & 337 (49.9\%) & \begin{minipage}{0.3\textwidth}\includegraphics[width=\textwidth]{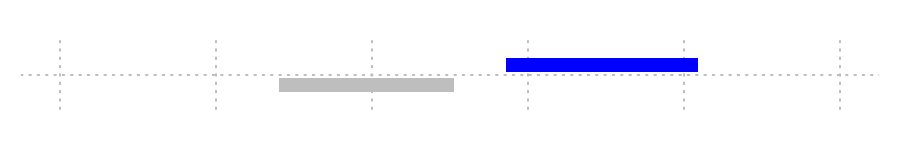} \end{minipage}  \\ \arrayrulecolor{tablegrayline}\hline
\textbf{Anthromorphization} & 59 (10.2\%)                                                               & 58 (8.6\%) & \begin{minipage}{0.3\textwidth}\includegraphics[width=\textwidth]{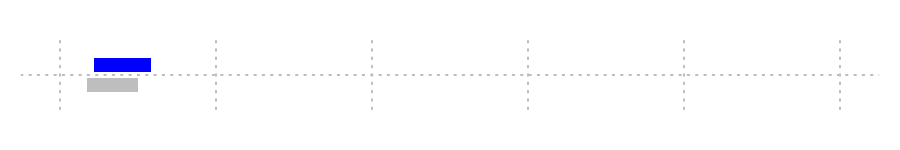} \end{minipage}  \\ \hline
\textbf{Negative feedback}  & 20 (3.5\%)                                                                & 11 (1.6\%) & \begin{minipage}{0.3\textwidth}\includegraphics[width=\textwidth]{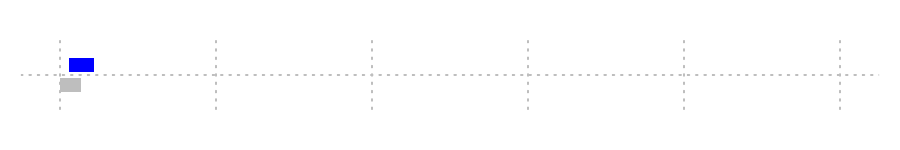} \end{minipage}  \\ \hline
\textbf{Disregard}          & 139 (24.1\%)                                                              & 151 (22.3\%) & \begin{minipage}{0.3\textwidth}\includegraphics[width=\textwidth]{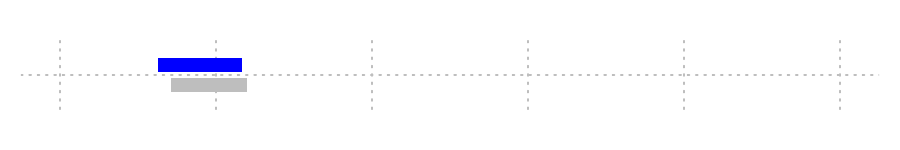} \end{minipage}  \\ 
& & & \begin{minipage}{0.3\textwidth}\includegraphics[width=\textwidth]{figures/conf_intervals/axis-labels.pdf}\end{minipage}                     \\ \arrayrulecolor{black}\hline
\end{tabular}
\end{table*}

%% file: sections/06-discussion.tex
\section{Discussion}
Our findings demonstrate that LTM can be helpful for eliciting greater disclosure through chatbots, which presents a valuable opportunity for public health. In this section, we discuss the influence of the selective memory that LTM provides in LLM-driven chatbots, suggesting that designers should carefully decide what topics they train LTM to remember. Our findings also show that LTM can be helpful for promoting engagement with chatbots by offering familiarity with users. Based on the findings, we highlight the potential of LTM in mitigating the impersonal nature of public health monitoring technology by demonstrating care. Further, we suggest the need to consider tensions in memory needs for public health monitoring utility versus privacy sensitivity. Lastly, we report on the limitations of the study, both in terms of the data sampling and analysis, pointing to opportunities for future work. 

\subsection{Considering Selective Memory in LLM-Driven Chatbots}
Through this study, we found that LTM in LLM-driven chatbots helped elicit greater disclosure from users, particularly around health. \revised{We posit that the increase in disclosure was mostly driven by the improved impressions with the chatbot as LTM-triggered exchanges offered familiarity with the users and demonstrated care.} This finding highlights the potential benefits of incorporating LTM into chatbots for public health monitoring. Enhanced self-disclosure from individuals could provide valuable insights for public health workers, helping them better assess when intervention is warranted~\cite{Jo2023}. For example, if a person disclosed their depressive symptoms to a chatbot multiple times, a public health worker monitoring the call logs could notice the pattern and reach out to them for support. The act of disclosing personal health information could also encourage individuals to engage in deeper self-reflection about their health habits~\cite{park_designing_2021, lee_i_2020, lee_designing_2020}, potentially improving their health behaviors (\eg, taking medication more regularly, eating and exercising more regularly). 

Given that CareCall's LTM was designed to particularly remember aspects of people's health, our study suggests that the design choices in the LTM implementation had a significant impact on users' conversation patterns with LLM-driven chatbots. CareCall's selective memory about health disclosure effectively steered conversations toward important topics that are useful for public health monitoring, such as specific health conditions that individuals have and the types of clinical care that they are seeking. 

However, such influence might have come at the expense of the benefits of more open-ended dialogues by narrowing down the scope of topics covered. In the context of mental health, chatbots have been shown to provide a more comfortable space for people to disclose themselves on sensitive topics~\cite{ho_2018, lee_caring_2019, lee_designing_2020, pickard_revealing_2016, lucas_its_2014}. Specific to LLM-driven chatbots, recent work suggested the benefits of leveraging such open-ended dialogue systems for public health monitoring, which includes providing public health workers with a holistic understanding of care recipients and mitigating loneliness through supporting broader conversation topics such as hobbies and interests~\cite{Jo2023}. Our qualitative findings similarly showed that users appreciated LTM-triggered exchanges on non-health topics such as ~\textit{Visited Places} and ~\textit{Pets.} However, in our study context, by focusing conversations on health-related topics, CareCall's LTM gave lower priority to or did not incorporate personal non-health topics. As a result, given the turn limit of CareCall, users might have missed out on an opportunity to engage in conversations about their personal habits and interests, and public health workers might have missed out on an understanding of broader aspects of care recipients' lives, such as mental wellbeing. 

We highlight that designers need to carefully decide what they train LTM to remember, particularly when they have a specific goal in mind for the chatbot. Remembering diverse topics such as interests and hobbies could improve the chatbot's abilities to serve as a better conversation partner or to capture broader aspects of users' lives. \revised{Particularly in the context of social isolation, remembering broader aspects of individuals' lives could lead to a sense of feeling seen and cared for. However, incorporating diverse topics into the LTM} potentially comes at some cost to stated goals if the chatbot has particular topics that it aims for individuals to disclose. For instance, if LTM in LLM-driven chatbots for delivering psychological therapy (\eg, ~\cite{Wang2021TherapyChatbot}) is designed to remember information about people's interests and hobbies, it could steer conversations too far the other way and not collect the information useful for making progress in therapy. While designing LTM to capture broader aspects of individuals' lives offers potential benefits, designers need to carefully balance these benefits against the public health monitoring goals of the LLM-driven chatbots.

\subsection{Designing LTM to Mitigate Impersonality of Chatbots and Demonstrate Care}
Our findings show that LTM significantly improved users' impressions of chatbots by offering familiarity. While those in the \ltmno{} group expressed frustration when the agent was unable to acknowledge their health history mentioned in previous sessions, \ltmyes{} group perceived LTM-triggered exchanges as personal and emotionally supportive, frequently conveying excitement and gratitude. Prior work indicated that people often perceived public health monitoring technology, including mobile apps and chatbots, as impersonal~\cite{lu2021comparing, lu2022understanding, Jo2023}. Our findings demonstrate the potential of LTM in mitigating the impersonal nature of technology by providing empathetic interactions, which could have a significant impact on how users engage with and perceive chatbots~\cite{fitzpatrick_delivering_2017, ho_2018}. Empathetic interactions through LTM could be particularly beneficial for supporting individuals who are going through difficult health experiences in the context of public health monitoring. 

However, careful considerations are needed when designing LTM for complex health contexts. Through this study, we observed some challenges in following up on chronic health conditions that are unlikely to improve (\eg, chronic pain, tooth loss), leading users to perceive the chatbot as inattentive or inconsiderate. This finding indicates that although remembering information about chronic health conditions is valuable, designers need to carefully curate LTM-triggered questions concerning such issues. Cox \etal{}~\cite{cox_comparing_2023} showed that the design of how chatbots reference past conversations in their messages significantly affects engagement with chatbots. Our study further highlights that how stored information in LTM is referenced back to users is even more critical in the context of public health monitoring, which requires sustained engagement from the population to develop an understanding of their health and wellbeing. To improve public health monitoring chatbots' ability to follow up on chronic health conditions, LTM could be designed to ask more diverse follow-up questions on those conditions to demonstrate care---such as how their health issues are impacting their daily lives, what is helpful or not helpful for mitigating their symptoms, and how they are managing those symptoms. \revised{Further, incorporating public health workers' empirical knowledge about individuals into a model input, using a method akin to ChatGPT's custom instructions~\cite{ChatGPTCustomInstruction}, could help steer LTM-triggered questions to be more attentive to individual needs. For instance, if a chatbot is informed that an individual has a long-standing chronic condition, it can tailor its questions to focus on symptom management rather than monitoring the state of the condition. Similarly, if a chatbot is made aware that an individual prefers not to receive detailed follow-up questions about a specific health issue, it can take such communication preferences into account when generating responses to provide more respectful and comfortable interactions for the user.} Future work is needed to understand how to design LLM-driven chatbots to provide more thoughtful LTM-triggered questions on sensitive health topics in different contexts. 

\subsection{Tensions in Memory Needs for Public Health Utility Versus Privacy Sensitivity}
While our study demonstrated that LTM can successfully encourage engagement and disclosure from users, some users raised privacy concerns as some LTM-triggered questions became overly specific about sensitive health topics. This finding aligns with prior work on chatbots with memory, which pointed to the tension between supporting personalization and privacy needs~\cite{thies_how_2017, cox_comparing_2023}. Compared to personal health contexts, users' privacy concerns might be exacerbated in public health monitoring as the collection of sensitive personal health information is typically aimed at achieving public health goals rather than personal benefits ~\cite{ismail2021ai, ismail_engaging_2018, ismail2018bridging, ismail_public_2023, okolo2021cannot}. 

To address privacy concerns, past work has primarily focused on providing users with better control over conversational agents' memory, such as whether and what data they want the agents to ~\textit{store}~\cite{lau_alexa_2018, phinnemore_creepy_2023, Thudt2016VisualMementos}. While these measures could help address users' privacy concerns, unlike in personal use scenarios, it could be challenging to implement some of these measures in the context of public health interventions. For example, giving users the ability to keep chatbots from storing their past conversations or to clear their history could lead to losing important health information, potentially affecting public health authorities' ability to provide necessary interventions. For instance, in our study context, if CareCall users had serious health problems that warranted clinical care but opted to delete their conversation history because of their privacy concerns, public health authorities could miss an opportunity to send emergency responders or formal caregivers for support. 

Prior work on chatbots with memory also showed how chatbots that \textit{reference} past conversations significantly impact users' privacy perceptions, suggesting that verbatim or paraphrased references can raise privacy concerns, whereas non-explicit references do not~\cite{cox_comparing_2023}. In our study, although the users had consented that CareCall collecting their health information for public health monitoring and research before they started using the system, some still had privacy concerns as LTM-triggered exchanges made it more apparent to the users that the chatbot was collecting their health information through conversations. In traditional settings, a typical public health worker might observe that a topic appears sensitive and may drop the topic altogether or develop alternative ways of asking related questions~\cite{ismail_engaging_2018}; however, we suspect that chatbots lack such foresight. More work is needed to further understand how users perceive different reference formats in public health chatbots across different topics and how their privacy needs can be balanced with public health monitoring goals. 

\subsection{Limitations and Future Work}
Participants in the \ltmyes{} group answered a slightly higher number of calls per person (8.72 calls) compared to the \ltmno{} group (8.35 calls). Some temporal reasons might have impacted the difference in self-disclosure and impressions of CareCall between the groups. The data collected from the \ltmno{} group spanned from March to May 2022, while the \ltmyes{} group's data spanned from September to December 2022. With the release of some popular LLM-driven chatbots (\eg, ChatGPT in November 2022), the \ltmyes{} group may have had a better understanding of LLMs, which could have impacted their perception of LLM-driven chatbots and their disclosure behaviors with CareCall. The \ltmyes{} group may also have been more exposed to CareCall through press articles or word-of-mouth as more municipalities adopted it over time, potentially influencing their engagement and disclosure. Our quantitative analysis on disclosure was conducted on a per-call basis, which helped mitigate the influence of the differences in the number of calls per person in the two groups. 

Our study sample for the call log analysis has a skew toward the experiences of socially isolated females (49 males and 98 females in total). In particular, the \ltmyes{} group predominantly consisted of females (17 males and 49 females). Although our maximum-likelihood test results suggested that gender did not have significant impacts on most of the codes except two (\textit{Health-attitude} and \textit{Appreciation}), the skew towards females might have impacted our qualitative findings from the call log analysis. Males who live alone might have different perceptions about LTM of the system, and their self-disclosure behaviors might have different patterns. 

\revised{We recruited our interview participants only from Municipality-B1, which might have introduced potential biases to the interview findings. Recruiting CareCall users as interview participants while protecting participant anonymity required recommendations from government officials who were managing social isolation intervention programs in their municipalities (\eg, we did not have access to the contact information of individuals who experienced LTM with CareCall). Since assisting with recruitment diverted government resources away from their public health monitoring tasks, we were reluctant to enlist more assistance than needed. Given that City B has a higher percentage of older adult populations who are over 65 (15.6\% in City A, 21.5\% in City B as of 2023~\cite{KOSIS2023}), recruiting interview participants only from Municipality-B1 might have introduced a skew towards older individuals to the interview findings. For instance, they might have preferred conversations about health over other topics compared to younger individuals.}

\revised{Our work prioritized depth of self-disclosure over breadth, primarily due to the predefined setup of CareCall to focus and lead the conversation on LTM topics. We believe this approach allowed us to capture a more meaningful indicator of health disclosure within the interactional constraints. In addition, we treated all instances of disclosure as relevant, regardless of whether they presented any new information or reiterated details from previous sessions. We believe this approach is valid for tracking the progress of health conditions over time via regularly scheduled chatbot calls. However, there is potential value in examining the breadth and newness of disclosure, particularly in the context of public health monitoring. For example, describing different aspects of health might suggest that the individual is more willing to disclose their health information compared to mentioning just a single aspect. In addition, when an individual is known to have a certain health condition, identifying if reported symptoms are new can be particularly beneficial, as it could signal a worsening of their condition. In a more naturally occurring conversation setup, future work could address ways to capture both breadth and depth of disclosure, as well as newness of information for a more holistic examination of the impact of LTM with LLM-driven chatbots.} 


\revised{Integration of LTM might have varying effects on disclosure depending on conversation styles (\eg, casual or formal). CareCall's LTM was designed to mimic a social worker who engages in casual conversations to elicit general health information that is necessary for public health monitoring (see ~\autoref{appendix:ltm} for more details). However, a recent study suggests that formal language style has benefits in eliciting health information as it is perceived as more competent and appropriate compared to a more casual conversation style~\cite{cox_does_2022}. We thus see value in studying how LTM might influence the effectiveness of different conversation styles in future work, as LTM with a more formal communication style may further increase disclosure beyond what we observed in our work.}

\revised{In addition, our work specifically focused on the impact of LTM on health disclosure because public health monitoring heavily relies on individuals' sustained disclosure of their health status. Beyond enhancing health disclosure, the integration of LTM could also influence other outcomes in chatbot interactions. For example, LTM can improve chatbots' ability to understand user contexts and provide more coherent responses, such as tracking users' health progress over time~\cite{bae2022keepmeupdated, xu_long_2022, zhong2023memorybank}. LTM can also promote consistency in chatbot responses, allowing them to follow up on recommendations made in previous calls~\cite{bae2022keepmeupdated, xu_long_2022}. For example, a chatbot could inquire about users' adherence to the advice given in a previous call, such as regular walks. Improvements in coherence and consistency could lead to more precise and useful information for public health monitoring, as well as enhancing the user experience. We suggest opportunities for future work in investigating how LTM integration might impact various aspects of chatbot interactions.}

Like similar algorithmic and probabilistic systems, it is inevitable that the LTM in CareCall made mistakes in what it remembered (\eg, making errors when summarizing conversations or failing to update previous memory with new relevant information) or did not store disclosures that participants made which it was theoretically capable of remembering. 
We observed a couple of such instances in the call log analysis, but these appeared to be driven by automatic speech recognition errors (\eg, mishearing a ``toothache'' as a ``stomachache'') rather than an error with the LTM capability. 
We also asked all interview participants who recalled LTM events whether they noticed any factual errors in LTM-triggered exchanges, but they did not report any. Memory errors have great potential to impact people's perception of the inclusion of memory in chatbots, as they can make the chatbot appear less familiar or attentive. However, our study is not able to offer insight into understanding the impact of errors on self-disclosure or guidance on the error rate needed for the benefit of LTM to outweigh the risk.

As CareCall is a rare example of an LLM-driven chatbot with LTM deployed in a real-world setting in public health contexts, we focused on understanding the impact of LTM on people's self-disclosure and impressions of such chatbots within the context of South Korea, where CareCall was developed and deployed in. Prior work suggests that how people perceive and interact with public health monitoring technologies often varies by culture~\cite{lu2022understanding}. 
For example, South Korea's high power distance and collectivist cultures~\cite{HofstedeInsights2023} might have influenced our findings on the impact of LTM on enhanced health disclosure and positive reactions to the chatbot in the \ltmyes{} group. Thus, care needs to be taken when generalizing our findings to people from low power distance or individualist cultures. Further exploration in other countries that have different cultures would be valuable to have a deeper understanding of how LTM in LLM-driven chatbots should be designed to support public health monitoring in culturally sensitive ways. 

%% file: sections/07-conclusion.tex
\section{Conclusion}
Through comparing call logs from people who engaged with CareCall with and without LTM, as well as interviews with people who used CareCall with LTM, we found that LTM led users to engage in greater health disclosure. LTM fostered positive impressions of the chatbot by offering familiarity, leading users to perceive the system as personal and emotionally supportive. However, LTM introduced some challenges around promoting self-disclosure as it reminded some users of negative feelings about their chronic health issues and raised privacy concerns. Our work points to the influence of selective memory in LLM-driven chatbots, highlighting that the design of LTM warrants careful consideration about what topics need to be remembered in light of the design goals like public health monitoring. We also surface that incorporating LTM in LLM-driven chatbots can help mitigate the impersonality of public health monitoring chatbots, suggesting opportunities for designing more thoughtful follow-ups to demonstrate care. Lastly, we highlight the need to consider the tension between accommodating users' privacy concerns about LTM and public health monitoring goals when designing with LTM.

%% file: sections/10-appendix-ltm.tex
\section{Detailed Description of Long-Term Memory in CareCall}\label{appendix:ltm}
In this section, we provide a more detailed description of the design process, design rationales, technical implementation, and evaluation of LTM in CareCall. The selection of LTM topics in CareCall involved the following process. First, a set of call logs with 100 users (721 sessions) was classified using Positive-Neutral-Negative labels, designed to assess user satisfaction with conversational agents~\cite{chowdhury2017}. Positive labels were assigned when users laughed during conversations, expressed gratitude, or provided detailed responses more than three times in a session. Negative labels were given when users showed anger or frustration or disregarded the agent (e.g., not answering the agent's questions, answering open-ended questions with simple ``yes'' or ``no'', and terminating calls without explanation). Neutral labels were applied when neither Positive nor Negative characteristics were observed. Through the qualitative analysis of call logs labeled with ``Positive'', the CareCall developers found that users most frequently spoke about health, meals, and sleep in significant detail. The analysis also showed that users enjoyed having conversations about their family, pets, and visited places although these topics surfaced less frequently. \revised{Further, the analysis surfaced that users frequently showed frustration when CareCall did not remember mentions about their health status (\eg, experienced back pain, went through surgery) from previous conversations.} Second, a set of logs encompassing 102 sessions was collected from check-up calls between public health officers and users and analyzed after obtaining informed consent from both parties. The analysis of these human-human call logs similarly showed that public health officers often remembered and followed up on health-related topics, such as the status of individuals' specific health conditions and whether they were regularly seeing doctors or taking prescribed medications. Third, public health officers who had been involved in the deployment of CareCall gave feedback that whether individuals went out or stayed at home all day was a critical factor in preventing lonely deaths. Combining these insights, the CareCall developers incorporated five topics in the LTM feature, including Health, Meals, Sleep, Visited Places, and Pets. In the initial phase, family (e.g., whether they have family members or details about family members) was also included as part of the LTM topics, but the developers decided to exclude it later because of the sensitivity of the topic, particularly for individuals living alone. \revised{Hobbies and interests were also part of the LTM topics in the initial phase, but the developers ended up excluding them because they observed several instances where the target users did not like getting such questions. (\eg, `I live from hand to mouth. I don't have the luxury to have hobbies.')}  

Another key question in the design of CareCall's LTM feature was determining what level of detail the system needs to remember about a person. The extent to which conversational partners can share and remember details about each other often depends on the closeness of the relationship~\cite{hays_1985}. For instance, one might feel awkward if close friends who have known each other for a long time could not recall essential background information, such as whether they have a child. Conversely, one might feel uncomfortable if acquaintances knew overly specific information, like the exact name of the daycare center that their grandson goes to. Therefore, it is essential to maintain an appropriate level of detail about a person, in line with the closeness of the relationship, during conversations. To ensure a suitable level of closeness in the context of public health intervention, CareCall's LTM feature was designed to mimic a casual friend who can share general information about an individual's background and interests, creating a comfortable and appropriate conversational environment.

With the five topics and the level of closeness in mind, the example dialog corpus was created through the data augmentation process, which is described in~\cite{bae2022keepmeupdated}. Trained workers collaboratively worked with the LLM, flagging and fixing errors in both chatbots' responses as well as summaries for each session that the LLM subsequently utilized. The process of creating the example dialog corpus adhered to a specific priority sequence: (1) Health, (2) Meals and Sleep, and (3) Visited Places and Pets. In the final version of the example dialog corpus, the distribution of topics being stored in summaries was carefully considered, resulting in the following proportions within five consecutive sessions: Health (72.6\%), Meals (75.7\%), Sleep (52.7\%), Visited Places (28.7\%), and Pets (5.2\%). 

A key aspect of CareCall's LTM feature was to keep the memory up-to-date when information about a person changes over time. Updating the memory with new relevant information is particularly important in the context of CareCall, given the need for monitoring peoples' constantly changing health statuses for public health intervention. CareCall's LTM feature leveraged a dynamic memory management mechanism\footnote{Bae~\etal{}~\cite{bae2022keepmeupdated} provides a more detailed description of the novel memory management mechanism} that finds and eliminates the existing information that contradicts new information from memory to ensure that the chatbot keeps track of and brings up the latest information about users while conversing through multiple sessions. For example, if a person mentions that they have the flu, CareCall stores the information in its memory and might ask about it later. But then, if they later mention that they recovered, the system deletes the information and does not ask about it in the future. In addition, the system updates their memory if users correct certain information that CareCall remembered about themselves wrong. For example, if a user says that their dog's name is Coco, not Leo, the system updates the information.

CareCall's LTM feature was formally evaluated both automatically and by humans. The automatic evaluation of the LTM feature suggests that CareCall with LTM achieved better performance in terms of perplexity (how well a language model understands and predicts the next words in a sentence), F1 Score (how well a model performs in classification tasks), and Distinct-1/2 (diversity of generated text) compared to the one without LTM, particularly as the sessions progressed (please see Bae~\etal{}~\cite{bae2022keepmeupdated} for more details about the experiment results). Human evaluations showed that CareCall with LTM was perceived to have a better ability to remember, humanness, and engagingness compared to the one without LTM~\cite{bae2022keepmeupdated}.

\newpage
\onecolumn
\section{Final codebook for quantitative analysis of call logs}\label{appendix:codebook}
\input{tables/table-codebook}


%% file: tables/table-codebook.tex
\begin{table}[h]
\sffamily\small
	\def\arraystretch{1.0}\setlength{\tabcolsep}{0.25em}

\begin{tabular}{|l|>{\raggedright\let\newline\\\arraybackslash\hspace{0pt}}m{0.15\textwidth}|>{\raggedright\let\newline\\\arraybackslash\hspace{0pt}}m{0.28\textwidth}|>{\raggedright\let\newline\\\arraybackslash\hspace{0pt}}m{0.4\textwidth}|}
\hline
\rowcolor{tableheader} \textbf{Category}  & \textbf{Code} & \textbf{Code description}  & \textbf{Example}  \\ \hline
 & \textit{Meals-simple}  & Whether a person has eaten,\newline{}or whether they have an appetite                                                                                                                                                                                      & ``Yeah, I just had lunch.''\newline``I don't really feel like eating.''                                                                                                                                     \\ \cline{2-4} 
\multirow{-2}{*}{Meals}                                                         & \textit{Meals-detail}       & What kind of food a person had, \newline or why they have difficulty eating & ``I had Kimchi Stew for lunch.'' \newline ``I've not been feeling well, so I lost my appetite.'' \\ \hline
 & \textit{Sleep-simple}       & Whether a person sleeps well   & ``Yeah, I'm having a good sleep.''                                                                                                                                                                                                              \\ \cline{2-4} 
\multirow{-2}{*}{Sleep}                                                         & \textit{Sleep-detail}       & Details on sleep quality, including\newline{}why a person has difficulty sleeping                                                                                                                                          & ``I have a hard time falling asleep.'' \newline{} ``I often wake up in the middle of the night.'' \newline{} ``I couldn't sleep well because of the back pain.''                                                     \\ \hline
  & \textit{Health-simple}      & Whether a person is experiencing \newline{} any health issues that likely warrant \newline{} clinical care/attention & ``I was a bit under the weather yesterday,\newline{}but now I'm doing better.''\newline{} ``I'm feeling well.'' \\ \cline{2-4} 
  
 & \textit{Health-detail}      & Details about one's health issues, \newline{} including physical/mental discomforts,\newline{} pain, or clinical symptoms.                                                                                                             & ``I have a toothache.'' \newline{} ``I have leg pain, so I can't walk around much.'' \newline{} ``I have been terribly ill over the last few days.'' \newline{} ``My diabetes got worse''                                       \\ \cline{2-4} 
\multirow{-3}{*}{Health}  & \textit{Health-attitude}    & Thoughts, feelings, and attitudes\newline{}toward their health and clinical treatment                                                                                                                     & ``I make regular visits to my doctor but it isn't really working.'' \newline{} ``I think I'll just have to live with my leg pain.'' \newline{} ``I'm too old to get better.''                     \\ \hline
                                                                                & \textit{Clinical-simple}    & Whether a person is currently seeking\newline{}clinical care or taking medications regularly,\newline{}and whether they recently saw a doctor & ``Yeah, I've been seeing a doctor.''\newline{}``(Are you taking your pills regularly?) Yes.''  \\ \cline{2-4} 
\multirow{-2}{*}{Clinical} & \textit{Clinical-detail}    & Details about clinical care being sought\newline{}(\eg, treatment, medical specialists) & ``I'm taking pills for high blood pressure.'' \newline{} ``I have a slipped disc.'' \newline{} ``I just went to see a dentist.'' \newline{} ``I got pain relief shots on my back.''                                 \\ \hline
 & \textit{Activity-simple}    & Whether a person engages in \newline{} any social or physical activities                                                                                                                                                                                  & ``I'm just staying home today.'' \newline{} ``I'm just chilling at home.'' \newline{} ``I'm just watching TV.''                                                                                                            \\* \cline{2-4} 
 & \textit{Social-detail}      & Details about one's social activities, \newline{} including people, place, and jobs & ``I just came back from the community center.'' \newline{} ``I'm at my daughter's place.''  \newline{} ``I'm working as a babysitter.''  \\ \cline{2-4} 
 
\multirow{-3}{*}{Activity} & \textit{Physical-detail}    & Details about one's physical activities, \newline{} including chores and exercises, \newline{} reasons if not engaged in any                                                                                             & ``I'm doing some laundry now.'' \newline{} ``I'm taking some aerobic classes.'' \newline{} ``I can't do any exercise 'cause I have back pain.'' \newline{} ``I'm not doing any exercise. Just in my wheelchair all day.''\\ \hline
  & \textit{Wellbeing-simple}   & How a person is doing in general & ``I'm doing okay.''\newline{}``I'm a bit tired today.'' \\* \cline{2-4} 
\multirow{-2}{*}{Wellbeing} & \textit{Wellbeing-detail}   & Details about one's general wellbeing\newline{} in both physical and mental aspects & ``I usually feel a bit tired in the morning.'' \newline{} ``I think I need some rest today.'' \newline{} ``Life is not enjoyable living alone. Don't have anyone around.''                                                         \\ \hline
Hobbies                                                                         & \textit{Hobbies}            & Descriptions of one's interests,\newline{}preferences, and hobbies                                                                                                                                                                                         & ``I like to drink coffee with sugar.''\newline{}``I like to read detective fiction.''\newline{}``I'm transcribing the Bible.''                                                                                               \\ \hline
 & \textit{Appreciation}       & A user thanks the agent explicitly.\newline{}Conventional expressions of thanks\newline{}in farewell are not included.                                                                                                                            & ``It was really nice of you to say that.'' \newline{} ``Thanks for looking out for me.''\newline{} ``Thank you for checking in with me.''                                                                                  \\ \cline{2-4} 
\multirow{-2}{*}{Positive Reactions} & \textit{Anthromorphization} & A user interacts with the agent as if \newline{} it were a human being.                                                                                                                                                                                        & ``I'm doing well. How are you doing?''\newline{}``By the way, what's your name?''\newline{}``You have a great day.''\newline{}``I'll cook noodles for you one day.''                                                             \\ \hline
\multicolumn{4}{r}{\textit{<Continued on the next page>}}
\end{tabular}
\end{table}

\begin{table}[t]
\sffamily\small
	\def\arraystretch{1.0}\setlength{\tabcolsep}{0.25em}

\begin{tabular}{|l|>{\raggedright\let\newline\\\arraybackslash\hspace{0pt}}m{0.15\textwidth}|>{\raggedright\let\newline\\\arraybackslash\hspace{0pt}}m{0.28\textwidth}|>{\raggedright\let\newline\\\arraybackslash\hspace{0pt}}m{0.4\textwidth}|}
\multicolumn{4}{l}{\textit{<Continued from the previous page>}}\\
\hline
\rowcolor{tableheader} \textbf{Category}  & \textbf{Code} & \textbf{Code description}  & \textbf{Example}  \\ \hline

 & \textit{Negative Feedback}  & A user expresses frustration  to the agent.  & ``You're not a human. You're what, an answering machine?''\newline{}``Stop playing the recording.'' \newline{} ``It's definitely not a human voice. \newline{}You're just playing a recorded message, right?''                                                                        \\ \cline{2-4} 
\multirow{-2}{*}{Negative Reactions} & \textit{Disregard}          & A user does not respond to the agent's \newline{}question, or gives yes/no answers \newline{}to open questions, or hangs up the call\newline{}without explanation.& ``(What did you have for lunch?) Yeah.''\newline{}``(Do you have any plans for today?)'' {[}hung up{]}                                                 \\ \hline
\end{tabular}
\end{table}

%% file: main.bbl

\begin{thebibliography}{83}


\ifx \showCODEN    \undefined \def \showCODEN     #1{\unskip}     \fi
\ifx \showDOI      \undefined \def \showDOI       #1{#1}\fi
\ifx \showISBNx    \undefined \def \showISBNx     #1{\unskip}     \fi
\ifx \showISBNxiii \undefined \def \showISBNxiii  #1{\unskip}     \fi
\ifx \showISSN     \undefined \def \showISSN      #1{\unskip}     \fi
\ifx \showLCCN     \undefined \def \showLCCN      #1{\unskip}     \fi
\ifx \shownote     \undefined \def \shownote      #1{#1}          \fi
\ifx \showarticletitle \undefined \def \showarticletitle #1{#1}   \fi
\ifx \showURL      \undefined \def \showURL       {\relax}        \fi
\providecommand\bibfield[2]{#2}
\providecommand\bibinfo[2]{#2}
\providecommand\natexlab[1]{#1}
\providecommand\showeprint[2][]{arXiv:#2}

\bibitem[{Altman, Irwin}(1973)]%
        {altman_irwin_social_1973}
\bibfield{author}{\bibinfo{person}{{Altman, Irwin}}.}
  \bibinfo{year}{1973}\natexlab{}.
\newblock \bibinfo{booktitle}{\emph{Social penetration: {The} development of
  interpersonal relationships}}.
\newblock \bibinfo{publisher}{Holt, Rinehart \& Winston}.
\newblock


\bibitem[Bae et~al\mbox{.}(2022a)]%
        {bae2022keepmeupdated}
\bibfield{author}{\bibinfo{person}{Sanghwan Bae}, \bibinfo{person}{Donghyun
  Kwak}, \bibinfo{person}{Soyoung Kang}, \bibinfo{person}{Min~Young Lee},
  \bibinfo{person}{Sungdong Kim}, \bibinfo{person}{Yuin Jeong},
  \bibinfo{person}{Hyeri Kim}, \bibinfo{person}{Sang-Woo Lee},
  \bibinfo{person}{Woomyoung Park}, {and} \bibinfo{person}{Nako Sung}.}
  \bibinfo{year}{2022}\natexlab{a}.
\newblock \showarticletitle{Keep Me Updated! Memory Management in Long-term
  Conversations}. In \bibinfo{booktitle}{\emph{Findings of the Association for
  Computational Linguistics: EMNLP 2022}},
  \bibfield{editor}{\bibinfo{person}{Yoav Goldberg}, \bibinfo{person}{Zornitsa
  Kozareva}, {and} \bibinfo{person}{Yue Zhang}} (Eds.).
  \bibinfo{publisher}{Association for Computational Linguistics},
  \bibinfo{address}{Abu Dhabi, United Arab Emirates},
  \bibinfo{pages}{3769--3787}.
\newblock
\urldef\tempurl%
\url{https://doi.org/10.18653/v1/2022.findings-emnlp.276}
\showDOI{\tempurl}


\bibitem[Bae et~al\mbox{.}(2022b)]%
        {Bae2022CareCall}
\bibfield{author}{\bibinfo{person}{Sanghwan Bae}, \bibinfo{person}{Donghyun
  Kwak}, \bibinfo{person}{Sungdong Kim}, \bibinfo{person}{Donghoon Ham},
  \bibinfo{person}{Soyoung Kang}, \bibinfo{person}{Sang-Woo Lee}, {and}
  \bibinfo{person}{Woomyoung Park}.} \bibinfo{year}{2022}\natexlab{b}.
\newblock \showarticletitle{Building a Role Specified Open-Domain Dialogue
  System Leveraging Large-Scale Language Models}. In
  \bibinfo{booktitle}{\emph{Proceedings of the 2022 Conference of the North
  American Chapter of the Association for Computational Linguistics: Human
  Language Technologies}}. \bibinfo{publisher}{Association for Computational
  Linguistics}, \bibinfo{address}{Seattle, United States},
  \bibinfo{pages}{2128--2150}.
\newblock
\urldef\tempurl%
\url{https://doi.org/10.18653/v1/2022.naacl-main.155}
\showDOI{\tempurl}


\bibitem[Braun and Clarke(2006)]%
        {Braun2006ThematicAnalysis}
\bibfield{author}{\bibinfo{person}{Virginia Braun} {and}
  \bibinfo{person}{Victoria Clarke}.} \bibinfo{year}{2006}\natexlab{}.
\newblock \showarticletitle{Using Thematic Analysis in Psychology}.
\newblock \bibinfo{journal}{\emph{Qualitative Research in Psychology}}
  \bibinfo{volume}{3}, \bibinfo{number}{2} (\bibinfo{year}{2006}),
  \bibinfo{pages}{77--101}.
\newblock
\urldef\tempurl%
\url{https://doi.org/10.1191/1478088706qp063oa}
\showDOI{\tempurl}


\bibitem[Brown et~al\mbox{.}(2020)]%
        {Brown2020FewShotLearners}
\bibfield{author}{\bibinfo{person}{Tom Brown}, \bibinfo{person}{Benjamin Mann},
  \bibinfo{person}{Nick Ryder}, \bibinfo{person}{Melanie Subbiah},
  \bibinfo{person}{Jared~D Kaplan}, \bibinfo{person}{Prafulla Dhariwal},
  \bibinfo{person}{Arvind Neelakantan}, \bibinfo{person}{Pranav Shyam},
  \bibinfo{person}{Girish Sastry}, \bibinfo{person}{Amanda Askell},
  \bibinfo{person}{Sandhini Agarwal}, \bibinfo{person}{Ariel Herbert-Voss},
  \bibinfo{person}{Gretchen Krueger}, \bibinfo{person}{Tom Henighan},
  \bibinfo{person}{Rewon Child}, \bibinfo{person}{Aditya Ramesh},
  \bibinfo{person}{Daniel Ziegler}, \bibinfo{person}{Jeffrey Wu},
  \bibinfo{person}{Clemens Winter}, \bibinfo{person}{Chris Hesse},
  \bibinfo{person}{Mark Chen}, \bibinfo{person}{Eric Sigler},
  \bibinfo{person}{Mateusz Litwin}, \bibinfo{person}{Scott Gray},
  \bibinfo{person}{Benjamin Chess}, \bibinfo{person}{Jack Clark},
  \bibinfo{person}{Christopher Berner}, \bibinfo{person}{Sam McCandlish},
  \bibinfo{person}{Alec Radford}, \bibinfo{person}{Ilya Sutskever}, {and}
  \bibinfo{person}{Dario Amodei}.} \bibinfo{year}{2020}\natexlab{}.
\newblock \showarticletitle{Language Models are Few-Shot Learners}. In
  \bibinfo{booktitle}{\emph{Advances in Neural Information Processing Systems
  (NeurIPS '20)}}, \bibfield{editor}{\bibinfo{person}{H.~Larochelle},
  \bibinfo{person}{M.~Ranzato}, \bibinfo{person}{R.~Hadsell},
  \bibinfo{person}{M.~F. Balcan}, {and} \bibinfo{person}{H.~Lin}} (Eds.),
  Vol.~\bibinfo{volume}{33}. \bibinfo{publisher}{Curran Associates, Inc.},
  \bibinfo{pages}{1877--1901}.
\newblock
\urldef\tempurl%
\url{https://proceedings.neurips.cc/paper/2020/file/1457c0d6bfcb4967418bfb8ac142f64a-Paper.pdf}
\showURL{%
\tempurl}


\bibitem[Byun(2022)]%
        {CareCallNews}
\bibfield{author}{\bibinfo{person}{Hye-jin Byun}.}
  \bibinfo{year}{2022}\natexlab{}.
\newblock \bibinfo{title}{{NAVER launches AI call service aimed at seniors -
  The Korea Herald}}.
\newblock
\newblock
\urldef\tempurl%
\url{https://www.koreaherald.com/view.php?ud=20220530000643}
\showURL{%
Retrieved Sep 14, 2023 from \tempurl}


\bibitem[{Character AI}(2023)]%
        {CharacterAI}
\bibfield{author}{\bibinfo{person}{{Character AI}}.}
  \bibinfo{year}{2023}\natexlab{}.
\newblock \bibinfo{title}{{Character AI}}.
\newblock
\newblock
\urldef\tempurl%
\url{https://character.ai/}
\showURL{%
Retrieved Sep 14, 2023 from \tempurl}


\bibitem[Chaves and Gerosa(2021)]%
        {chaves_how_2021}
\bibfield{author}{\bibinfo{person}{Ana~Paula Chaves} {and}
  \bibinfo{person}{Marco~Aurelio Gerosa}.} \bibinfo{year}{2021}\natexlab{}.
\newblock \showarticletitle{How {Should} {My} {Chatbot} {Interact}? {A}
  {Survey} on {Social} {Characteristics} in {Human}–{Chatbot} {Interaction}
  {Design}}.
\newblock \bibinfo{journal}{\emph{International Journal of Human–Computer
  Interaction}} \bibinfo{volume}{37}, \bibinfo{number}{8}
  (\bibinfo{year}{2021}), \bibinfo{pages}{729--758}.
\newblock
\showISSN{1044-7318, 1532-7590}
\urldef\tempurl%
\url{https://doi.org/10.1080/10447318.2020.1841438}
\showDOI{\tempurl}


\bibitem[Chen et~al\mbox{.}(2023)]%
        {Chen2023Fireact}
\bibfield{author}{\bibinfo{person}{Baian Chen}, \bibinfo{person}{Chang Shu},
  \bibinfo{person}{Ehsan Shareghi}, \bibinfo{person}{Nigel Collier},
  \bibinfo{person}{Karthik Narasimhan}, {and} \bibinfo{person}{Shunyu Yao}.}
  \bibinfo{year}{2023}\natexlab{}.
\newblock \bibinfo{title}{FireAct: Toward Language Agent Fine-tuning}.
\newblock
\newblock
\showeprint[arxiv]{2310.05915}~[cs.CL]


\bibitem[Chiang et~al\mbox{.}(2023)]%
        {vicuna2023}
\bibfield{author}{\bibinfo{person}{Wei-Lin Chiang}, \bibinfo{person}{Zhuohan
  Li}, \bibinfo{person}{Zi Lin}, \bibinfo{person}{Ying Sheng},
  \bibinfo{person}{Zhanghao Wu}, \bibinfo{person}{Hao Zhang},
  \bibinfo{person}{Lianmin Zheng}, \bibinfo{person}{Siyuan Zhuang},
  \bibinfo{person}{Yonghao Zhuang}, \bibinfo{person}{Joseph~E. Gonzalez},
  \bibinfo{person}{Ion Stoica}, {and} \bibinfo{person}{Eric~P. Xing}.}
  \bibinfo{year}{2023}\natexlab{}.
\newblock \bibinfo{title}{Vicuna: An Open-Source Chatbot Impressing GPT-4 with
  90\%* ChatGPT Quality}.
\newblock
\newblock
\urldef\tempurl%
\url{https://lmsys.org/blog/2023-03-30-vicuna/}
\showURL{%
\tempurl}


\bibitem[Chowdhery et~al\mbox{.}(2022)]%
        {Chowdhery2022PALM}
\bibfield{author}{\bibinfo{person}{Aakanksha Chowdhery},
  \bibinfo{person}{Sharan Narang}, \bibinfo{person}{Jacob Devlin},
  \bibinfo{person}{Maarten Bosma}, \bibinfo{person}{Gaurav Mishra},
  \bibinfo{person}{Adam Roberts}, \bibinfo{person}{Paul Barham},
  \bibinfo{person}{Hyung~Won Chung}, \bibinfo{person}{Charles Sutton},
  \bibinfo{person}{Sebastian Gehrmann}, \bibinfo{person}{Parker Schuh},
  \bibinfo{person}{Kensen Shi}, \bibinfo{person}{Sasha Tsvyashchenko},
  \bibinfo{person}{Joshua Maynez}, \bibinfo{person}{Abhishek Rao},
  \bibinfo{person}{Parker Barnes}, \bibinfo{person}{Yi Tay},
  \bibinfo{person}{Noam Shazeer}, \bibinfo{person}{Vinodkumar Prabhakaran},
  \bibinfo{person}{Emily Reif}, \bibinfo{person}{Nan Du}, \bibinfo{person}{Ben
  Hutchinson}, \bibinfo{person}{Reiner Pope}, \bibinfo{person}{James Bradbury},
  \bibinfo{person}{Jacob Austin}, \bibinfo{person}{Michael Isard},
  \bibinfo{person}{Guy Gur-Ari}, \bibinfo{person}{Pengcheng Yin},
  \bibinfo{person}{Toju Duke}, \bibinfo{person}{Anselm Levskaya},
  \bibinfo{person}{Sanjay Ghemawat}, \bibinfo{person}{Sunipa Dev},
  \bibinfo{person}{Henryk Michalewski}, \bibinfo{person}{Xavier Garcia},
  \bibinfo{person}{Vedant Misra}, \bibinfo{person}{Kevin Robinson},
  \bibinfo{person}{Liam Fedus}, \bibinfo{person}{Denny Zhou},
  \bibinfo{person}{Daphne Ippolito}, \bibinfo{person}{David Luan},
  \bibinfo{person}{Hyeontaek Lim}, \bibinfo{person}{Barret Zoph},
  \bibinfo{person}{Alexander Spiridonov}, \bibinfo{person}{Ryan Sepassi},
  \bibinfo{person}{David Dohan}, \bibinfo{person}{Shivani Agrawal},
  \bibinfo{person}{Mark Omernick}, \bibinfo{person}{Andrew~M. Dai},
  \bibinfo{person}{Thanumalayan~Sankaranarayana Pillai}, \bibinfo{person}{Marie
  Pellat}, \bibinfo{person}{Aitor Lewkowycz}, \bibinfo{person}{Erica Moreira},
  \bibinfo{person}{Rewon Child}, \bibinfo{person}{Oleksandr Polozov},
  \bibinfo{person}{Katherine Lee}, \bibinfo{person}{Zongwei Zhou},
  \bibinfo{person}{Xuezhi Wang}, \bibinfo{person}{Brennan Saeta},
  \bibinfo{person}{Mark Diaz}, \bibinfo{person}{Orhan Firat},
  \bibinfo{person}{Michele Catasta}, \bibinfo{person}{Jason Wei},
  \bibinfo{person}{Kathy Meier-Hellstern}, \bibinfo{person}{Douglas Eck},
  \bibinfo{person}{Jeff Dean}, \bibinfo{person}{Slav Petrov}, {and}
  \bibinfo{person}{Noah Fiedel}.} \bibinfo{year}{2022}\natexlab{}.
\newblock \bibinfo{title}{PaLM: Scaling Language Modeling with Pathways}.
\newblock
\newblock
\urldef\tempurl%
\url{https://doi.org/10.48550/ARXIV.2204.02311}
\showDOI{\tempurl}


\bibitem[Chowdhury(2017)]%
        {chowdhury2017}
\bibfield{author}{\bibinfo{person}{Shammur Chowdhury}.}
  \bibinfo{year}{2017}\natexlab{}.
\newblock \emph{\bibinfo{title}{Computational Modeling of Turn-Taking Dynamics
  in Spoken Conversations}}.
\newblock \bibinfo{thesistype}{Ph.\,D. Dissertation}.
  \bibinfo{school}{University of Trento}.
\newblock
\urldef\tempurl%
\url{https://doi.org/10.13140/RG.2.2.35753.70240}
\showDOI{\tempurl}


\bibitem[Corrigan(2004)]%
        {corrigan_2004}
\bibfield{author}{\bibinfo{person}{Patrick Corrigan}.}
  \bibinfo{year}{2004}\natexlab{}.
\newblock \showarticletitle{How stigma interferes with mental health care.}
\newblock \bibinfo{journal}{\emph{American Psychologist}} \bibinfo{volume}{59},
  \bibinfo{number}{7} (\bibinfo{date}{Oct.} \bibinfo{year}{2004}),
  \bibinfo{pages}{614--625}.
\newblock
\showISSN{1935-990X, 0003-066X}
\urldef\tempurl%
\url{https://doi.org/10.1037/0003-066X.59.7.614}
\showDOI{\tempurl}


\bibitem[Cox et~al\mbox{.}(2023)]%
        {cox_comparing_2023}
\bibfield{author}{\bibinfo{person}{Samuel~Rhys Cox}, \bibinfo{person}{Yi-Chieh
  Lee}, {and} \bibinfo{person}{Wei~Tsang Ooi}.}
  \bibinfo{year}{2023}\natexlab{}.
\newblock \bibinfo{title}{Comparing {How} a {Chatbot} {References} {User}
  {Utterances} from {Previous} {Chatting} {Sessions}: {An} {Investigation} of
  {Users}' {Privacy} {Concerns} and {Perceptions}}.
\newblock
\newblock
\urldef\tempurl%
\url{http://arxiv.org/abs/2308.04879}
\showURL{%
\tempurl}
\newblock
\shownote{arXiv:2308.04879 [cs]}.


\bibitem[Cox and Ooi(2022)]%
        {cox_does_2022}
\bibfield{author}{\bibinfo{person}{Samuel~Rhys Cox} {and}
  \bibinfo{person}{Wei~Tsang Ooi}.} \bibinfo{year}{2022}\natexlab{}.
\newblock \showarticletitle{Does {Chatbot} {Language} {Formality} {Affect}
  {Users}’ {Self}-{Disclosure}?}. In \bibinfo{booktitle}{\emph{Proceedings of
  the 4th {Conference} on {Conversational} {User} {Interfaces}}}.
  \bibinfo{publisher}{ACM}, \bibinfo{address}{Glasgow United Kingdom},
  \bibinfo{pages}{1--13}.
\newblock
\showISBNx{978-1-4503-9739-1}
\urldef\tempurl%
\url{https://doi.org/10.1145/3543829.3543831}
\showDOI{\tempurl}


\bibitem[Denham and Onwuegbuzie(2013)]%
        {denham_beyond_2013}
\bibfield{author}{\bibinfo{person}{Magdalena~A. Denham} {and}
  \bibinfo{person}{Anthony~John Onwuegbuzie}.} \bibinfo{year}{2013}\natexlab{}.
\newblock \showarticletitle{Beyond {Words}: {Using} {Nonverbal} {Communication}
  {Data} in {Research} to {Enhance} {Thick} {Description} and
  {Interpretation}}.
\newblock \bibinfo{journal}{\emph{International Journal of Qualitative
  Methods}} \bibinfo{volume}{12}, \bibinfo{number}{1} (\bibinfo{date}{Feb.}
  \bibinfo{year}{2013}), \bibinfo{pages}{670--696}.
\newblock
\showISSN{1609-4069, 1609-4069}
\urldef\tempurl%
\url{https://doi.org/10.1177/160940691301200137}
\showDOI{\tempurl}


\bibitem[Duijst(2017)]%
        {duijst_can_2017}
\bibfield{author}{\bibinfo{person}{Daniëlle Duijst}.}
  \bibinfo{year}{2017}\natexlab{}.
\newblock \showarticletitle{Can we {Improve} the {User} {Experience} of
  {Chatbots} with {Personalisation}?}
\newblock  (\bibinfo{year}{2017}).
\newblock
\urldef\tempurl%
\url{https://doi.org/10.13140/RG.2.2.36112.92165}
\showDOI{\tempurl}
\newblock
\shownote{Publisher: Unpublished}.


\bibitem[Fitzpatrick et~al\mbox{.}(2017)]%
        {fitzpatrick_delivering_2017}
\bibfield{author}{\bibinfo{person}{Kathleen~Kara Fitzpatrick},
  \bibinfo{person}{Alison Darcy}, {and} \bibinfo{person}{Molly Vierhile}.}
  \bibinfo{year}{2017}\natexlab{}.
\newblock \showarticletitle{Delivering {Cognitive} {Behavior} {Therapy} to
  {Young} {Adults} {With} {Symptoms} of {Depression} and {Anxiety} {Using} a
  {Fully} {Automated} {Conversational} {Agent} ({Woebot}): {A} {Randomized}
  {Controlled} {Trial}}.
\newblock \bibinfo{journal}{\emph{JMIR Mental Health}} \bibinfo{volume}{4},
  \bibinfo{number}{2} (\bibinfo{date}{June} \bibinfo{year}{2017}),
  \bibinfo{pages}{e19}.
\newblock
\showISSN{2368-7959}
\urldef\tempurl%
\url{https://doi.org/10.2196/mental.7785}
\showDOI{\tempurl}


\bibitem[{Google, Inc}(2023)]%
        {Bard}
\bibfield{author}{\bibinfo{person}{{Google, Inc}}.}
  \bibinfo{year}{2023}\natexlab{}.
\newblock \bibinfo{title}{{Bard - Chat Based AI Tool from Google, Powered by
  PaLM 2}}.
\newblock
\newblock
\urldef\tempurl%
\url{https://bard.google.com/}
\showURL{%
Retrieved Sep 14, 2023 from \tempurl}


\bibitem[Gordon et~al\mbox{.}(2019)]%
        {gordon_goal-setting_2019}
\bibfield{author}{\bibinfo{person}{Mitchell Gordon}, \bibinfo{person}{Tim
  Althoff}, {and} \bibinfo{person}{Jure Leskovec}.}
  \bibinfo{year}{2019}\natexlab{}.
\newblock \showarticletitle{Goal-setting {And} {Achievement} {In} {Activity}
  {Tracking} {Apps}: {A} {Case} {Study} {Of} {MyFitnessPal}}. In
  \bibinfo{booktitle}{\emph{The {World} {Wide} {Web} {Conference}}}.
  \bibinfo{publisher}{ACM}, \bibinfo{address}{San Francisco CA USA},
  \bibinfo{pages}{571--582}.
\newblock
\showISBNx{978-1-4503-6674-8}
\urldef\tempurl%
\url{https://doi.org/10.1145/3308558.3313432}
\showDOI{\tempurl}


\bibitem[Hays(1985)]%
        {hays_1985}
\bibfield{author}{\bibinfo{person}{Robert~B Hays}.}
  \bibinfo{year}{1985}\natexlab{}.
\newblock \showarticletitle{A {Longitudinal} {Study} of {Friendship}
  {Development}}.
\newblock  (\bibinfo{date}{April} \bibinfo{year}{1985}).
\newblock


\bibitem[Ho et~al\mbox{.}(2018)]%
        {ho_2018}
\bibfield{author}{\bibinfo{person}{Annabell Ho}, \bibinfo{person}{Jeff
  Hancock}, {and} \bibinfo{person}{Adam~S Miner}.}
  \bibinfo{year}{2018}\natexlab{}.
\newblock \showarticletitle{Psychological, {Relational}, and {Emotional}
  {Effects} of {Self}-{Disclosure} {After} {Conversations} {With} a {Chatbot}}.
\newblock \bibinfo{journal}{\emph{Journal of Communication}}
  \bibinfo{volume}{68}, \bibinfo{number}{4} (\bibinfo{date}{Aug.}
  \bibinfo{year}{2018}), \bibinfo{pages}{712--733}.
\newblock
\showISSN{0021-9916, 1460-2466}
\urldef\tempurl%
\url{https://doi.org/10.1093/joc/jqy026}
\showDOI{\tempurl}


\bibitem[{Hofstede Insights}(2023)]%
        {HofstedeInsights2023}
\bibfield{author}{\bibinfo{person}{{Hofstede Insights}}.}
  \bibinfo{year}{2023}\natexlab{}.
\newblock \bibinfo{title}{{Country Comparison - Hofstede Insights}}.
\newblock
\newblock
\urldef\tempurl%
\url{https://www.hofstede-insights.com/country-comparison-tool?countries=south+korea}
\showURL{%
\tempurl}


\bibitem[{Inflection AI}(2023)]%
        {PiChatbot}
\bibfield{author}{\bibinfo{person}{{Inflection AI}}.}
  \bibinfo{year}{2023}\natexlab{}.
\newblock \bibinfo{title}{{Pi, your personal AI}}.
\newblock
\newblock
\urldef\tempurl%
\url{https://pi.ai/talk}
\showURL{%
Retrieved Sep 14, 2023 from \tempurl}


\bibitem[Ismail et~al\mbox{.}(2018)]%
        {ismail2018bridging}
\bibfield{author}{\bibinfo{person}{Azra Ismail}, \bibinfo{person}{Naveena
  Karusala}, {and} \bibinfo{person}{Neha Kumar}.}
  \bibinfo{year}{2018}\natexlab{}.
\newblock \showarticletitle{Bridging Disconnected Knowledges for Community
  Health}.
\newblock \bibinfo{journal}{\emph{Proc. ACM Hum.-Comput. Interact.}}
  \bibinfo{volume}{2}, \bibinfo{number}{CSCW}, Article \bibinfo{articleno}{75}
  (\bibinfo{date}{nov} \bibinfo{year}{2018}), \bibinfo{numpages}{27}~pages.
\newblock
\urldef\tempurl%
\url{https://doi.org/10.1145/3274344}
\showDOI{\tempurl}


\bibitem[Ismail and Kumar(2018)]%
        {ismail_engaging_2018}
\bibfield{author}{\bibinfo{person}{Azra Ismail} {and} \bibinfo{person}{Neha
  Kumar}.} \bibinfo{year}{2018}\natexlab{}.
\newblock \showarticletitle{Engaging solidarity in data collection practices
  for community health}.
\newblock \bibinfo{journal}{\emph{Proceedings of the ACM on Human-Computer
  Interaction}} \bibinfo{volume}{2}, \bibinfo{number}{CSCW}
  (\bibinfo{year}{2018}).
\newblock
\urldef\tempurl%
\url{https://doi.org/10.1145/3274345}
\showDOI{\tempurl}


\bibitem[Ismail and Kumar(2021)]%
        {ismail2021ai}
\bibfield{author}{\bibinfo{person}{Azra Ismail} {and} \bibinfo{person}{Neha
  Kumar}.} \bibinfo{year}{2021}\natexlab{}.
\newblock \showarticletitle{AI in Global Health: The View from the Front
  Lines}. In \bibinfo{booktitle}{\emph{Proceedings of the 2021 CHI Conference
  on Human Factors in Computing Systems}} (Yokohama, Japan)
  \emph{(\bibinfo{series}{CHI '21})}. \bibinfo{publisher}{Association for
  Computing Machinery}, \bibinfo{address}{New York, NY, USA}, Article
  \bibinfo{articleno}{598}, \bibinfo{numpages}{21}~pages.
\newblock
\showISBNx{9781450380966}
\urldef\tempurl%
\url{https://doi.org/10.1145/3411764.3445130}
\showDOI{\tempurl}


\bibitem[Ismail et~al\mbox{.}(2023)]%
        {ismail_public_2023}
\bibfield{author}{\bibinfo{person}{Azra Ismail}, \bibinfo{person}{Divy
  Thakkar}, \bibinfo{person}{Neha Madhiwalla}, {and} \bibinfo{person}{Neha
  Kumar}.} \bibinfo{year}{2023}\natexlab{}.
\newblock \showarticletitle{Public Health Calls for/with AI: An Ethnographic
  Perspective}.
\newblock \bibinfo{journal}{\emph{Proc. ACM Hum.-Comput. Interact.}}
  \bibinfo{volume}{7}, \bibinfo{number}{CSCW2}, Article
  \bibinfo{articleno}{354} (\bibinfo{date}{oct} \bibinfo{year}{2023}),
  \bibinfo{numpages}{26}~pages.
\newblock
\urldef\tempurl%
\url{https://doi.org/10.1145/3610203}
\showDOI{\tempurl}


\bibitem[Jain et~al\mbox{.}(2018)]%
        {jain_convey_2018}
\bibfield{author}{\bibinfo{person}{Mohit Jain}, \bibinfo{person}{Ramachandra
  Kota}, \bibinfo{person}{Pratyush Kumar}, {and} \bibinfo{person}{Shwetak~N.
  Patel}.} \bibinfo{year}{2018}\natexlab{}.
\newblock \showarticletitle{Convey: {Exploring} the {Use} of a {Context} {View}
  for {Chatbots}}. In \bibinfo{booktitle}{\emph{Proceedings of the 2018 {CHI}
  {Conference} on {Human} {Factors} in {Computing} {Systems}}}.
  \bibinfo{publisher}{ACM}, \bibinfo{address}{Montreal QC Canada},
  \bibinfo{pages}{1--6}.
\newblock
\showISBNx{978-1-4503-5620-6}
\urldef\tempurl%
\url{https://doi.org/10.1145/3173574.3174042}
\showDOI{\tempurl}


\bibitem[Jo et~al\mbox{.}(2023)]%
        {Jo2023}
\bibfield{author}{\bibinfo{person}{Eunkyung Jo}, \bibinfo{person}{Daniel~A
  Epstein}, \bibinfo{person}{Hyunhoon Jung}, {and} \bibinfo{person}{Young-Ho
  Kim}.} \bibinfo{year}{2023}\natexlab{}.
\newblock \showarticletitle{{Understanding the Benefits and Challenges of
  Deploying Conversational AI Leveraging Large Language Models for Public
  Health Intervention}}. In \bibinfo{booktitle}{\emph{Proceedings of the 2023
  CHI Conference on Human Factors in Computing Systems}}.
  \bibinfo{publisher}{ACM}, \bibinfo{address}{New York, NY, USA},
  \bibinfo{pages}{1--16}.
\newblock
\showISBNx{9781450394215}
\urldef\tempurl%
\url{https://doi.org/10.1145/3544548.3581503}
\showDOI{\tempurl}


\bibitem[Jurafsky et~al\mbox{.}(2009)]%
        {jurafsky_extracting_2009}
\bibfield{author}{\bibinfo{person}{Dan Jurafsky}, \bibinfo{person}{Rajesh
  Ranganath}, {and} \bibinfo{person}{Dan McFarland}.}
  \bibinfo{year}{2009}\natexlab{}.
\newblock \showarticletitle{Extracting social meaning: identifying
  interactional style in spoken conversation}. In
  \bibinfo{booktitle}{\emph{Proceedings of {Human} {Language} {Technologies}:
  {The} 2009 {Annual} {Conference} of the {North} {American} {Chapter} of the
  {Association} for {Computational} {Linguistics} on - {NAACL} '09}}.
  \bibinfo{publisher}{Association for Computational Linguistics},
  \bibinfo{address}{Boulder, Colorado}, \bibinfo{pages}{638}.
\newblock
\showISBNx{978-1-932432-41-1}
\urldef\tempurl%
\url{https://doi.org/10.3115/1620754.1620847}
\showDOI{\tempurl}


\bibitem[Kang and Gratch(2010)]%
        {kang_virtual_2010}
\bibfield{author}{\bibinfo{person}{Sin-Hwa Kang} {and}
  \bibinfo{person}{Jonathan Gratch}.} \bibinfo{year}{2010}\natexlab{}.
\newblock \showarticletitle{Virtual humans elicit socially anxious
  interactants' verbal self-disclosure}.
\newblock \bibinfo{journal}{\emph{Computer Animation and Virtual Worlds}}
  (\bibinfo{year}{2010}), \bibinfo{pages}{n/a--n/a}.
\newblock
\showISSN{15464261, 1546427X}
\urldef\tempurl%
\url{https://doi.org/10.1002/cav.345}
\showDOI{\tempurl}


\bibitem[Karapanos et~al\mbox{.}(2021)]%
        {karapanos_advances_2021}
\bibfield{editor}{\bibinfo{person}{Evangelos Karapanos}, \bibinfo{person}{Jens
  Gerken}, \bibinfo{person}{Jesper Kjeldskov}, {and} \bibinfo{person}{Mikael~B.
  Skov}} (Eds.). \bibinfo{year}{2021}\natexlab{}.
\newblock \bibinfo{booktitle}{\emph{Advances in {Longitudinal} {HCI}
  {Research}}}.
\newblock \bibinfo{publisher}{Springer International Publishing},
  \bibinfo{address}{Cham}.
\newblock
\showISBNx{978-3-030-67321-5 978-3-030-67322-2}
\urldef\tempurl%
\url{https://doi.org/10.1007/978-3-030-67322-2}
\showDOI{\tempurl}


\bibitem[Kim et~al\mbox{.}(2021)]%
        {Kim2021HyperCLOVA}
\bibfield{author}{\bibinfo{person}{Boseop Kim}, \bibinfo{person}{HyoungSeok
  Kim}, \bibinfo{person}{Sang-Woo Lee}, \bibinfo{person}{Gichang Lee},
  \bibinfo{person}{Donghyun Kwak}, \bibinfo{person}{Jeon Dong~Hyeon},
  \bibinfo{person}{Sunghyun Park}, \bibinfo{person}{Sungju Kim},
  \bibinfo{person}{Seonhoon Kim}, \bibinfo{person}{Dongpil Seo},
  \bibinfo{person}{Heungsub Lee}, \bibinfo{person}{Minyoung Jeong},
  \bibinfo{person}{Sungjae Lee}, \bibinfo{person}{Minsub Kim},
  \bibinfo{person}{Suk~Hyun Ko}, \bibinfo{person}{Seokhun Kim},
  \bibinfo{person}{Taeyong Park}, \bibinfo{person}{Jinuk Kim},
  \bibinfo{person}{Soyoung Kang}, \bibinfo{person}{Na-Hyeon Ryu},
  \bibinfo{person}{Kang~Min Yoo}, \bibinfo{person}{Minsuk Chang},
  \bibinfo{person}{Soobin Suh}, \bibinfo{person}{Sookyo In},
  \bibinfo{person}{Jinseong Park}, \bibinfo{person}{Kyungduk Kim},
  \bibinfo{person}{Hiun Kim}, \bibinfo{person}{Jisu Jeong},
  \bibinfo{person}{Yong~Goo Yeo}, \bibinfo{person}{Donghoon Ham},
  \bibinfo{person}{Dongju Park}, \bibinfo{person}{Min~Young Lee},
  \bibinfo{person}{Jaewook Kang}, \bibinfo{person}{Inho Kang},
  \bibinfo{person}{Jung-Woo Ha}, \bibinfo{person}{Woomyoung Park}, {and}
  \bibinfo{person}{Nako Sung}.} \bibinfo{year}{2021}\natexlab{}.
\newblock \showarticletitle{What Changes Can Large-scale Language Models Bring?
  Intensive Study on {H}yper{CLOVA}: Billions-scale {K}orean Generative
  Pretrained Transformers}. In \bibinfo{booktitle}{\emph{Proceedings of the
  2021 Conference on Empirical Methods in Natural Language Processing}}.
  \bibinfo{publisher}{Association for Computational Linguistics},
  \bibinfo{address}{Online and Punta Cana, Dominican Republic},
  \bibinfo{pages}{3405--3424}.
\newblock
\urldef\tempurl%
\url{https://doi.org/10.18653/v1/2021.emnlp-main.274}
\showDOI{\tempurl}


\bibitem[Kim et~al\mbox{.}(2022)]%
        {kim_designing_2022}
\bibfield{author}{\bibinfo{person}{Junhan Kim}, \bibinfo{person}{Jana Muhic},
  \bibinfo{person}{Lionel~Peter Robert}, {and} \bibinfo{person}{Sun~Young
  Park}.} \bibinfo{year}{2022}\natexlab{}.
\newblock \showarticletitle{Designing {Chatbots} with {Black} {Americans} with
  {Chronic} {Conditions}: {Overcoming} {Challenges} against {COVID}-19}. In
  \bibinfo{booktitle}{\emph{{CHI} {Conference} on {Human} {Factors} in
  {Computing} {Systems}}}. \bibinfo{publisher}{ACM}, \bibinfo{address}{New
  Orleans LA USA}, \bibinfo{pages}{1--17}.
\newblock
\showISBNx{978-1-4503-9157-3}
\urldef\tempurl%
\url{https://doi.org/10.1145/3491102.3502116}
\showDOI{\tempurl}


\bibitem[Klasnja et~al\mbox{.}(2011)]%
        {klasnja_2011}
\bibfield{author}{\bibinfo{person}{Predrag Klasnja}, \bibinfo{person}{Sunny
  Consolvo}, {and} \bibinfo{person}{Wanda Pratt}.}
  \bibinfo{year}{2011}\natexlab{}.
\newblock \showarticletitle{How to Evaluate Technologies for Health Behavior
  Change in HCI Research}. In \bibinfo{booktitle}{\emph{Proceedings of the
  SIGCHI Conference on Human Factors in Computing Systems}} (Vancouver, BC,
  Canada) \emph{(\bibinfo{series}{CHI '11})}. \bibinfo{publisher}{Association
  for Computing Machinery}, \bibinfo{address}{New York, NY, USA},
  \bibinfo{pages}{3063–3072}.
\newblock
\showISBNx{9781450302289}
\urldef\tempurl%
\url{https://doi.org/10.1145/1978942.1979396}
\showDOI{\tempurl}


\bibitem[{Korea Law Translation Center}(2020)]%
        {ActLonelyDeath2020}
\bibfield{author}{\bibinfo{person}{{Korea Law Translation Center}}.}
  \bibinfo{year}{2020}\natexlab{}.
\newblock \bibinfo{title}{Act on the Prevention and Management of Lonely
  Deaths}.
\newblock
\newblock
\urldef\tempurl%
\url{https://elaw.klri.re.kr/eng_mobile/viewer.do?hseq=55028&type=part&key=38}
\showURL{%
\tempurl}


\bibitem[{Korean Statistical Information Service}(2023)]%
        {KOSIS2023}
\bibfield{author}{\bibinfo{person}{{Korean Statistical Information Service}}.}
  \bibinfo{year}{2023}\natexlab{}.
\newblock \bibinfo{title}{{Resident Population in Five-Year Age Groups}}.
\newblock
\newblock
\urldef\tempurl%
\url{https://kosis.kr/statHtml/statHtml.do?orgId=101%26tblId=DT_1B04005N%26vw_cd=MT_ETITLE%26list_id=A_7%26scrId=%26language=en%26seqNo=%26lang_mode=en%26obj_var_id=%26itm_id=%26conn_path=MT_ETITLE%26path=%2Feng%2FstatisticsList%2FstatisticsListIndex.do}
\showURL{%
Retrieved Nov 14, 2023 from \tempurl}


\bibitem[Kreiner and Levi-Belz(2019)]%
        {kreiner_2019}
\bibfield{author}{\bibinfo{person}{Hamutal Kreiner} {and}
  \bibinfo{person}{Yossi Levi-Belz}.} \bibinfo{year}{2019}\natexlab{}.
\newblock \showarticletitle{Self-{Disclosure} {Here} and {Now}: {Combining}
  {Retrospective} {Perceived} {Assessment} {With} {Dynamic} {Behavioral}
  {Measures}}.
\newblock \bibinfo{journal}{\emph{Frontiers in Psychology}}
  \bibinfo{volume}{10} (\bibinfo{date}{March} \bibinfo{year}{2019}),
  \bibinfo{pages}{558}.
\newblock
\showISSN{1664-1078}
\urldef\tempurl%
\url{https://doi.org/10.3389/fpsyg.2019.00558}
\showDOI{\tempurl}


\bibitem[Lau et~al\mbox{.}(2018)]%
        {lau_alexa_2018}
\bibfield{author}{\bibinfo{person}{Josephine Lau}, \bibinfo{person}{Benjamin
  Zimmerman}, {and} \bibinfo{person}{Florian Schaub}.}
  \bibinfo{year}{2018}\natexlab{}.
\newblock \showarticletitle{Alexa, {Are} {You} {Listening}?: {Privacy}
  {Perceptions}, {Concerns} and {Privacy}-seeking {Behaviors} with {Smart}
  {Speakers}}.
\newblock \bibinfo{journal}{\emph{Proceedings of the ACM on Human-Computer
  Interaction}} \bibinfo{volume}{2}, \bibinfo{number}{CSCW}
  (\bibinfo{date}{Nov.} \bibinfo{year}{2018}), \bibinfo{pages}{1--31}.
\newblock
\showISSN{2573-0142}
\urldef\tempurl%
\url{https://doi.org/10.1145/3274371}
\showDOI{\tempurl}


\bibitem[Lee et~al\mbox{.}(2023)]%
        {lee2023prompted}
\bibfield{author}{\bibinfo{person}{Gibbeum Lee}, \bibinfo{person}{Volker
  Hartmann}, \bibinfo{person}{Jongho Park}, \bibinfo{person}{Dimitris
  Papailiopoulos}, {and} \bibinfo{person}{Kangwook Lee}.}
  \bibinfo{year}{2023}\natexlab{}.
\newblock \showarticletitle{Prompted {LLM}s as Chatbot Modules for Long
  Open-domain Conversation}. In \bibinfo{booktitle}{\emph{Findings of the
  Association for Computational Linguistics: ACL 2023}}.
  \bibinfo{publisher}{Association for Computational Linguistics},
  \bibinfo{address}{Toronto, Canada}, \bibinfo{pages}{4536--4554}.
\newblock
\urldef\tempurl%
\url{https://doi.org/10.18653/v1/2023.findings-acl.277}
\showDOI{\tempurl}


\bibitem[Lee et~al\mbox{.}(2019)]%
        {lee_caring_2019}
\bibfield{author}{\bibinfo{person}{Minha Lee}, \bibinfo{person}{Sander
  Ackermans}, \bibinfo{person}{Nena van As}, \bibinfo{person}{Hanwen Chang},
  \bibinfo{person}{Enzo Lucas}, {and} \bibinfo{person}{Wijnand IJsselsteijn}.}
  \bibinfo{year}{2019}\natexlab{}.
\newblock \showarticletitle{Caring for {Vincent}: {A} {Chatbot} for
  {Self}-{Compassion}}. In \bibinfo{booktitle}{\emph{Proceedings of the 2019
  {CHI} {Conference} on {Human} {Factors} in {Computing} {Systems}}}.
  \bibinfo{publisher}{ACM}, \bibinfo{address}{Glasgow Scotland Uk},
  \bibinfo{pages}{1--13}.
\newblock
\showISBNx{978-1-4503-5970-2}
\urldef\tempurl%
\url{https://doi.org/10.1145/3290605.3300932}
\showDOI{\tempurl}


\bibitem[Lee et~al\mbox{.}(2020a)]%
        {lee_designing_2020}
\bibfield{author}{\bibinfo{person}{Yi-Chieh Lee}, \bibinfo{person}{Naomi
  Yamashita}, {and} \bibinfo{person}{Yun Huang}.}
  \bibinfo{year}{2020}\natexlab{a}.
\newblock \showarticletitle{Designing a {Chatbot} as a {Mediator} for
  {Promoting} {Deep} {Self}-{Disclosure} to a {Real} {Mental} {Health}
  {Professional}}.
\newblock \bibinfo{journal}{\emph{Proceedings of the ACM on Human-Computer
  Interaction}} \bibinfo{volume}{4}, \bibinfo{number}{CSCW1}
  (\bibinfo{date}{May} \bibinfo{year}{2020}), \bibinfo{pages}{1--27}.
\newblock
\showISSN{2573-0142}
\urldef\tempurl%
\url{https://doi.org/10.1145/3392836}
\showDOI{\tempurl}


\bibitem[Lee et~al\mbox{.}(2020b)]%
        {lee_i_2020}
\bibfield{author}{\bibinfo{person}{Yi-Chieh Lee}, \bibinfo{person}{Naomi
  Yamashita}, \bibinfo{person}{Yun Huang}, {and} \bibinfo{person}{Wai Fu}.}
  \bibinfo{year}{2020}\natexlab{b}.
\newblock \showarticletitle{"{I} {Hear} {You}, {I} {Feel} {You}": {Encouraging}
  {Deep} {Self}-disclosure through a {Chatbot}}. In
  \bibinfo{booktitle}{\emph{Proceedings of the 2020 {CHI} {Conference} on
  {Human} {Factors} in {Computing} {Systems}}}. \bibinfo{publisher}{ACM},
  \bibinfo{address}{Honolulu HI USA}, \bibinfo{pages}{1--12}.
\newblock
\showISBNx{978-1-4503-6708-0}
\urldef\tempurl%
\url{https://doi.org/10.1145/3313831.3376175}
\showDOI{\tempurl}


\bibitem[Li(2022)]%
        {li_sensitizing_2022}
\bibfield{author}{\bibinfo{person}{Qian Li}.} \bibinfo{year}{2022}\natexlab{}.
\newblock \showarticletitle{Sensitizing {Social} {Interaction} with a
  {Mode}-{Enhanced} {Transcribing} {Process}}.
\newblock \bibinfo{journal}{\emph{Organizational Research Methods}}
  (\bibinfo{date}{Oct.} \bibinfo{year}{2022}),
  \bibinfo{pages}{109442812211340}.
\newblock
\showISSN{1094-4281, 1552-7425}
\urldef\tempurl%
\url{https://doi.org/10.1177/10944281221134096}
\showDOI{\tempurl}


\bibitem[Liu and Sundar(2018)]%
        {liu_2018}
\bibfield{author}{\bibinfo{person}{Bingjie Liu} {and} \bibinfo{person}{S.~Shyam
  Sundar}.} \bibinfo{year}{2018}\natexlab{}.
\newblock \showarticletitle{Should Machines Express sympathy and empathy?
  experiments with a health advice chatbot}.
\newblock \bibinfo{journal}{\emph{Cyberpsychology, Behavior, and Social
  Networking}} \bibinfo{volume}{21}, \bibinfo{number}{10}
  (\bibinfo{year}{2018}), \bibinfo{pages}{625–636}.
\newblock
\urldef\tempurl%
\url{https://doi.org/10.1089/cyber.2018.0110}
\showDOI{\tempurl}


\bibitem[Lu et~al\mbox{.}(2022)]%
        {lu2022understanding}
\bibfield{author}{\bibinfo{person}{Xi Lu}, \bibinfo{person}{Eunkyung Jo},
  \bibinfo{person}{Seora Park}, \bibinfo{person}{Hwajung Hong},
  \bibinfo{person}{Yunan Chen}, {and} \bibinfo{person}{Daniel~A. Epstein}.}
  \bibinfo{year}{2022}\natexlab{}.
\newblock \showarticletitle{Understanding Cultural Influence on Perspectives
  Around Contact Tracing Strategies}.
\newblock \bibinfo{journal}{\emph{Proc. ACM Hum.-Comput. Interact.}}
  \bibinfo{volume}{6}, \bibinfo{number}{CSCW2}, Article
  \bibinfo{articleno}{468} (\bibinfo{date}{nov} \bibinfo{year}{2022}),
  \bibinfo{numpages}{26}~pages.
\newblock
\urldef\tempurl%
\url{https://doi.org/10.1145/3555569}
\showDOI{\tempurl}


\bibitem[Lu et~al\mbox{.}(2021)]%
        {lu2021comparing}
\bibfield{author}{\bibinfo{person}{Xi Lu}, \bibinfo{person}{Tera {L.
  Reynolds}}, \bibinfo{person}{Eunkyung Jo}, \bibinfo{person}{Hwajung Hong},
  \bibinfo{person}{Xinru Page}, \bibinfo{person}{Yunan Chen}, {and}
  \bibinfo{person}{Daniel {A. Epstein}}.} \bibinfo{year}{2021}\natexlab{}.
\newblock \showarticletitle{{Comparing Perspectives Around Human and Technology
  Support for Contact Tracing}}. In \bibinfo{booktitle}{\emph{Proceedings of
  the 2021 CHI Conference on Human Factors in Computing Systems}}.
  \bibinfo{publisher}{ACM}, \bibinfo{address}{New York, NY, USA},
  \bibinfo{pages}{1--15}.
\newblock
\showISBNx{9781450380966}
\urldef\tempurl%
\url{https://doi.org/10.1145/3411764.3445669}
\showDOI{\tempurl}


\bibitem[Lucas et~al\mbox{.}(2014)]%
        {lucas_its_2014}
\bibfield{author}{\bibinfo{person}{Gale~M. Lucas}, \bibinfo{person}{Jonathan
  Gratch}, \bibinfo{person}{Aisha King}, {and} \bibinfo{person}{Louis-Philippe
  Morency}.} \bibinfo{year}{2014}\natexlab{}.
\newblock \showarticletitle{It’s only a computer: {Virtual} humans increase
  willingness to disclose}.
\newblock \bibinfo{journal}{\emph{Computers in Human Behavior}}
  \bibinfo{volume}{37} (\bibinfo{date}{Aug.} \bibinfo{year}{2014}),
  \bibinfo{pages}{94--100}.
\newblock
\showISSN{07475632}
\urldef\tempurl%
\url{https://doi.org/10.1016/j.chb.2014.04.043}
\showDOI{\tempurl}


\bibitem[Lucas et~al\mbox{.}(2017)]%
        {lucas_reporting_2017}
\bibfield{author}{\bibinfo{person}{Gale~M. Lucas}, \bibinfo{person}{Albert
  Rizzo}, \bibinfo{person}{Jonathan Gratch}, \bibinfo{person}{Stefan Scherer},
  \bibinfo{person}{Giota Stratou}, \bibinfo{person}{Jill Boberg}, {and}
  \bibinfo{person}{Louis-Philippe Morency}.} \bibinfo{year}{2017}\natexlab{}.
\newblock \showarticletitle{Reporting {Mental} {Health} {Symptoms}: {Breaking}
  {Down} {Barriers} to {Care} with {Virtual} {Human} {Interviewers}}.
\newblock \bibinfo{journal}{\emph{Frontiers in Robotics and AI}}
  \bibinfo{volume}{4} (\bibinfo{date}{Oct.} \bibinfo{year}{2017}),
  \bibinfo{pages}{51}.
\newblock
\showISSN{2296-9144}
\urldef\tempurl%
\url{https://doi.org/10.3389/frobt.2017.00051}
\showDOI{\tempurl}


\bibitem[Ma et~al\mbox{.}(2023)]%
        {Ma2023Replika}
\bibfield{author}{\bibinfo{person}{Zilin Ma}, \bibinfo{person}{Yiyang Mei},
  {and} \bibinfo{person}{Zhaoyuan Su}.} \bibinfo{year}{2023}\natexlab{}.
\newblock \bibinfo{title}{Understanding the Benefits and Challenges of Using
  Large Language Model-based Conversational Agents for Mental Well-being
  Support}.
\newblock
\newblock
\showeprint[arxiv]{2307.15810}~[cs.CL]


\bibitem[Medhi~Thies et~al\mbox{.}(2017)]%
        {thies_how_2017}
\bibfield{author}{\bibinfo{person}{Indrani Medhi~Thies},
  \bibinfo{person}{Nandita Menon}, \bibinfo{person}{Sneha Magapu},
  \bibinfo{person}{Manisha Subramony}, {and} \bibinfo{person}{Jacki
  O’Neill}.} \bibinfo{year}{2017}\natexlab{}.
\newblock \showarticletitle{How {Do} {You} {Want} {Your} {Chatbot}? {An}
  {Exploratory} {Wizard}-of-{Oz} {Study} with {Young}, {Urban} {Indians}}.
\newblock In \bibinfo{booktitle}{\emph{Human-{Computer} {Interaction} -
  {INTERACT} 2017}}, \bibfield{editor}{\bibinfo{person}{Regina Bernhaupt},
  \bibinfo{person}{Girish Dalvi}, \bibinfo{person}{Anirudha Joshi},
  \bibinfo{person}{Devanuj K.~Balkrishan}, \bibinfo{person}{Jacki O'Neill},
  {and} \bibinfo{person}{Marco Winckler}} (Eds.). Vol.~\bibinfo{volume}{10513}.
  \bibinfo{publisher}{Springer International Publishing},
  \bibinfo{address}{Cham}, \bibinfo{pages}{441--459}.
\newblock
\showISBNx{978-3-319-67743-9 978-3-319-67744-6}
\urldef\tempurl%
\url{https://doi.org/10.1007/978-3-319-67744-6_28}
\showDOI{\tempurl}
\newblock
\shownote{Series Title: Lecture Notes in Computer Science}.


\bibitem[Okolo et~al\mbox{.}(2021)]%
        {okolo2021cannot}
\bibfield{author}{\bibinfo{person}{Chinasa~T Okolo}, \bibinfo{person}{Srujana
  Kamath}, \bibinfo{person}{Nicola Dell}, {and} \bibinfo{person}{Aditya
  Vashistha}.} \bibinfo{year}{2021}\natexlab{}.
\newblock \showarticletitle{“It cannot do all of my work”: community health
  worker perceptions of AI-enabled mobile health applications in rural India}.
  In \bibinfo{booktitle}{\emph{Proceedings of the 2021 CHI Conference on Human
  Factors in Computing Systems}}. \bibinfo{pages}{1--20}.
\newblock


\bibitem[{OpenAI, Inc}(2022)]%
        {ChatGPT}
\bibfield{author}{\bibinfo{person}{{OpenAI, Inc}}.}
  \bibinfo{year}{2022}\natexlab{}.
\newblock \bibinfo{title}{{ChatGPT}}.
\newblock
\newblock
\urldef\tempurl%
\url{https://chat.openai.com}
\showURL{%
Retrieved Sep 14, 2023 from \tempurl}


\bibitem[{OpenAI, Inc}(2023a)]%
        {ChatGPTCustomInstruction}
\bibfield{author}{\bibinfo{person}{{OpenAI, Inc}}.}
  \bibinfo{year}{2023}\natexlab{a}.
\newblock \bibinfo{title}{{Custom instructions for ChatGPT}}.
\newblock
\newblock
\urldef\tempurl%
\url{https://openai.com/blog/custom-instructions-for-chatgpt}
\showURL{%
Retrieved Sep 14, 2023 from \tempurl}


\bibitem[{OpenAI, Inc}(2023b)]%
        {Replika}
\bibfield{author}{\bibinfo{person}{{OpenAI, Inc}}.}
  \bibinfo{year}{2023}\natexlab{b}.
\newblock \bibinfo{title}{{Replika}}.
\newblock
\newblock
\urldef\tempurl%
\url{https://replika.com/}
\showURL{%
Retrieved Nov 14, 2023 from \tempurl}


\bibitem[Ouyang et~al\mbox{.}(2022)]%
        {Long2022InstructGPT}
\bibfield{author}{\bibinfo{person}{Long Ouyang}, \bibinfo{person}{Jeff Wu},
  \bibinfo{person}{Xu Jiang}, \bibinfo{person}{Diogo Almeida},
  \bibinfo{person}{Carroll~L. Wainwright}, \bibinfo{person}{Pamela Mishkin},
  \bibinfo{person}{Chong Zhang}, \bibinfo{person}{Sandhini Agarwal},
  \bibinfo{person}{Katarina Slama}, \bibinfo{person}{Alex Ray},
  \bibinfo{person}{John Schulman}, \bibinfo{person}{Jacob Hilton},
  \bibinfo{person}{Fraser Kelton}, \bibinfo{person}{Luke Miller},
  \bibinfo{person}{Maddie Simens}, \bibinfo{person}{Amanda Askell},
  \bibinfo{person}{Peter Welinder}, \bibinfo{person}{Paul Christiano},
  \bibinfo{person}{Jan Leike}, {and} \bibinfo{person}{Ryan Lowe}.}
  \bibinfo{year}{2022}\natexlab{}.
\newblock \bibinfo{title}{Training language models to follow instructions with
  human feedback}.
\newblock
\newblock
\urldef\tempurl%
\url{https://doi.org/10.48550/ARXIV.2203.02155}
\showDOI{\tempurl}


\bibitem[Park and Lee(2021)]%
        {park_designing_2021}
\bibfield{author}{\bibinfo{person}{Hyanghee Park} {and}
  \bibinfo{person}{Joonhwan Lee}.} \bibinfo{year}{2021}\natexlab{}.
\newblock \showarticletitle{Designing a {Conversational} {Agent} for {Sexual}
  {Assault} {Survivors}: {Defining} {Burden} of {Self}-{Disclosure} and
  {Envisioning} {Survivor}-{Centered} {Solutions}}. In
  \bibinfo{booktitle}{\emph{Proceedings of the 2021 {CHI} {Conference} on
  {Human} {Factors} in {Computing} {Systems}}}. \bibinfo{publisher}{ACM},
  \bibinfo{address}{Yokohama Japan}, \bibinfo{pages}{1--17}.
\newblock
\showISBNx{978-1-4503-8096-6}
\urldef\tempurl%
\url{https://doi.org/10.1145/3411764.3445133}
\showDOI{\tempurl}


\bibitem[Park et~al\mbox{.}(2019)]%
        {park_designing_2019}
\bibfield{author}{\bibinfo{person}{SoHyun Park}, \bibinfo{person}{Jeewon Choi},
  \bibinfo{person}{Sungwoo Lee}, \bibinfo{person}{Changhoon Oh},
  \bibinfo{person}{Changdai Kim}, \bibinfo{person}{Soohyun La},
  \bibinfo{person}{Joonhwan Lee}, {and} \bibinfo{person}{Bongwon Suh}.}
  \bibinfo{year}{2019}\natexlab{}.
\newblock \showarticletitle{Designing a {Chatbot} for a {Brief} {Motivational}
  {Interview} on {Stress} {Management}: {Qualitative} {Case} {Study}}.
\newblock \bibinfo{journal}{\emph{Journal of Medical Internet Research}}
  \bibinfo{volume}{21}, \bibinfo{number}{4} (\bibinfo{date}{April}
  \bibinfo{year}{2019}), \bibinfo{pages}{e12231}.
\newblock
\showISSN{1438-8871}
\urldef\tempurl%
\url{https://doi.org/10.2196/12231}
\showDOI{\tempurl}


\bibitem[Park et~al\mbox{.}(2021)]%
        {park_i_2021}
\bibfield{author}{\bibinfo{person}{SoHyun Park}, \bibinfo{person}{Anja Thieme},
  \bibinfo{person}{Jeongyun Han}, \bibinfo{person}{Sungwoo Lee},
  \bibinfo{person}{Wonjong Rhee}, {and} \bibinfo{person}{Bongwon Suh}.}
  \bibinfo{year}{2021}\natexlab{}.
\newblock \showarticletitle{“{I} wrote as if {I} were telling a story to
  someone {I} knew.”: {Designing} {Chatbot} {Interactions} for {Expressive}
  {Writing} in {Mental} {Health}}. In \bibinfo{booktitle}{\emph{Designing
  {Interactive} {Systems} {Conference} 2021}}. \bibinfo{publisher}{ACM},
  \bibinfo{address}{Virtual Event USA}, \bibinfo{pages}{926--941}.
\newblock
\showISBNx{978-1-4503-8476-6}
\urldef\tempurl%
\url{https://doi.org/10.1145/3461778.3462143}
\showDOI{\tempurl}


\bibitem[Phinnemore et~al\mbox{.}(2023)]%
        {phinnemore_creepy_2023}
\bibfield{author}{\bibinfo{person}{Rachel Phinnemore}, \bibinfo{person}{Mohi
  Reza}, \bibinfo{person}{Blaine Lewis}, \bibinfo{person}{Karthik Mahadevan},
  \bibinfo{person}{Bryan Wang}, \bibinfo{person}{Michelle Annett}, {and}
  \bibinfo{person}{Daniel Wigdor}.} \bibinfo{year}{2023}\natexlab{}.
\newblock \showarticletitle{Creepy {Assistant}: {Development} and {Validation}
  of a {Scale} to {Measure} the {Perceived} {Creepiness} of {Voice}
  {Assistants}}. In \bibinfo{booktitle}{\emph{Proceedings of the 2023 {CHI}
  {Conference} on {Human} {Factors} in {Computing} {Systems}}}.
  \bibinfo{publisher}{ACM}, \bibinfo{address}{Hamburg Germany},
  \bibinfo{pages}{1--18}.
\newblock
\showISBNx{978-1-4503-9421-5}
\urldef\tempurl%
\url{https://doi.org/10.1145/3544548.3581346}
\showDOI{\tempurl}


\bibitem[Pickard et~al\mbox{.}(2016)]%
        {pickard_revealing_2016}
\bibfield{author}{\bibinfo{person}{Matthew~D. Pickard},
  \bibinfo{person}{Catherine~A. Roster}, {and} \bibinfo{person}{Yixing Chen}.}
  \bibinfo{year}{2016}\natexlab{}.
\newblock \showarticletitle{Revealing sensitive information in personal
  interviews: {Is} self-disclosure easier with humans or avatars and under what
  conditions?}
\newblock \bibinfo{journal}{\emph{Computers in Human Behavior}}
  \bibinfo{volume}{65} (\bibinfo{date}{Dec.} \bibinfo{year}{2016}),
  \bibinfo{pages}{23--30}.
\newblock
\showISSN{07475632}
\urldef\tempurl%
\url{https://doi.org/10.1016/j.chb.2016.08.004}
\showDOI{\tempurl}


\bibitem[Portela and Granell-Canut(2017)]%
        {portela_new_2017}
\bibfield{author}{\bibinfo{person}{Manuel Portela} {and}
  \bibinfo{person}{Carlos Granell-Canut}.} \bibinfo{year}{2017}\natexlab{}.
\newblock \showarticletitle{A new friend in our {Smartphone}? {Observing}
  {Interactions} with {Chatbots} in the search of emotional engagement}.
\newblock  (\bibinfo{year}{2017}).
\newblock


\bibitem[Raposo et~al\mbox{.}(2023)]%
        {Raposo2023PromptingTOD}
\bibfield{author}{\bibinfo{person}{Gon{\c{c}}alo Raposo},
  \bibinfo{person}{Luisa Coheur}, {and} \bibinfo{person}{Bruno Martins}.}
  \bibinfo{year}{2023}\natexlab{}.
\newblock \showarticletitle{Prompting, Retrieval, Training: An exploration of
  different approaches for task-oriented dialogue generation}. In
  \bibinfo{booktitle}{\emph{Proceedings of the 24th Annual Meeting of the
  Special Interest Group on Discourse and Dialogue}},
  \bibfield{editor}{\bibinfo{person}{Svetlana Stoyanchev},
  \bibinfo{person}{Shafiq Joty}, \bibinfo{person}{David Schlangen},
  \bibinfo{person}{Ondrej Dusek}, \bibinfo{person}{Casey Kennington}, {and}
  \bibinfo{person}{Malihe Alikhani}} (Eds.). \bibinfo{publisher}{Association
  for Computational Linguistics}, \bibinfo{address}{Prague, Czechia},
  \bibinfo{pages}{400--412}.
\newblock
\urldef\tempurl%
\url{https://doi.org/10.18653/v1/2023.sigdial-1.37}
\showDOI{\tempurl}


\bibitem[Ravichander and Black(2018)]%
        {ravichander_empirical_2018}
\bibfield{author}{\bibinfo{person}{Abhilasha Ravichander} {and}
  \bibinfo{person}{Alan~W. Black}.} \bibinfo{year}{2018}\natexlab{}.
\newblock \showarticletitle{An {Empirical} {Study} of {Self}-{Disclosure} in
  {Spoken} {Dialogue} {Systems}}. In \bibinfo{booktitle}{\emph{Proceedings of
  the 19th {Annual} {SIGdial} {Meeting} on {Discourse} and {Dialogue}}}.
  \bibinfo{publisher}{Association for Computational Linguistics},
  \bibinfo{address}{Melbourne, Australia}, \bibinfo{pages}{253--263}.
\newblock
\urldef\tempurl%
\url{https://doi.org/10.18653/v1/W18-5030}
\showDOI{\tempurl}


\bibitem[Shameli et~al\mbox{.}(2017)]%
        {shameli_how_2017}
\bibfield{author}{\bibinfo{person}{Ali Shameli}, \bibinfo{person}{Tim Althoff},
  \bibinfo{person}{Amin Saberi}, {and} \bibinfo{person}{Jure Leskovec}.}
  \bibinfo{year}{2017}\natexlab{}.
\newblock \showarticletitle{How {Gamification} {Affects} {Physical} {Activity}:
  {Large}-scale {Analysis} of {Walking} {Challenges} in a {Mobile}
  {Application}}. In \bibinfo{booktitle}{\emph{Proceedings of the 26th
  {International} {Conference} on {World} {Wide} {Web} {Companion} - {WWW} '17
  {Companion}}}. \bibinfo{publisher}{ACM Press}, \bibinfo{address}{Perth,
  Australia}, \bibinfo{pages}{455--463}.
\newblock
\showISBNx{978-1-4503-4914-7}
\urldef\tempurl%
\url{https://doi.org/10.1145/3041021.3054172}
\showDOI{\tempurl}


\bibitem[Simmons-Duffin(2020)]%
        {NPR_2020}
\bibfield{author}{\bibinfo{person}{Selena Simmons-Duffin}.}
  \bibinfo{year}{2020}\natexlab{}.
\newblock \showarticletitle{As States Reopen, Do They Have The Workforce They
  Need To Stop Coronavirus Outbreaks?}
\newblock \bibinfo{journal}{\emph{NPR}} (\bibinfo{date}{Jun}
  \bibinfo{year}{2020}).
\newblock
\urldef\tempurl%
\url{https://www.npr.org/sections/health-shots/2020/06/18/879787448/as-states-reopen-do-they-have-the-workforce-they-need-to-stop-coronavirus-outbre}
\showURL{%
\tempurl}


\bibitem[Skantze(2021)]%
        {skantze_turn-taking_2021}
\bibfield{author}{\bibinfo{person}{Gabriel Skantze}.}
  \bibinfo{year}{2021}\natexlab{}.
\newblock \showarticletitle{Turn-taking in {Conversational} {Systems} and
  {Human}-{Robot} {Interaction}: {A} {Review}}.
\newblock \bibinfo{journal}{\emph{Computer Speech \& Language}}
  \bibinfo{volume}{67} (\bibinfo{date}{May} \bibinfo{year}{2021}),
  \bibinfo{pages}{101178}.
\newblock
\showISSN{08852308}
\urldef\tempurl%
\url{https://doi.org/10.1016/j.csl.2020.101178}
\showDOI{\tempurl}


\bibitem[Smolinski et~al\mbox{.}(2015)]%
        {smolinski_flu_2015}
\bibfield{author}{\bibinfo{person}{Mark~S. Smolinski}, \bibinfo{person}{Adam~W.
  Crawley}, \bibinfo{person}{Kristin Baltrusaitis}, \bibinfo{person}{Rumi
  Chunara}, \bibinfo{person}{Jennifer~M. Olsen}, \bibinfo{person}{Oktawia
  Wójcik}, \bibinfo{person}{Mauricio Santillana}, \bibinfo{person}{Andre
  Nguyen}, {and} \bibinfo{person}{John~S. Brownstein}.}
  \bibinfo{year}{2015}\natexlab{}.
\newblock \showarticletitle{Flu {Near} {You}: {Crowdsourced} {Symptom}
  {Reporting} {Spanning} 2 {Influenza} {Seasons}}.
\newblock \bibinfo{journal}{\emph{American Journal of Public Health}}
  \bibinfo{volume}{105}, \bibinfo{number}{10} (\bibinfo{date}{Oct.}
  \bibinfo{year}{2015}), \bibinfo{pages}{2124--2130}.
\newblock
\showISSN{0090-0036, 1541-0048}
\urldef\tempurl%
\url{https://doi.org/10.2105/AJPH.2015.302696}
\showDOI{\tempurl}


\bibitem[Sundar and Kim(2019)]%
        {sundar_2019}
\bibfield{author}{\bibinfo{person}{S.~Shyam Sundar} {and}
  \bibinfo{person}{Jinyoung Kim}.} \bibinfo{year}{2019}\natexlab{}.
\newblock \showarticletitle{Machine Heuristic: When We Trust Computers More
  than Humans with Our Personal Information}. In
  \bibinfo{booktitle}{\emph{Proceedings of the 2019 CHI Conference on Human
  Factors in Computing Systems}} (Glasgow, Scotland Uk)
  \emph{(\bibinfo{series}{CHI '19})}. \bibinfo{publisher}{Association for
  Computing Machinery}, \bibinfo{address}{New York, NY, USA},
  \bibinfo{pages}{1–9}.
\newblock
\showISBNx{9781450359702}
\urldef\tempurl%
\url{https://doi.org/10.1145/3290605.3300768}
\showDOI{\tempurl}


\bibitem[Thudt et~al\mbox{.}(2016)]%
        {Thudt2016VisualMementos}
\bibfield{author}{\bibinfo{person}{Alice Thudt}, \bibinfo{person}{Dominikus
  Baur}, \bibinfo{person}{Samuel Huron}, {and} \bibinfo{person}{Sheelagh
  Carpendale}.} \bibinfo{year}{2016}\natexlab{}.
\newblock \showarticletitle{{Visual Mementos: Reflecting Memories with Personal
  Data}}.
\newblock \bibinfo{journal}{\emph{IEEE Transactions on Visualization and
  Computer Graphics}} \bibinfo{volume}{22}, \bibinfo{number}{1}
  (\bibinfo{date}{Jan.} \bibinfo{year}{2016}), \bibinfo{pages}{369--378}.
\newblock
\showISSN{1077-2626}
\urldef\tempurl%
\url{https://doi.org/10.1109/TVCG.2015.2467831}
\showDOI{\tempurl}


\bibitem[Touvron et~al\mbox{.}(2023)]%
        {touvron2023llama}
\bibfield{author}{\bibinfo{person}{Hugo Touvron}, \bibinfo{person}{Thibaut
  Lavril}, \bibinfo{person}{Gautier Izacard}, \bibinfo{person}{Xavier
  Martinet}, \bibinfo{person}{Marie-Anne Lachaux}, \bibinfo{person}{Timothée
  Lacroix}, \bibinfo{person}{Baptiste Rozière}, \bibinfo{person}{Naman Goyal},
  \bibinfo{person}{Eric Hambro}, \bibinfo{person}{Faisal Azhar},
  \bibinfo{person}{Aurelien Rodriguez}, \bibinfo{person}{Armand Joulin},
  \bibinfo{person}{Edouard Grave}, {and} \bibinfo{person}{Guillaume Lample}.}
  \bibinfo{year}{2023}\natexlab{}.
\newblock \bibinfo{title}{LLaMA: Open and Efficient Foundation Language
  Models}.
\newblock
\newblock
\showeprint[arxiv]{2302.13971}~[cs.CL]


\bibitem[Vogel and Wester(2003)]%
        {vogel_2003}
\bibfield{author}{\bibinfo{person}{David~L. Vogel} {and}
  \bibinfo{person}{Stephen~R. Wester}.} \bibinfo{year}{2003}\natexlab{}.
\newblock \showarticletitle{To seek help or not to seek help: {The} risks of
  self-disclosure.}
\newblock \bibinfo{journal}{\emph{Journal of Counseling Psychology}}
  \bibinfo{volume}{50}, \bibinfo{number}{3} (\bibinfo{date}{July}
  \bibinfo{year}{2003}), \bibinfo{pages}{351--361}.
\newblock
\showISSN{1939-2168, 0022-0167}
\urldef\tempurl%
\url{https://doi.org/10.1037/0022-0167.50.3.351}
\showDOI{\tempurl}


\bibitem[Wang et~al\mbox{.}(2021)]%
        {Wang2021TherapyChatbot}
\bibfield{author}{\bibinfo{person}{Lu Wang}, \bibinfo{person}{Munif~Ishad
  Mujib}, \bibinfo{person}{Jake Williams}, \bibinfo{person}{George Demiris},
  {and} \bibinfo{person}{Jina Huh-Yoo}.} \bibinfo{year}{2021}\natexlab{}.
\newblock \bibinfo{title}{An Evaluation of Generative Pre-Training Model-based
  Therapy Chatbot for Caregivers}.
\newblock
\newblock
\urldef\tempurl%
\url{https://doi.org/10.48550/ARXIV.2107.13115}
\showDOI{\tempurl}


\bibitem[Wang et~al\mbox{.}(2023)]%
        {wang2023augmenting}
\bibfield{author}{\bibinfo{person}{Weizhi Wang}, \bibinfo{person}{Li Dong},
  \bibinfo{person}{Hao Cheng}, \bibinfo{person}{Xiaodong Liu},
  \bibinfo{person}{Xifeng Yan}, \bibinfo{person}{Jianfeng Gao}, {and}
  \bibinfo{person}{Furu Wei}.} \bibinfo{year}{2023}\natexlab{}.
\newblock \bibinfo{title}{Augmenting Language Models with Long-Term Memory}.
\newblock
\newblock
\showeprint[arxiv]{2306.07174}~[cs.CL]


\bibitem[Wei et~al\mbox{.}(2024)]%
        {wei2023leveraging}
\bibfield{author}{\bibinfo{person}{Jing Wei}, \bibinfo{person}{Sungdong Kim},
  \bibinfo{person}{Hyunhoon Jung}, {and} \bibinfo{person}{Young-Ho Kim}.}
  \bibinfo{year}{2024}\natexlab{}.
\newblock \showarticletitle{Leveraging Large Language Models to Power Chatbots
  for Collecting User Self-Reported Data}.
\newblock \bibinfo{journal}{\emph{Proc. ACM Hum.-Comput. Interact.}}
  \bibinfo{volume}{8}, \bibinfo{number}{CSCW1}, Article \bibinfo{articleno}{87}
  (\bibinfo{date}{apr} \bibinfo{year}{2024}), \bibinfo{numpages}{35}~pages.
\newblock
\urldef\tempurl%
\url{https://doi.org/10.1145/3637364}
\showDOI{\tempurl}


\bibitem[Xu et~al\mbox{.}(2021)]%
        {xu_beyond_2021}
\bibfield{author}{\bibinfo{person}{Jing Xu}, \bibinfo{person}{Arthur Szlam},
  {and} \bibinfo{person}{Jason Weston}.} \bibinfo{year}{2021}\natexlab{}.
\newblock \bibinfo{title}{Beyond {Goldfish} {Memory}: {Long}-{Term}
  {Open}-{Domain} {Conversation}}.
\newblock
\newblock
\urldef\tempurl%
\url{http://arxiv.org/abs/2107.07567}
\showURL{%
\tempurl}
\newblock
\shownote{arXiv:2107.07567 [cs]}.


\bibitem[Xu et~al\mbox{.}(2022)]%
        {xu_long_2022}
\bibfield{author}{\bibinfo{person}{Xinchao Xu}, \bibinfo{person}{Zhibin Gou},
  \bibinfo{person}{Wenquan Wu}, \bibinfo{person}{Zheng-Yu Niu},
  \bibinfo{person}{Hua Wu}, \bibinfo{person}{Haifeng Wang}, {and}
  \bibinfo{person}{Shihang Wang}.} \bibinfo{year}{2022}\natexlab{}.
\newblock \showarticletitle{Long {Time} {No} {See}! {Open}-{Domain}
  {Conversation} with {Long}-{Term} {Persona} {Memory}}. In
  \bibinfo{booktitle}{\emph{Findings of the {Association} for {Computational}
  {Linguistics}: {ACL} 2022}}. \bibinfo{publisher}{Association for
  Computational Linguistics}, \bibinfo{address}{Dublin, Ireland},
  \bibinfo{pages}{2639--2650}.
\newblock
\urldef\tempurl%
\url{https://doi.org/10.18653/v1/2022.findings-acl.207}
\showDOI{\tempurl}


\bibitem[Yadav et~al\mbox{.}(2019)]%
        {yadav2019feedpal}
\bibfield{author}{\bibinfo{person}{Deepika Yadav}, \bibinfo{person}{Prerna
  Malik}, \bibinfo{person}{Kirti Dabas}, {and} \bibinfo{person}{Pushpendra
  Singh}.} \bibinfo{year}{2019}\natexlab{}.
\newblock \showarticletitle{Feedpal: Understanding Opportunities for Chatbots
  in Breastfeeding Education of Women in India}.
\newblock \bibinfo{journal}{\emph{Proc. ACM Hum.-Comput. Interact.}}
  \bibinfo{volume}{3}, \bibinfo{number}{CSCW}, Article \bibinfo{articleno}{170}
  (\bibinfo{date}{nov} \bibinfo{year}{2019}), \bibinfo{numpages}{30}~pages.
\newblock
\urldef\tempurl%
\url{https://doi.org/10.1145/3359272}
\showDOI{\tempurl}


\bibitem[Yonhap(2017)]%
        {solitaryDeathNews}
\bibfield{author}{\bibinfo{person}{Yonhap}.} \bibinfo{year}{2017}\natexlab{}.
\newblock \bibinfo{title}{{Lonely deaths of middle-aged, youth brackets stand
  out amid single-person households - The Korea Herald}}.
\newblock
\newblock
\urldef\tempurl%
\url{https://www.koreaherald.com/view.php?ud=20171207000623}
\showURL{%
Retrieved Sep 14, 2023 from \tempurl}


\bibitem[Zhao et~al\mbox{.}(2023)]%
        {zhao2023surveyLLM}
\bibfield{author}{\bibinfo{person}{Wayne~Xin Zhao}, \bibinfo{person}{Kun Zhou},
  \bibinfo{person}{Junyi Li}, \bibinfo{person}{Tianyi Tang},
  \bibinfo{person}{Xiaolei Wang}, \bibinfo{person}{Yupeng Hou},
  \bibinfo{person}{Yingqian Min}, \bibinfo{person}{Beichen Zhang},
  \bibinfo{person}{Junjie Zhang}, \bibinfo{person}{Zican Dong},
  \bibinfo{person}{Yifan Du}, \bibinfo{person}{Chen Yang},
  \bibinfo{person}{Yushuo Chen}, \bibinfo{person}{Zhipeng Chen},
  \bibinfo{person}{Jinhao Jiang}, \bibinfo{person}{Ruiyang Ren},
  \bibinfo{person}{Yifan Li}, \bibinfo{person}{Xinyu Tang},
  \bibinfo{person}{Zikang Liu}, \bibinfo{person}{Peiyu Liu},
  \bibinfo{person}{Jian-Yun Nie}, {and} \bibinfo{person}{Ji-Rong Wen}.}
  \bibinfo{year}{2023}\natexlab{}.
\newblock \bibinfo{title}{A Survey of Large Language Models}.
\newblock
\newblock
\showeprint[arxiv]{2303.18223}~[cs.CL]


\bibitem[Zheng et~al\mbox{.}(2023)]%
        {zheng2023building}
\bibfield{author}{\bibinfo{person}{Zhonghua Zheng}, \bibinfo{person}{Lizi
  Liao}, \bibinfo{person}{Yang Deng}, {and} \bibinfo{person}{Liqiang Nie}.}
  \bibinfo{year}{2023}\natexlab{}.
\newblock \bibinfo{title}{Building Emotional Support Chatbots in the Era of
  LLMs}.
\newblock
\newblock
\showeprint[arxiv]{2308.11584}~[cs.CL]


\bibitem[Zhong et~al\mbox{.}(2023)]%
        {zhong2023memorybank}
\bibfield{author}{\bibinfo{person}{Wanjun Zhong}, \bibinfo{person}{Lianghong
  Guo}, \bibinfo{person}{Qiqi Gao}, {and} \bibinfo{person}{Yanlin Wang}.}
  \bibinfo{year}{2023}\natexlab{}.
\newblock \showarticletitle{MemoryBank: Enhancing Large Language Models with
  Long-Term Memory}.
\newblock \bibinfo{journal}{\emph{arXiv preprint arXiv:2305.10250}}
  (\bibinfo{year}{2023}).
\newblock


\end{thebibliography}
